\documentclass{article}
\usepackage[utf8]{inputenc}
\usepackage{arxiv}
\usepackage{amsmath,amsfonts, mathtools}
\usepackage{enumitem}
\usepackage{url}
\usepackage{blkarray}
\usepackage{graphicx}
\usepackage{caption}
\usepackage{subcaption}
\usepackage{algorithm}
\usepackage{algpseudocode}
\usepackage{booktabs}
\usepackage{lscape}
\usepackage{longtable}
\usepackage{array}
\usepackage{authblk}
\usepackage[titletoc,title]{appendix}
\usepackage{lscape}

\newcolumntype{L}{>{\centering\arraybackslash}m{1.5cm}}
\newcolumntype{N}{>{\centering\arraybackslash}m{2.5cm}}
\newcolumntype{M}{>{\centering\arraybackslash}m{3.5cm}}

\usepackage[hidelinks]{hyperref}

\algnewcommand{\IfThenElse}[3]{% \IfThenElse{<if>}{<then>}{<else>}
  \State \algorithmicif\ #1\ \algorithmicthen\ #2\ \algorithmicelse\ #3}

\usepackage[sorting=none]{biblatex}
\providecommand{\keywords}[1]
{
  \small	
  \textbf{\textit{Keywords---}} #1
}

\title{Metareview--informed Explainable Cytokine Storm Detection during CAR--T cell Therapy}

\author[ ,1,2]{Alex Bogatu\thanks{Corresponding author: alex.bogatu@manchester.ac.uk}}
\author[ ,2]{Magdalena Wysocka\thanks{The first two authors contributed equally.}}
\author[1,2]{Oskar Wysocki}
\author[3]{Holly Butterworth}
\author[2]{D\'{o}nal Landers}
\author[3]{Elaine Kilgour}
\author[1,2,4]{Andr\'{e} Freitas}

\affil[1]{Department of Computer Science, The University of Manchester}
\affil[2]{digital Experimental Cancer Medicine Team, Cancer Biomarker Centre,
CRUK Manchester Institute, University of Manchester}
\affil[3]{Cancer Biomarker Centre,
CRUK Manchester Institute, University of Manchester}
\affil[4]{Idiap Research Institute, Switzerland}
\date{} % clear date
\begin{document}

\maketitle

\begin{abstract}

Cytokine release syndrome (CRS), also known as cytokine storm, is one of the most consequential adverse effects of chimeric antigen receptor therapies that have shown promising results in cancer treatment. When emerging, CRS could be identified by the analysis of specific cytokine and chemokine profiles that tend to exhibit similarities across patients. In this paper, we exploit these similarities using machine learning algorithms and set out to pioneer a meta--review informed method for the identification of CRS based on specific cytokine peak concentrations and evidence from previous clinical studies. We argue that such methods could support clinicians in analyzing suspect cytokine profiles by matching them against CRS knowledge from past clinical studies, with the ultimate aim of swift CRS diagnosis. During evaluation with real--world CRS clinical data, we emphasize the potential of our proposed method of producing interpretable results, in addition to being effective in identifying the onset of cytokine storm.

\end{abstract}


\section{Introduction} 

Scientific advances in cancer immunology, genetic engineering and cell manufacturing have recently resulted in a paradigm shift in the field of cancer treatment. In addition to traditional anticancer agents, patient-specific cell immunotherapies based on genetically engineered T cells have emerged \cite{brentjensCD19TargetedCellsRapidly2013, gruppChimericAntigenReceptor2013, maudeChimericAntigenReceptor2014, kochenderferChemotherapyRefractoryDiffuseLarge2015, davilaEfficacyToxicityManagement2014}. For example, the use of chimeric antigen receptors (CARs) targeting low-risk malignant diseases have showed promising therapeutic potential. \cite{juneChimericAntigenReceptor2018}. 
However, despite the clinical benefits observed in many patients, the use of CAR--T cells may lead to severe adverse events that are directly related to the induction of strong immune system effector responses. The toxicity can range from minor, such as fever or fatigue, to life-threatening, such as shock or dysfunction of major organ systems. \cite{morrisCytokineReleaseSyndrome2022}. 
%Significant immune-mediated toxicities such as cytokine release syndrome (CRS) and immune effector cell-associated neurotoxicity syndrome (ICANS; commonly referred as neurotoxicity) are observed in more than a third and half of patients with B cell acute lymphoblastic leukaemia, respectively \cite{pennisiComparingCARTcell2020}. 

Cytokine release syndrome (CRS) has been shown to be the most significant adverse event of T cell-engaging therapies, mainly CD19-targeted CAR--T cells \cite{morganCaseReportSerious2010, porterChimericAntigenReceptor2011, brudnoToxicitiesChimericAntigen2016, xiaoMechanismsCytokineRelease2021}. In these cases, CRS has been reported with a frequency of up to 100\% after infusion \cite{liuNovelDominantnegativePD12021, sangPhaseIITrial2020}, while up to 67\% of patients developed severe CRS \cite{fryCD22targetedCARCells2018, huPotentAntileukemiaActivities2017, porterChimericAntigenReceptor2015}. More specific, the interaction between CAR--T cells and tumour cells activates the release of host cells, especially macrophages, by distorting the cytokine network. The released cytokines then induce activation of endothelial cells, contributing to the constitutional symptoms associated with CRS \cite{morrisCytokineReleaseSyndrome2022, xiaoMechanismsCytokineRelease2021}. 
% CRS usually manifests with fever and systemic symptoms within a few days after CAR--T cells infusion. In severe cases of CRS, other features of a systemic inflammatory response such as tachycardia, hypotension or organ dysfunction are observed. 
%The onset of neurotoxicity (usually manifested by encephalopathy, aphasia followed by seizures) may be delayed and may or may not be preceded by clinically significant CRS. In most patients, both CRS and ICANS are reversible without persistent neurological deficits if symptoms and signs are quickly recognized and treated. 

In most patients, CRS is reversible without persistent neurological deficits if symptoms and signs are quickly recognized and treated. For example, a meta-analysis of 2,592 patients from 84 eligible studies revealed that CRS mortality was less than 1\% \cite{leiTreatmentRelatedAdverseEvents2021}. Furthermore, although CRS may develop at different times depending on the CAR--Type, the cytokine profiles observed in the patient's serum are often similar in terms of peak cytokine and chemokine levels \cite{teacheyIdentificationPredictiveBiomarkers2016, giavridisCARCellInduced2018, shimabukuro-vornhagenCytokineReleaseSyndrome2018, brudnoToxicitiesChimericAntigen2016, oluwoleBedsideClinicalReview2016}. This uncovers an opportunity to employ computerised semi-automatic methods to detect CRS and monitor its progression based on specific biomarkers. Such methods could decisively contribute to the safety of current CAR--based treatments and improve overall survival, especially given that most patients treated with CD19 CAR--T cells develop some level of CRS \cite{pennisiComparingCARTcell2020}. In this paper we aim to seize this opportunity and pioneer a framework for explainability enablement and knowledge integration for machine learning (ML) algorithms that could assist clinicians in identifying patients that develop cytokine storm scenarios and in making appropriate adaptive decisions. The proposed methods combine active monitoring of specific cytokine biomarkers with a systematic integration of evidence available from previous clinical studies regarding the concentration levels of the said cytokines in CRS patients. During evaluation, the proposed methods have proven effective achieving an accuracy of up to 94\% when evaluated with real--world clinical data. Furthermore, while CRS is the clinical focus in this paper, we argue that the proposed framework could be adapted to other clinical scenarios as well. To the best of our knowledge this is the first explainable and knowledge-based CRS prediction method.

% Therefore, the understanding of CRS could have broad implications for other cytokine-mediated systemic inflammatory diseases. 
% Sepsis often resembles CRS and is the cause of mortality in this patient population \cite{singerThirdInternationalConsensus2016, shimabukuro-vornhagenCytokineReleaseSyndrome2018}. However the basis of treatment for both diseases is different.

% The overview of the proposed model is depicted Fig. \ref{fig:single_prediction_diagram}.

% \begin{figure}[htb!]
% \centering
% \includegraphics[width= .9\textwidth]{figures/single_prediction_diagram.pdf}
% \caption{metareview--informed model and the process of generation of a prediction for a single input data point $x$}
% \label{fig:single_prediction_diagram}
% \end{figure}

\section{Related work} 

The cytokine levels collected during clinical trials were used in the several studies of dynamic interactions of cytokines. Yiu et al. \cite{yiuDynamicsCytokineStorm2012} used the cytokine levels of six subjects which experienced a `cytokine storm'. Their study was not predictive in nature but analytic with respect to the response histories of cytokines modeled by a set of time-invariant linear ordinary differential equations that illustrate time-dependent coupled interactions among the studies cytokines. Other studies, such as Teachey at el. \cite{teacheyIdentificationPredictiveBiomarkers2016}, proposed ML--based predictive models to detect early CRS. Using regression modeling and decision trees they predicted which patients would develop severe CRS with a signature composed of three cytokines. They developed 16 predictive models using different combinations of cohorts. However, they did not model for an adults only cohort. For a pediatric cohort, they accurately predicted CRS based on model including IFN--$\gamma$, IL13 and MIP1a with sensitivity 100\% and specificity 96\%. In turn, Hay et al. \cite{hayKineticsBiomarkersSevere2017} modeled the relationship between the peak CAR--T--cell counts in blood and the occurrence of toxicity or disease response by using classification--tree modeling to design an algorithm to predict grade $\geq4$ CRS. They also scored high for a model including a single serum cytokine concentration of MCP--1 measured in the subset of patients with fever within a certain period of time to detection (sensitivity 100\%; specificity 95\%).

\section{An overview of data and methods} \label{sec:overview}

The proposed method in this paper has been conducted in the context of work supported by the Innovate Manchester Advanced Therapy Centre Hub \textit{iMATCH} programme\footnote{\url{https://www.theattcnetwork.co.uk/centres/imatch}}. For the construction of the proposed CRS prediction methods, in this paper, we rely on data from patients with refractory/relapsed hematological malignancies that underwent chemotherapy followed by CD19 CAR--T cell treatment at The Christie NHS Foundation Trust. Cases of CRS were identified by specialists based on version 5.0 of the National Cancer Institute Common Terminology Criteria for Adverse Events, which defines CRS as a clinical disorder involving fever, tachypnea, headache, tachycardia, hypotension, rash, and/or hypoxia caused by the release of cytokines \cite{leeCurrentconcepts2014}. The study was approved by the institutional review board of ethics commission and was conducted in accordance with the Good Clinical Practice guidelines. All patients provided a written informed consent. This paper uses clinical and laboratory data from 9 such advanced cell therapy patients.

%25 advanced cell therapy patients and 12 sepsis patients consecutively treated in the study. %The cytokine markers values from 9 patients from ATTACK cohort provided validation cohort for our models. 

% \subsection{}{Clinical laboratory parameters and serum biomarkers}

For each patient, data was collected regarding their epidemiological, clinical, laboratory, and treatment characteristics. The clinical laboratory variables included complete blood count, serum biochemical test results, coagulation profile, ferritin, CRP, and immunological test results, including serum cytokines, for up to 30 days post infusion. In our predictive analysis we rely on the serum cytokine concentrations as these biomarkers have been suggested to have CRS predictive power, as described in the next section, while other clinical features tend to be reactions to CRS onset. To this end, blood was collected at up to 17 time points from the 9 advanced cell therapy patients for analyses of plasma cytokine concentrations. The total number of samples was 102. Then, a 14 cytokine enzyme-linked immunosorbent assay (SP-X ELISA, Quanterix (formerly Aushon)) panel, validated to GCP, was established for rapid identification of changes in cytokine levels associated with cytokine storms. Four ELISA assays for the detection of 14 cytokines were validated (Table \ref{tab:ELISAassays}). A detailed description of the laboratory tests and cytokines is included in Supplemental Methods.  Finally, the proposed methods rely on the predictive power of ten cytokines and chemokines: \textit{Interleukin 2} (IL2), \textit{Interleukin 4} (IL4), \textit{Interleukin 6} (IL6), \textit{Interleukin 8} (IL8), \textit{Interleukin 10} (IL10), \textit{Interleukin 15} (IL15), \textit{Interleukin-2 receptor alpha} (IL2R$\alpha$), \textit{Tumour necrosis factor alpha} (TNF--$\alpha$), \textit{Interferon gamma} (IFN--$\gamma$), and \textit{Granulocyte-macrophage colony-stimulating factor} (GMCSF). These choices were motivated by the prevalence of the selected cytokines in CRS--relevant studies, as we now describe.

% \textbf{Search Strategy}

\subsection{Motivation of selected biomarkers}

Recent studies have identified several biomarkers that can predict the development of CRS following CAR--T cell therapy \cite{teacheyIdentificationPredictiveBiomarkers2016, davilaEfficacyToxicityManagement2014, shimabukuro-vornhagenCytokineReleaseSyndrome2018, maudeManagingCytokineRelease2014}. Consequently, at the core of our CRS predictive framework we use a group of the most studied cytokines to train an ML algorithm to determine the likelyhood of CRS onset based on their concentration levels. This cytokine group included effector cytokines such as IL2 and IL6, IFN--$\gamma$, and GMCSF, but also cytokines secreted by monocytes and / or macrophages, such as IL8 (chemokine), IL10, and TNF--$\alpha$ \cite{xuCytokineReleaseSyndrome2014, norelliMonocytederivedIL1IL62018}. IL6 is a core cytokine in CRS pathophysiology, which enhances T cell proliferation and B cell differentiation \cite{norelliMonocytederivedIL1IL62018, liuCytokineReleaseSyndrome2018}. Similarly, IFN--$\gamma$, secreted by activated T cells and tumor cells, is considered a strong contributor to CRS development and plays a key role in mobilizing CRS after CAR--T cell infusion. IFN--$\gamma$ also stimulates other immune cells, especially macrophages, which secrete proinflammatory cytokines, such as IL6, IL8, IL15, and TNF--$\alpha$ \cite{suInterferongRegulatesCellular2015, maudeManagingCytokineRelease2014}. The levels of homeostatic cytokines such as IL2 and IL15 may increase after conditioning therapy, which is administered prior to infusion of CD19--targeted CAR--T cells. In turn, a further increase in the level of cytokines is observed after infusion of CAR--T lymphocytes \cite{morrisCytokineReleaseSyndrome2022}. Therefore, hemostatic cytokines such as IL2 and IL15 were also included in the model. As there is no consensus on the contribution of serum biochemical parameters (e.g., C--reactive protein (CRP) or ferritin), to the prediction of CRS severity, the algorithm was limited to the inflammatory cytokines mentioned above \cite{porterChimericAntigenReceptor2015, leeCurrentconcepts2014, davilaEfficacyToxicityManagement2014, teacheyIdentificationPredictiveBiomarkers2016, huPotentAntileukemiaActivities2017}.

\subsection{Metareview--informed prediction}

Often, the type of data measurements such as the ones described above are challenging to obtain and researchers investigating applications of ML algorithms to diagnosis and treatment are faced with a data scarcity problem. In this paper, we set out to address this by incorporating the (potentially vast) space of existing relevant studies on CRS diagnosis and treatment in the decision making process of our proposed method. To this end, we started by searching several electronic bibliographic databases (e.g., PubMed, the Cochrane Database of Systematic Reviews, Cochrane Central Register of Controlled Trials and Web of Science, etc.) for relevant studies published between Jan 1, 2010, and March 1, 2022. We used the following Mesh terms and Entry terms, `$\textsf{car-t}$' or `$\textsf{CART}$' or `$\textsf{chimeric antigen receptor}$', and `$\textsf{cytokine release syndrome}$' or `$\textsf{CRS}$'. The same search was repeated just before the final experimental analysis for completeness.

\begin{figure}[t]
\centering
\includegraphics[width= .7\textwidth]{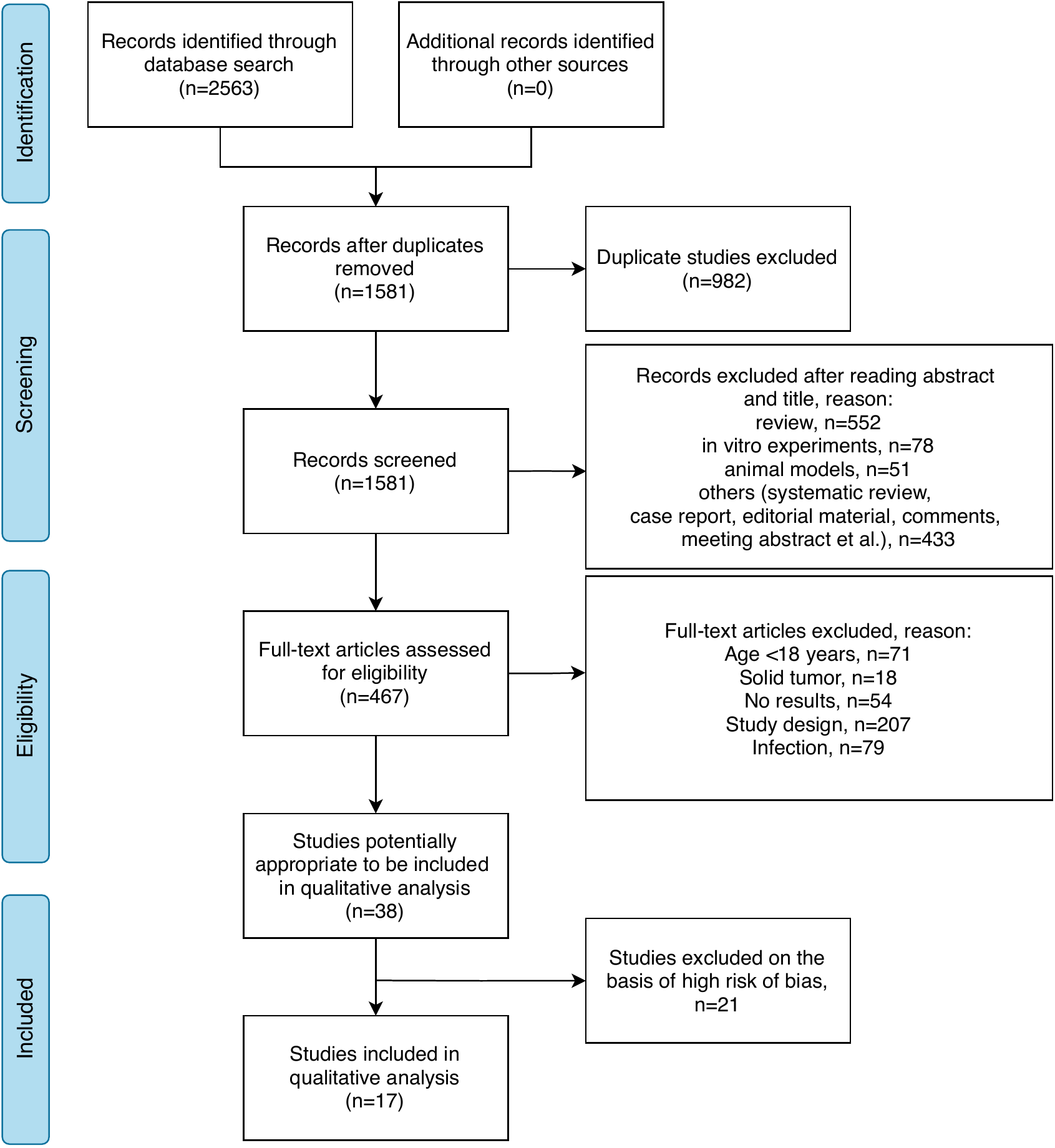}
\caption{The workflow summarizing screening and selection of studies included in the knowledge base.}
\label{fig:Workflow_v3}
\end{figure}

From the search results, we selected CRS--focused clinical trials that reported clinical data on the cytokine and chemokine levels of adult patients (i.e., age $\geq18$ years) with relapsed or refractory haematological malignancies that were treated with CAR--T--cell therapy. We extracted values for peak blood levels of cytokines and chemokines associated with CRS of any grade at any time after CAR--T--cell infusion reported in the selected studies. 
% Both single-centre and multi-institutional trials were included. The most recently updated results of each study were analysed, either from published articles. Furthermore, we also searched the reference lists of published trials and the relevant review articles. At last we only included the clinical trials of CAR-T cells in adult patients with haematological malignancies that reported CRS. 
We excluded studies published in languages other than English, studies with insufficient data (i.e., studies where cytokine levels after CAR--T--cell therapy were not reported, irrelevant studies, or where full texts were not available), and studies on fewer than three patients. Similarly, all pre--clinical studies, review articles, meta-analyses or studies performed on animals and cell lines were excluded. The selection criteria resulted in seventeen studies\footnote{Each of the selected studies is from the existing state of the art and not performed by any of the authors.} listed in Table \ref{tab:StudiesCharacteristics}. An overview of the workflow leading to the selection of these seventeen studies is depicted in the Fig \ref{fig:Workflow_v3}.

From the selected studies we manually curated statistical information about the observed post--infusion peak concentration values of the studied cytokines and chemokines, such as mean, median, range, inter--quartile--range(IQR) and study population size. We focused on the reported peak values during the first month after infusion. Collectively, we refer to the extracted cytokine statistics as the \textit{Knowledge Base} (KB) and we formally define it in Section \ref{sec:method}.

\subsection{Explainable prediction}

% Summary of results - to be filled in later.

The integration of literature knowledge introduced above can offer a distinctive multi--perspective view in the CRS diagnosis process, as we further describe in Section \ref{subsec: prediction}. Additionally, it enables an explanatory dimension for the ML algorithms underlying the predictive process. This is valuable because such a dimension can attenuate the reluctance of adopting ML--based methods in practice often exacerbated by their `black--box nature". Specifically, our proposed framework uses the integrated studies to compute a plausible explanation for a given predictive decision (i.e., CRS/NO CRS) using an abductive reasoning \cite{lipton2003inference} process based on first order logic. We describe this process in detail in Section \ref{subsec:expl}.

In this paper we, therefore, pioneer a practical framework for ML--based CRS diagnosis characterized by three main contributions:
\begin{itemize}
% [leftmargin=*]
    \item A \textit{metareview--informed} diagnosis process that maximizes the use of external available information, in the form of a Knowledge Base (KB) and in addition to observed patient data, with a view to overcome the data scarcity problem often present in healthcare predictive scenarios.
    \item An ML--based \textit{multi--perspective} diagnosis process that integrates the information from the above--mentioned KB into various machine learning algorithms (i.e., the framework is flexible and not bound to one particular algorithm) to offer predictions with a diversity of viewpoints.
    \item \textit{Abductive} diagnosis process that backtracks a give prediction to the KB elements that support or refute it with a view to making the prediction result explainable.
\end{itemize}
In addition, our proposed system can easily be applied to a multitude of diagnosis scenarios when both external and case--specific data is available.

For the rest of this paper we refer to our proposed framework as \textit{M2--CRS} for \textbf{M}etareview--informed \textbf{M}ulti--perspective CRS prediction.

\section{Method} \label{sec:method} 

Cytokine release syndrome (CRS) is the most common side effect associated with CAR--T cell therapy. The severity of the CRS, which is determined based on clinical symptoms, vital signs, and organ dysfunction, influences the management of CRS. However, the variability of general symptoms makes them not ideal candidates for an accurate CRS assessment. Therefore, specific biomarkers are needed to closely monitor patients at risk and receive timely prophylactic treatment. We have briefly described the specific biomarkers we consider in this paper and the methods by which their concentrations have been obtained from real--world patients in Section \ref{sec:overview}. However, there is a recurrent limitation in applying machine learning or even statistical methods in the filed of medicine and diagnostics when it comes to data availability - even more so in the rare context of CRS as a CAR--T therapy side effect. So it is impractical to satisfy the requirements of many ML algorithms in terms of size of exemplar/training data (typically $>$10 or $>$15 events per variable \cite{vanderploegModernModellingTechniques2014}). An immediate consequence of this data scarcity problem is that predictive algorithms cannot achieve the reach of generality and intepretability required by the field of diagnostics. 

Returning to our \textit{iMATCH} research, introduced in Section \ref{sec:overview}, we have extracted a total of 102 samples from 9 patients at 17 different time points. Our core goal here is to use these measurements to train ML--algorithms so that clinicians can be machine--assisted in diagnosing future occurrences of CRS. Due to reasons discussed above, these available samples were insufficient to enable many of the selected algorithms to converge, as we further show in Section \ref{sec:eval}. To mitigate this challenge, in this section we define a knowledge base (KB) of statistical biomedical facts extracted from the selected studies described in Figure \ref{fig:Workflow_v3}. We use this KB to augment our 102 samples of real--world measurements.

\subsection{KB formal definition} \label{subsec:kb_def} 

We collectively refer to the collection of predictive biomarkers used in our proposed method as \textit{biomarkers}. We denote each one generically as $b$, its measured value as $\dot{b}$, and all biomarkers collectively as $B$. More formally:
% Similarly, we denote each study--specific parameter individually as $p$, its value as $\dot{p}$ and all parameters collectively as $P$. 

\begin{equation} \label{eq:KB}
\begin{aligned}
    KB &\coloneqq B \rightarrow \mathcal{S} \\
    \mathcal{S} &: S \rightarrow Q^{\perp} \\
    Q^{\perp} &= Q \cup \{\perp\}; Q_j = \bigcup_{i=1,k}(\dot{med}_i^j, \dot{min}_i^j, \dot{max}_i^j)
\end{aligned}
\end{equation}

Intuitively, $KB$ is defined as a mapping from the collection of biomarkers $B$ to the studies where summary statistics about the observed concentration values of these biomarkers have been reported. Similarly, the right--hand--side of this mapping (i.e., $\mathcal{S}$) is itself a mapping from a set of study identifiers $S$ (e.g., the study names) to the summary statistics each study reports. Here, by summary statistics we mean median, minimum, and maximum (or some approximated values thereof, such as mean instead of median and inter--quartile range instead of range) of peak biomarker concentrations observed in the context of CRS, computed over the measurements of the participants to the reporting study. Concretely, in Equation \ref{eq:KB}, $(\dot{med}_i^j, \dot{min}_i^j, \dot{max}_i^j)$ denotes the triple of summary statistics reported in study $S_j$ for biomarker $b_i$.

\subsection{KB integration} \label{subsec:kb_integ}

Given the extracted $KB$ defined in the previous subsection, the goal of our knowledge integration process is twofold: (i) to extend the reach of CRS predictive algorithms beyond the level of generalization given by often scarce training/exemplar data, and (ii) to enable the explainability of CRS decisions in the form of evidence--based abductive reasoning, i.e., inference to the best available explanation.

% However, with respect to (ii) above, the infusion of external domain knowledge in the form of $KB$ opens the possibility for explainable predictions with reference to $KB$.

In practice, given $KB$ as defined by Equations \ref{eq:KB}, and $k$ biomarker concentration values for a new subject denoted as a vector $\Vec{M} = [\dot{b}_1, \ldots \dot{b}_k]$, the integration of knowledge from $KB$ could be performed through a \textit{multi--perspective scaling of each biomarker value in $\Vec{M}$ with respect to each study in $KB$}. This scaling is grounded on a function $f(\dot{b}_i, Q_j)$ that quantifies the degree of similarity between the given measurement value of $b_i$ and the statistical values reported in a study $S_j$. Therefore, the collection of measurements of a new subject could be extended from a vectorized view (i.e., $\Vec{M}$), to a matrix view of biomarker value similarities with respect to $KB$. This extension could result from Algorithm \ref{alg:scale} with its result summarized by Equation \ref{eq:matrix}.

\begin{algorithm}[t]
	\caption{$KB$ integration}
	\begin{flushleft}
	\textbf{Input}: Knowledge base $KB$, $\Vec{M} = [\dot{b}_1, \ldots \dot{b}_k]$, a scaling function $f$. \\
 	\textbf{Output}: a matrix representation of $\Vec{M}$.
	\end{flushleft}
	\begin{algorithmic}[1]
		\Function{ScaleBiomarkers}{}
		\State $\mathbb{M} \gets [[\ ]]$ // An empty matrix
		\ForAll{$\dot{b}_i \in \Vec{M}\ \&\&\ (b_i, \mathcal{S}_i) \in KB\ \&\&\ i \in \{1 \ldots k\}$}
		    \ForAll{$(S_j, Q_j) \in \mathcal{S}_i\ \&\&\ j \in \{1 \ldots e\}$}
		        \State $\mathbb{M}[j][i] \gets f(\dot{b}_i, Q_j)$
		    \EndFor
		\EndFor
		 \State \Return $\mathbb{M}$
		\EndFunction
	\end{algorithmic}
\label{alg:scale}
\end{algorithm}

\begin{equation} \label{eq:matrix}
    \mathbb{M} =
    \bordermatrix{ & b_1 & b_2 & \ldots & b_k \cr
        S_1 & f(\dot{b}_1, Q_1) & f(\dot{b}_2, Q_1) & \ldots & f(\dot{b}_k, Q_1) \cr
        S_2 & f(\dot{b}_1, Q_2) & f(\dot{b}_2, Q_2) & \ldots & f(\dot{b}_k, Q_2) \cr
        \vdots & \vdots  & \vdots & \vdots & \vdots \cr
        S_e & f(\dot{b}_1, Q_e) & f(\dot{b}_2, Q_e) & \ldots & f(\dot{b}_k, Q_e) \cr 
        } \qquad
\end{equation}

Intuitively, assuming $e$ studies in $S$, each observed biomarker measurement is expanded to $e$ normalized measurements, one for each study in $S$. This is where we denote the potential for our proposed $KB$ integration to overcome the data scarcity problem. For example, in our case, the 109 biomarker concentration measurements are expanded to 1853 samples through normalization with respect to the 17 studies from our $KB$. Additionally, when trained on the new extended dataset of biomarker measurements, ML algorithms will model the biomarker correlations not only within the space defined by the available data, but relative to past discoveries as well. In other words, $KB$ integration could enable ML algorithms to capture interactions between biomarkers insufficiently exhibited otherwise.

In Equation \ref{eq:matrix} the normalization of biomarker measurements is performed through a function $f$. We now explore various choices for $f$.

\subsubsection{Binary scaling} \label{subsubsec:binary}

Given a new value measurement for a biomarker $b_i$, the binary scaling variant is built on a simple definition of $f$ as: 
\begin{equation} \label{eq:binary}
    f(\dot{b}_i, Q_j) = \begin{cases}
        1 & \text{$\dot{min}_i^j \leq \dot{b}_i \leq \dot{max}_i^j$ or $\dot{b}_i > \dot{max}_i^j$}\\
        0 & \text{otherwise}
    \end{cases}
\end{equation}
where $\dot{min}_i^j$ and $\dot{max}_i^j$ are the minimum and maximum peak concentration values, resp., reported for biomarker $b_i$ in study $S_j$ during CRS. In practice, observed values for some biomarker $b_i$ similar to the abnormal concentrations reported during CRS in study $S_j$ will be normalized to $1$ and to $0$ otherwise.

\subsubsection{Continuous scaling} \label{subsubsec:cont}

Given a new value measurement for a biomarker $b_i$, the continuous scaling variant is built on a simple definition of $f$ as: 

\begin{equation} \label{eq:continuous}
    f(\dot{b}_i, Q_j) = \begin{cases}
        0 & \text{$\dot{b}_i \leq \dot{min}_i^j$} \\
        \frac{\dot{b}_i - \dot{min}_i^j}{\dot{max}_i^j-\dot{min}_i^j} & \text{$\dot{min}_i^j \leq \dot{b}_i \leq \dot{max}_i^j$} \\
        1 & \text{$\dot{b}_i \geq \dot{max}_i^j$}
    \end{cases}
\end{equation}

% \begin{equation} \label{eq:continuous}
    % f(\dot{b}_i, Q_j) = \frac{(\dot{b}_i - min(\dot{b}_i, \dot{min}_i^j))}{(max(\dot{b}_i, \dot{max}_i^j) - min(\dot{b}_i, \dot{min}_i^j))}
% \end{equation}

Equation \ref{eq:continuous} defines a min--max normalization process adjusted for out--of--range values. In practice, observed values for some biomarker $b_i$ similar to the abnormal concentrations reported during CRS in study $S_j$ will be normalized closer to $1$ and closer to $0$ otherwise. This normalization variant can be further reduced to the binary model by configuring value thresholds for each $b_i$. However, deciding on such thresholds is often hard and requires expert--level domain knowledge. As such, we do not rely on such thresholds in our method.

\subsubsection{Probabilistic scaling} \label{subsubsec:prob}

This variant relies on the assumption that each abnormal measurement value of $b_i$ that could signal CRS \textit{is drawn from some arbitrary probability distribution that can be approximated by a mixture of Gaussian distributions}. Concretely, given a biomarker $b_i$ and studies $S_1 \ldots S_e$ that analyze the values of $b_i$ in the context of CRS, the probabilistic variant considers a new measured value of $b_i$ as drawn from a Gaussian mixture model with at least $e$ components: $(\mathcal{N}(\dot{\mu}_i^1, \dot{\sigma}_i^1), \ldots \mathcal{N}(\dot{\mu}_i^e, \dot{\sigma}_i^e))$, i.e., each study that analyzes $b_i$ is used to synthesize the parameters of a Gaussian component in the resulting mixture. Using the formulas proposed by Hozo \textit{et al.} \cite{hozo2005estimating} we can approximate the values of each $\dot{\mu}_i^j$ and $\dot{\sigma}_i^j$ from their corresponding summary statistics $Q_i^j$. Then the probabilistic variant is built on a simple definition of $f$ as:

\begin{equation} \label{eq:prob}
    f(\dot{b}_i, Q_j) = \textsc{CDF}(\dot{b}_i, \dot{\mu}_i^j, \dot{\sigma}_i^j) 
\end{equation}

$\textsc{CDF}$ denotes the cumulative distribution function of $\mathcal{N}(\dot{\mu}_i^j, \dot{\sigma}_i^j)$ evaluated at $\dot{b}_i$. Intuitively, the closer a new concentration value $\dot{b}_i$ will be to the right hand side tail of $\mathcal{N}(\dot{\mu}_i^j, \dot{\sigma}_i^j)$, the higher the probability of $\dot{b}_i$ to signal CRS because there is a higher probability of the observed $\dot{b}_i$ to be similar to the abnormal biomarker value distribution derived from study $S_j$.

This variant can be further reduced to the binary model by configuring CDF probability thresholds for each $b_i$. Since this scaling variant is characterized by a probabilistic interpretation, in practice it is accessible to set informed threshold values, such as $0.5$, i.e, a probability of at least $50\%$. 

\textbf{Discussion}

\begin{figure}[t]
\centering
\includegraphics[width= .7\textwidth]{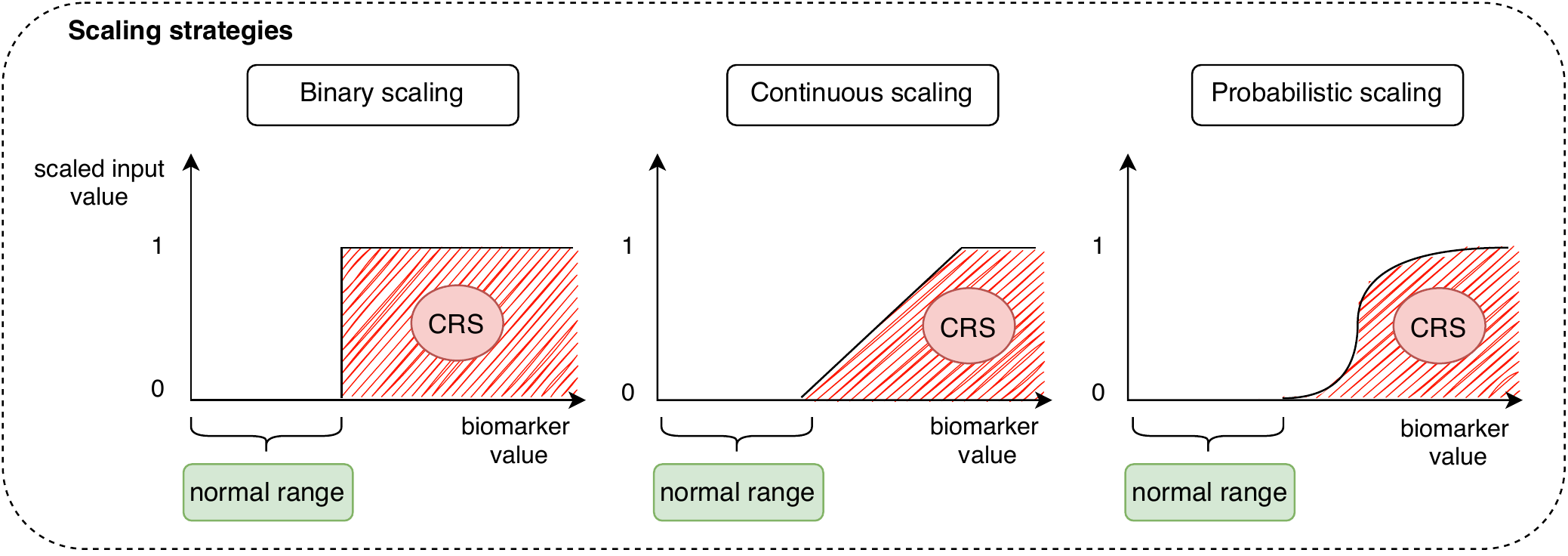}
\caption{Scaling visualization}
\label{fig:scaling_visualization}
\end{figure}

Having defined the $KB$ scaling alternatives in this section, Figure \ref{fig:scaling_visualization} depicts the overall distinction between them: the binary and continuous scaling approaches confer the biomarker representation space a linear interpretation, i.e., biomarker measurement values are \textit{linearly} combined to achieve a prediction, while the probabilistic scaling approach aims to combine the features in a non--linear, more expressive manner.

\subsubsection{KB missing value imputation}

Looking back at Equation \ref{eq:matrix}, the normalization process can uniformly be performed if and only if each biomarker $b_i$ is studied and reported in each $S_j$. However, in practice, not all studies include all biomarkers. Therefore, there is a need to impute missing values in our $KB$. During the application of \textit{M2--CRS} to CRS prediction we have used the following simple value imputation methods when statistics about some biomarker $b_i$ are missing from a study $S_j$:

\begin{itemize}
    \item $\dot{min}_i^j$ is given by \textit{the minimum of all \textbf{min} values reported by studies that include $b_i$}.
    \item $\dot{max}_i^j$ is given by \textit{the maximum of all \textbf{max} values reported by studies that include $b_i$}.
    \item $\dot{med}_i^j$ is given by \textit{the weighted average of all \textbf{med} values reported by studies that included $b_i$}, with the weights given by the numbers of participants to the respective study.
\end{itemize}

The technical evaluation described in Section \ref{sec:eval} has shown that with this simple imputation method our proposed multi--perspective data scaling leads to effective results for most of the tested ML algorithms.

\subsection{Multi--perspective CRS prediction strategies} \label{subsec: prediction}

Once each measurement of $b_i$ has been scaled using Algorithm \ref{alg:scale}, matrix $\mathbb{M}$ will consist of $e$ rows and $k$ columns with binary/continuous/probability values describing the potential for CRS signaled by each $b_i$ with respect to each study $S_j$. The resulting matrix, formally exemplified in Equation \ref{eq:matrix}, represents the underlying data structure of our multi--perspective approach to CRS prediction. In practice, given a collection of $k$ measurements for a patient, the resulting $e \times k$ matrix has to be reduced to a single predictive value, e.g., a probability of CRS/NO CRS. In describing this process, we start by considering the special case where  $e=1$, i.e., only one study in $KB$. Therefore, $\mathbb{M}$ is a vector, i.e., $\Vec{M}$, and the goal is to predict a CRS/NO CRS label for each new collection of measurements, $\Vec{M}$. This prediction process is governed by a function $g$, generically defined as $g(\Vec{M}) = \hat{q}$, with $\hat{q}$ a real value that quantifies the likelihood of CRS. 

The following are potential choices for $g$:

\begin{enumerate}
% [leftmargin=*,align=left]
    \item \textit{Data--driven prediction} - $g$ is initially unknown and synthesized through the use of an ML algorithm (e.g., random forests, linear regression, etc.) trained on a given dataset $\mathcal{D}$ consisting of exemplar $(\Vec{M}, q)$ pairs. The predicted value $\hat{q}$ can then be construed as a probability of CRS and, given a probability threshold (e.g., $50\%$), a CRS/NO CRS label can be associated.
    \item \textit{Aggregated prediction} - $g$ is a pre--defined aggregation function of $\Vec{M}$, e.g., average, min, max, and $\hat{q}$ is a real value result of applying $g$ on $\Vec{M}$. As already mentioned, setting thresholds in this case is more challenging and domain expertise may bee needed to interpret $\hat{q}$. We therefore focus on the data--driven prediction strategy and only mention the aggregated strategy as a potential alternative, albeit less clear on how to apply it in practice.
\end{enumerate}

Going back to the more general scenario where $e \geq 2$, i.e., there are at least two studies analyzing the $k$ biomarkers w.r.t. CRS in $KB$, consider the application of the data--drive single prediction approach (i.e., 1 from above) to each study, i.e., to each row of matrix $\mathbb{M}$ as illustrated below:

\[
\begin{blockarray}{cccccc} \label{eq:multiscore}
    & b_1 & b_2 & \ldots & b_k & g(\Vec{M_j}) \\
    \begin{block}{c(cccc) @{\hspace{20pt}} | @{\hspace{20pt}} [c]}
      S_1 & f(\dot{b}_1, Q_1) & f(\dot{b}_2, Q_1) & \ldots & f(\dot{b}_k, Q_1) & \hat{q_1} \\
      S_2 & f(\dot{b}_1, Q_2) & f(\dot{b}_2, Q_2) & \ldots & f(\dot{b}_k, Q_2) & \hat{q_2}\\
      \vdots & \vdots  & \vdots & \vdots & \vdots & \vdots \\
      S_e & f(\dot{b}_1, Q_e) & f(\dot{b}_2, Q_e) & \ldots & f(\dot{b}_k Q_e) & \hat{q_e}\\
    \end{block}
\end{blockarray}
\]

One immediate observation from the relation above is that different studies may disagree with respect to the prediction outcome. Therefore, a collective decision strategy is required in order to reason on CRS evidence originating from different, potentially conflicting studies. This is where we note, once again, the multi--perspective dimension of our proposed method. We observe that, in practice, the collective decision can be obtained by adopting various majority voting strategies specific to ensemble classification methods \cite{dietterich00}. Specifically, given $g(\Vec{M}) = [\hat{q}_1, \ldots \hat{q}_e]$, where $\hat{q}_e$ is a probability score returned by the ML model based on the inputs scaled according to study $S_j$, and $C(\hat{q}_j) = 
\begin{cases} 
0 & \hat{q}_j < 0.5 \\ 
1 & \hat{q}_j \geq 0.5 
\end{cases}$ a labeling function, we consider the following majority voting options: hard majority voting (HMV), weighted hard majority voting (wHMV), soft majority voting (SMV), weighted soft majority voting (wSMV), defined in Equation \ref{eq:mv}:

\begin{subequations} \label{eq:mv}
\begin{equation} \label{eq:hmv}
    L = mode(C(\hat{q}_1), \ldots C(\hat{q}_e))
\end{equation}

\begin{equation} \label{eq:whmv}
    L = arg\max_{c \in \{0, 1\}} \sum_{j=1}^e w_j(C(\hat{q}_j) = c)
\end{equation}

\begin{equation} \label{eq:smv}
    L = C\left(\frac{\sum_{j=1}^e \hat{q}_j}{e}\right)
\end{equation}

\begin{equation} \label{eq:wsmv}
    L = C\left(\frac{\sum_{j=1}^e w_j\hat{q}_j}{\sum_{j=1}^e w_j}\right)
\end{equation}
\end{subequations}

Intuitively, HMV, defined in Equation \ref{eq:hmv}, predicts the class label 1/0 (i.e., CRS/NO CRS) by simply counting the majority labels. wHMV, defined in Equation \ref{eq:whmv}, extends HMV by assigning weights to each study and predicts by choosing the label with the largest combined weight. In soft majority voting, SMV, defined in Equation \ref{eq:smv}, the class label is predicted by comparing the average probabilities of both classes across all studies. wSMV, defined in Equation \ref{eq:wsmv}, extends this strategy by assigning weights and performing a weighted average. In practice, we tested all four strategies and we report the results in Section \ref{sec:eval}. For the weighted schemes we used the study population sizes as weights.

% \begin{subequations} \label{eq:agg}
% \begin{equation} \label{eq:agg_mv}
%     L = sign(\sum_{j=1,e} (w_j\hat{q}_j) - \tau)
% \end{equation}

% \begin{equation} \label{eq:agg_proba}
%     L = sign(\frac{\sum_{j=1,e} (w_j\hat{q}_j)}{ \sum_{j=1,e} (w_j)} - \tau)
% \end{equation}
% \end{subequations}

% Intuitively, given $e$ weights $w_j \in [0, 1]$ that control the power each study in $KB$ has over the predictive decision, a CRS/NO CRS label is determined by the weighted sum aggregation of prediction scores $\hat{q}_j$ with respect to a pre--defined prediction threshold $\tau$. 

% In eq. \ref{eq:agg} $\tau\in\{0,1,..,e\}$, in eq. \ref{eq:agg2} $\tau\in(0, 1)$. 

% Let $\Vec{N} = [\ddot{p}_1, \ldots \ddot{p}_l]$ be the vector of parameter values associated with a new subject (e.g. patient sex, age), i.e., akin to $\Vec{M}$ but for parameters in $P$, and $\Vec{T_j} = [\dot{p}_1^j, \ldots \dot{p}_l^j]$ a vectorized representation of the parameter values characteristic to a study $S_j$. Then, the study--specific aggregation weights can be defined as the Euclidean distance (i.e., $L_2$ norm) between $\Vec{T_j}$ and $\Vec{N}$, as shown in Equation \ref{eq:agg_w}:

% \begin{equation} \label{eq:agg_w}
    % w_j = ||\Vec{T}_j - \Vec{N}||_2 = \sqrt{\sum_{l=1, r} (\dot{p}_l^j - \ddot{p}_l)}
% \end{equation}

Having defined the various collective decision options available, the overall CRS prediction strategy described in this section can be summarized by Algorithm \ref{alg:predict} where the input weight vector $\Vec{W}$ quantifies the influence of each study on the prediction, e.g., number of study participants, and function $\textsc{labeling}$ from line $6$ could be one of the functions defined in Equation \ref{eq:mv}.

\begin{algorithm}[t]
	\caption{CRS prediction}
	\begin{flushleft}
	\textbf{Input}: Matrix $\mathbb{M}$, prediction function $g$, Weight vectors $\Vec{W}$. \\
 	\textbf{Output}: a CRS/NO CRS label $L$.
	\end{flushleft}
	\begin{algorithmic}[1]
		\Function{Predict}{}
		\State $\Vec{PV} \gets [\ ]$ // Empty prediction vector
		\ForAll{$\Vec{M}_j \in \mathbb{M}\ \&\&\ j \in \{1 \ldots e\}$} // Row--wise iteration
		    \State $\Vec{PV}[j] \gets g(\Vec{M}_j)$
		\EndFor
		\State $L \gets \textsc{labeling}(\Vec{PV},\Vec{W})$
		\State \Return $L$
		\EndFunction
	\end{algorithmic}
\label{alg:predict}
\end{algorithm}

\subsection{Explainable CRS prediction} \label{subsec:expl}

The multi--perspective representation of CRS data defined in Equation \ref{eq:matrix} alos enables an abductive evidence--based reasoning aimed at explaining the predictions of CRS likelihood in new subjects. Using the elements of $KB$ introduced in Equations \ref{eq:KB} as the evidence, we formalize this abductive reasoning process using first order logic. To this end, we abuse the notation introduced in Equation \ref{eq:mv} and derive the labeling function $C(\hat{q}_j)$ into a unary predicate $\mathbb{C}_j(b)$ that describes whether or not $b$ signals CRS according to study $S_j$. 

Given a study $S_j$ and some biomarker $b$ with its associated CRS probability $\hat{q}_j$, $\mathbb{C}_j(b) = 
\begin{cases} 
True & C(\hat{q}_j) = 1 \\ 
False & C(\hat{q}_j) = 0 \\ 
\end{cases}$

Then, in the first order logic representation, we use $b$ generically as a variable to refer to any biomarker. When specificity is intended we use constants defined by the actual biomarker name, e.g., \textit{IL2, IL4}, etc. Then, given a collection of new biomarker measurements (i.e., for a new subject), we consider the binary--version matrix $\mathbb{M}$ \footnote{If a non--binary version of $f$ is used, $\mathbb{M}$ can be reduced to a binary version using pre--defined thresholds.} and define Algorithm \ref{alg:explain} to generate \textit{conjunctive explanations} that explain the reasoning behind the final prediction. This generative process is done by backtracking the prediction process to the individual biomarkers and studies that support or refute the final decision CRS/NO CRS. The process can, therefore, be seen as an inference to the best observable explanation, i.e., \textit{abductive inference}. The resulting explanations are expressed in first order logic and are easily transferable to natural language.

\begin{algorithm}[t]
	\caption{CRS prediction explanation}
	\begin{flushleft}
	\textbf{Input}: Binary biomarker matrix $\mathbb{M}$, CRS label $L$. \\
 	\textbf{Output}: an explanatory first order logic expression $\mathrm{E}$.
	\end{flushleft}
	\begin{algorithmic}[1]
		\Function{Explain}{}
		\State $expr \gets [\ ]$ // Empty list of biomarker expressions
		\ForAll{$\Vec{M} \in \mathbb{M}\ \&\&\ j \in \{1 \ldots e\}$} //Row--wise iteration
		    \If{$\forall \dot{b} \in \Vec{M}, \dot{b} > 0\ \&\&\ L == 1$} // Non--zero elements
		        \State $expr[j] \gets (\forall b~\mathbb{C}_j(b))$
		    \ElsIf {$\forall \dot{b} \in \Vec{M}, \dot{b} == 0\ \&\&\ L == 0$} // All--zero elements
		        \State $expr[j] \gets (\forall b~\neg \mathbb{C}_j(b))$
		    \Else
		        \State $sub\_expr \gets [\ ]$ //Empty list of sub--expressions
		        \ForAll{$\dot{b} \in \Vec{M}\ \&\&\ i \in \{1 \ldots k\}$}
		            \IfThenElse{$L == 1$}{$sub\_expr[i] \gets \exists b_i~\mathbb{C}_j(b_i))$}{$sub\_expr[i] \gets (\exists b_i~\neg \mathbb{C}_j(b_i))$}
		        \EndFor
		        \State $expr[j] \gets (\bigwedge\limits_{i=1}^k sub\_expr[i])$
		      %  \IfThenElse{$L == 1$}{$expr[j] \gets (\exists b~\mathbb{C}_j(b))$}{$expr[j] \gets (\exists b~\neg \mathbb{C}_j(b))$}
		    \EndIf
		\EndFor
		\State \Return $\bigwedge\limits_{j=1}^e expr[j]$
		\EndFunction
	\end{algorithmic}
\label{alg:explain}
\end{algorithm}

Intuitively, Algorithm \ref{alg:explain} returns a collection of study--wise explanatory biomarker expressions that are combined in one single conjunctive expression at the output. Concretely, if the intention is to explain a positive prediction (i.e., $L = 1$) then the explanation will contain expressions for studies where at least one biomarker is in the value range of CRS. The alternative for $L = 0$, i.e., biomarker not in CRS value range, is returned when the prediction to explain is negative.

\textbf{Explanation example}

As an example of the potential results produced by Algorithm \ref{alg:explain}, consider the following instantiation of the probabilistic scaling matrix introduced in Equation \ref{eq:matrix}, its associated binary version obtained by considering a study similarity threshold, say $0.5$, and the corresponding prediction vector obtained after classifying each row with some ML algorithm. 

\[
\begin{blockarray}{cccccccc}
    & IL\text{--}2 & IL\text{--}6 & TNF\text{--}\alpha & IL\text{--}2 & IL\text{--}6 & TNF\text{--}\alpha & g(\Vec{M_j}) \\
    \begin{block}{c(ccc) @{\hspace{20pt}} | @{\hspace{20pt}} (ccc) @{\hspace{20pt}} | @{\hspace{20pt}} [c]}
      S_1 & 0.19 & 0.001 & 0.0 & 0 & 0 & 0 & 0\\
      S_2 & 0.63 & 0.1 & 0.58 & 1 & 0 & 1 & 1 \\
      S_3 & 1.0 & 0.76 & 0.44 & 1 & 1 & 1 & 1 \\
    \end{block} \notag
\end{blockarray}
\]

When the above is sent as input to Algorithm \ref{alg:explain} the following results are produced:

\[
\begin{cases}
    (\exists ~ IL2 ~ \mathbb{C}_2(IL2)) \wedge (\exists ~ TNF\text{--}\alpha ~ \mathbb{C}_2(TNF\text{--}\alpha)) \wedge (\forall ~ b~\mathbb{C}_3(b)) & \text{if $L=1$} \\
    \forall ~ b~\neg \mathbb{C}_3(b) & \text{if $L=0$}
\end{cases}
\]

Translated into natural language, the first order logic expression above could become a CRS prediction explanation such as the one below.

\[
\begin{cases}
    \begin{aligned}
        & \text{(\textit{IL2} and \textit{TNF--}$\alpha$ have similar concentrations to the ones observed in CRS patients in study $S_2$)} \text{ AND }\\
        & \text{(all biomarkers have similar concentrations to the ones observed in CRS patients in study $S_3$)}
    \end{aligned} & \text{if $L=1$}) \\
    
    & \\
    
    \text{none of the biomarkers have similar concentrations to the ones observed in CRS patients in study $S_1$} & \text{if $L=0$}
\end{cases}
\]

Natural language translations such as the one above can be easily performed by following a short set of rules:

\begin{itemize}
    \item Constants in existential expressions such as \textit{IL2, TNF--$\alpha$}, i.e., biomarker names, become subjects of the resulting natural language sentence.
    \item Unary predicates $\mathbb{C}_j$ are replaced with \textit{have/don't have similar concentrations to the ones observed in CRS patients in study $j$}, depending on the presence of negation before the predicate.
    \item Variable $b$ in universal expressions becomes the subject of the natural language sentence in the form of a noun phrase: \textit{all/none of the biomarkers}, depending on the presence of negation before the following predicate.
\end{itemize}

\section{Evaluation} \label{sec:eval}

In this paper we introduce \textit{M2--CRS}, a metareview--informed, multi--perspective explainable framework for detecting CRS in patients undergoing CAR--T--cell therapy. In this section we evaluate our proposed methods in several scenarios by using different ML algorithms and majority voting strategies. We also analyze the benefit of integrating our $KB$ by comparing \textit{M2--CRS} against a purely data--driven version that only uses the 109 extracted data samples. Our overall hypotheses in this section are:

\begin{itemize}
    \item $H_1$: \textit{M2--CRS} is a flexible framework that can be used with several ML algorithms while achieving practical effectiveness.
    \item $H_2$: \textit{M2--CRS} leads to better performance than a pure data--driven approach by augmenting available datasets with metareview--informed statistics, i.e., overcomes the data scarcity problem.
    \item $H_3$: \textit{M2--CRS} extends the explanation capabilities of classical ML explainability methods, such as SHAP \cite{lundberg17SHAP}, to generates more expressive explanations.
\end{itemize}

Our analysis in this section is based on the model performance defined by the area under the receiver operating characteristic (AUC), precision, recall, F-1 score and accuracy. Additionally, we investigate biomarkers predictive contribution via SHAP explanation.

\subsection{Experimental setup} \label{subsec:setup}

\begin{figure}[t]
\centering
\includegraphics[width= .7\textwidth]{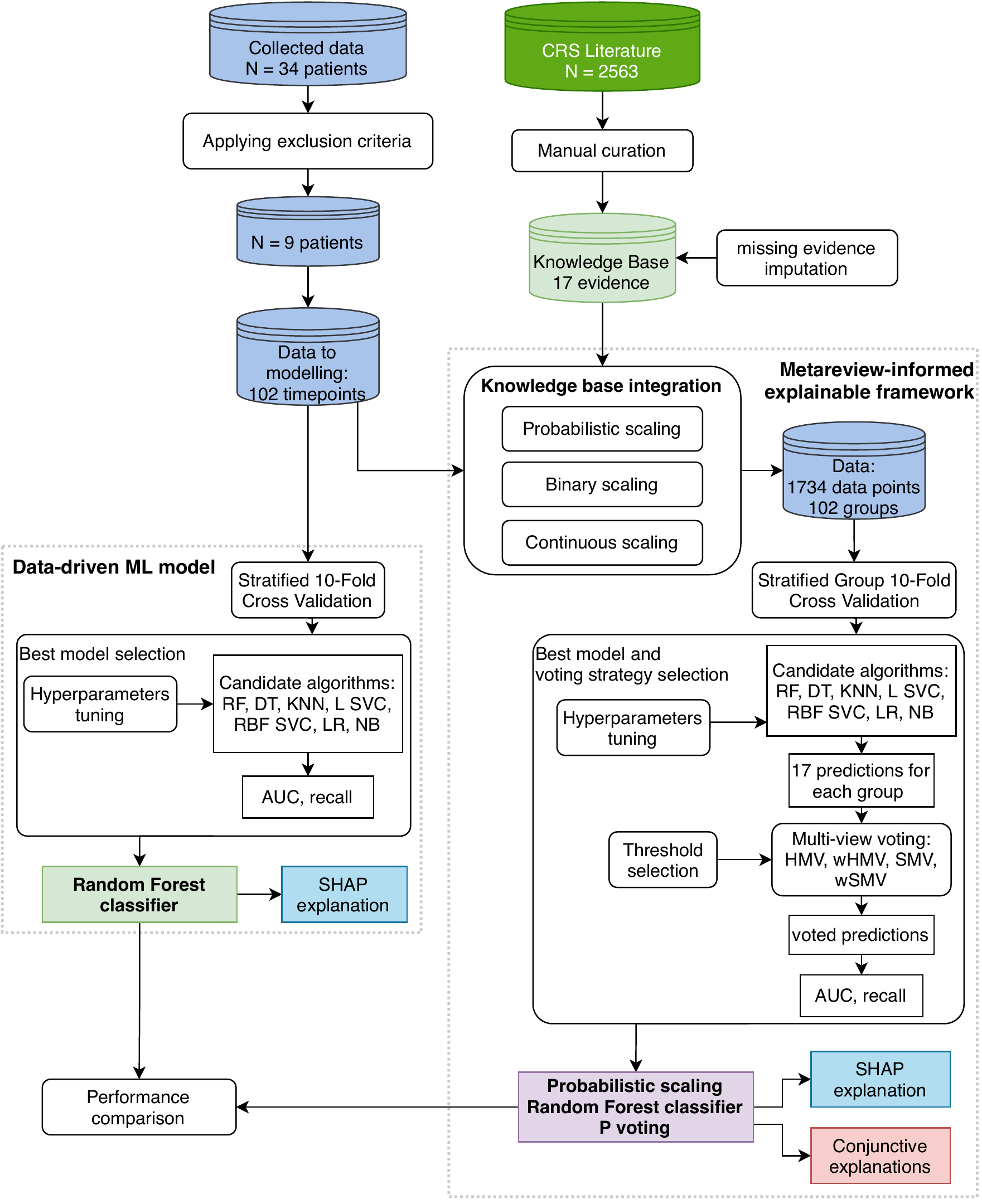}
\caption{Experimental setup for \textit{M2--CRS} evaluation}
\label{fig:model_derivation_diagram}
\end{figure}

The experimental workflow used in this evaluation is presented in Fig. \ref{fig:model_derivation_diagram}. I includes both the setup of our predictive baseline, described in the next section, and the setap of \textit{M2--CRS}. Broadly, the latter setup includes three main steps: (i) choosing a KB integration strategy; (ii) choosing and configuring an ML algorithm; (iii) choosing an ensemble classification strategy. When needed, the weights for (iii) have been given by the study cohort sizes. During (ii) we performed stratified group 10--fold cross validation, where groups are defined as the initial 102 measurements. This ensures that collections of scaled data points originating from the same real measurement vector are not split into train and test sets in a fold. As Random Forests proved to be the most effective ML algorithm, it has been our choice when performing further analysis, such as SHAP--based explanations generation.

% Three strategies for KB integration was investigated: binary, continuous and probabilistic scaling (see Sec. \ref{subsec:kb_integ}). ML algorithm was selected among the same as for \textit{data-driven ML model}. We  considered four multi-view voting strategies: hard majority voting (HMV, all weights set to 1), weighted hard majority voting (wHMV), soft majority voting (SMV, all weights set to 1), weighted soft majority voting (wSMV). %(Supp Fig.\ref{fig:voting_strategies_example}). 
% As the weights can be assigned to predictions reflecting any parameter, in our experiments we used cohort size $N$ in a study as the study weight (details in Suppl. Methods \ref{subsec:SuppleMethods}).

\subsection{Baselines}

We use a simple data-driven ML model that exercises the usual ML approach, where the model is solely fitted to the 109 extracted measurements, i.e., no $KB$ integration. During configuration, we first conduct a stratified 10--fold cross validation. As with \textit{M2--CRS}, we consider seven ML algorithms as potential classifiers: logistic regression (LR), support vector classifier (SVC) (RBF and linear), decision tree (DT), random forest (RF), K--nearest neighbor (KNN), Naive Bayes (NB) \cite{scikit-learn}. The optimal hyperparameters were selected based on AUC (see Suppl. Methods \ref{subsec:SuppleMethods}, Suppl. Tables \ref{subsec:SuppleTables}, Suppl. Figures \ref{subsec:SuppleFigures}).

With respect to the explanation capabilities, we use SHAP explanations \cite{lundberg17SHAP} as baseline to infer biomarkers contribution to each prediction and to compare against our $KB$--based explanations. Absolute SHAP values for each prediction from the 10 test sets (derived from the cross validation) were investigated for the overall feature importance. Additionally, we explored dependency plots as a post--hoc measure of the contribution of biomarkers to individual predictions.

\subsection{Results} \label{subsec:results}

In this section we present the main results obtained when applying \textit{M2--CRS} in practice. Specifically, we evaluate the performance of \textit{M2--CRS} under several scenarios defined by different combinations of ML algorithm + Majority voting strategy + Scaling method.

Table \ref{tab:PerformanceBinaryScalingMainText} lists the results obtain for all combinations of ML algorithms and ensemble scheme when binary scaling (defined in Section \ref{subsubsec:binary}) was used to integrate the $KB$ statistical information. We report the average results of each measure across 10--fold cross validation, together with their standard deviation (SD).

\begin{longtable}[c]{@{}lllll@{}}
\caption[tab]{Performance of \textit{M2--CRS} with \textbf{binary scaling} across several combinations of ML algorithms and voting schemes.} \\   
\label{tab:PerformanceBinaryScalingMainText}\\
\toprule
Model               & \begin{tabular}[c]{@{}l@{}}Majority Voting \\  strategy\end{tabular} & \begin{tabular}[c]{@{}l@{}}Precision \\ mean (SD)\end{tabular} & \begin{tabular}[c]{@{}l@{}}Recall\\ mean (SD)\end{tabular} & \begin{tabular}[c]{@{}l@{}}F1-score \\ mean (SD)\end{tabular} \\* \midrule
\endhead
\bottomrule
\endfoot
\endlastfoot
Random Forest       & HMV                                                                   & 0.832 (0.178)                                                  & 0.830 (0.204)                                               & 0.820 (0.172)                                                 \\
                    & wHMV                                                                  & 0.810 (0.163)                                                  & \textbf{0.900 (0.175)}                                                & 0.845 (0.145)                                                 \\
                    & SMV                                                                    & 0.857 (0.183)                                                  & 0.855 (0.172)                                              & 0.850 (0.161)                                                 \\
                    & wSMV                                                                   & 0.810 (0.163)                                                  & \textbf{0.900 (0.175)}                                                & 0.845 (0.145)                                                 \\
Decision Tree       & HMV                                                                   & 0.828 (0.178)                                                  & 0.835 (0.165)                                              & 0.826 (0.154)                                                 \\
                    & wHMV                                                                  & 0.860 (0.182)                                                  & 0.875 (0.177)                                              & \textbf{0.861 (0.161)}                                                 \\
                    & SMV                                                                    & 0.873 (0.188)                                                  & 0.855 (0.172)                                              & 0.859 (0.167)                                                 \\
                    & wSMV                                                                   & 0.873 (0.188)                                                  & 0.855 (0.172)                                              & 0.859 (0.167)                                                 \\
Nearest Neighbors   & HMV                                                                   & 0.857 (0.139)                                                  & 0.855 (0.172)                                              & 0.845 (0.135)                                                 \\
                    & wHMV                                                                  & 0.823 (0.141)                                                  & 0.880 (0.127)                                               & 0.846 (0.116)                                                 \\
                    & SMV                                                                    & \textbf{0.905 (0.130)}                                                  & 0.790 (0.207)                                               & 0.830 (0.149)                                                 \\
                    & wSMV                                                                   & 0.880 (0.133)                                                  & 0.790 (0.207)                                               & 0.820 (0.151)                                                 \\
Linear SVC          & HMV                                                                   & 0.783 (0.220)                                                  & 0.835 (0.165)                                              & 0.801 (0.182)                                                 \\
                    & wHMV                                                                  & 0.857 (0.165)                                                  & 0.745 (0.250)                                               & 0.778 (0.189)                                                 \\
                    & SMV                                                                    & 0.840 (0.192)                                                  & 0.745 (0.22)                                               & 0.776 (0.185)                                                 \\
                    & wSMV                                                                   & 0.863 (0.154)                                                  & 0.720 (0.262)                                               & 0.769 (0.198)                                                 \\
RBF SVC             & HMV                                                                   & 0.816 (0.182)                                                  & 0.855 (0.172)                                              & 0.830 (0.161)                                                 \\
                    & wHMV                                                                  & 0.838 (0.202)                                                  & 0.855 (0.172)                                              & 0.842 (0.180)                                                 \\
                    & SMV                                                                    & 0.848 (0.186)                                                  & 0.855 (0.172)                                              & 0.848 (0.171)                                                 \\
                    & wSMV                                                                   & 0.848 (0.186)                                                  & 0.830 (0.204)                                               & 0.829 (0.180)                                                 \\
Logistic Regression & HMV                                                                   & 0.813 (0.191)                                                  & 0.810 (0.196)                                               & 0.801 (0.175)                                                 \\
                    & wHMV                                                                  & 0.860 (0.198)                                                  & 0.770 (0.235)                                               & 0.799 (0.200)                                                 \\
                    & SMV                                                                    & 0.860 (0.198)                                                  & 0.770 (0.235)                                               & 0.799 (0.200)                                                 \\
                    & wSMV                                                                   & 0.843 (0.188)                                                  & 0.79 (0.207)                                               & 0.805 (0.178)                                                 \\
Naive Bayes         & HMV                                                                   & 0.769 (0.216)                                                  & 0.835 (0.165)                                              & 0.783 (0.163)                                                 \\
                    & wHMV                                                                  & 0.788 (0.198)                                                  & 0.810 (0.156)                                               & 0.782 (0.141)                                                 \\
                    & SMV                                                                    & 0.822 (0.178)                                                  & 0.720 (0.234)                                               & 0.755 (0.191)                                                 \\
                    & wSMV                                                                   & 0.822 (0.178)                                                  & 0.720 (0.234)                                               & 0.755 (0.191)                                                 \\* \bottomrule
\end{longtable}

Binary scaling leads to performances of up to 0.9 in recall when RF + wSMV/wHMV is the combination of ML algorithm + voting scheme of choice and of 0.861 in F1-score when DT + wHMV are used. Overall, most of the times the variations between different ML algorithm/voting strategy combinations are below 0.2, with RD, DT and KNN distinguishing as the best ML algorithm choices.

Similarly, Table \ref{tab:PerformanceContinuousScalingMainText} lists the results obtain for all combinations of ML algorithms and ensemble schemes when continuous scaling (defined in Section \ref{subsubsec:cont}) was used to integrate the $KB$ statistical information.

\begin{longtable}[c]{@{}lllll@{}}
\caption[tab]{Performance of \textit{M2--CRS} with \textbf{continuous scaling} across several combinations of ML algorithms and voting schemes.} \\   
\label{tab:PerformanceContinuousScalingMainText}\\
\toprule
Model               & \begin{tabular}[c]{@{}l@{}}Majority Voting \\ strategy\end{tabular} & \begin{tabular}[c]{@{}l@{}}Precision \\ mean (SD)\end{tabular} & \begin{tabular}[c]{@{}l@{}}Recall \\ mean (SD)\end{tabular} & \begin{tabular}[c]{@{}l@{}}F1-score \\ mean (SD)\end{tabular} \\* \midrule
\endhead
\bottomrule
\endfoot
\endlastfoot
Random Forest       & HMV                                                                   & \textbf{0.963 (0.078)}                                                  & 0.905 (0.123)                                               & \textbf{0.926 (0.066)}                                                 \\
                    & wHMV                                                                  & 0.938 (0.101)                                                  & 0.905 (0.123)                                               & 0.915 (0.084)                                                 \\
                    & SMV                                                                    & \textbf{0.963 (0.078)}                                                  & 0.905 (0.123)                                               & \textbf{0.926 (0.066)}                                                 \\
                    & wSMV                                                                   & \textbf{0.963 (0.078)}                                                  & 0.905 (0.123)                                               & \textbf{0.926 (0.066)}                                                 \\
Decision Tree       & HMV                                                                   & 0.853 (0.171)                                                 & \textbf{0.950 (0.105)}                                                & 0.889 (0.117)                                              \\
                    & wHMV                                                                  & 0.855 (0.169)                                               & \textbf{0.950 (0.105)}                                                & 0.897 (0.135)                                              \\
                    & SMV                                                                    & 0.861 (0.159)                                               & \textbf{0.950 (0.105)}                                                & 0.896 (0.111)                                              \\
                    & wSMV                                                                   & 0.855 (0.169)                                               & \textbf{0.950 (0.105)}                                                & 0.896 (0.135)                                              \\
Nearest Neighbors   & HMV                                                                   & 0.858 (0.103)                                                  & \textbf{0.950 (0.105)}                                                & 0.892 (0.045)                                                 \\
                    & wHMV                                                                  & 0.861 (0.128)                                                  & \textbf{0.950 (0.105)}                                                & 0.892 (0.065)                                                 \\
                    & SMV                                                                    & 0.861 (0.128)                                                  & \textbf{0.950 (0.105)}                                                & 0.892 (0.065)                                                 \\
                    & wSMV                                                                   & 0.861 (0.128)                                                  & \textbf{0.950 (0.105)}                                                & 0.892 (0.065)                                                 \\
Linear SVC          & HMV                                                                   & 0.938 (0.101)                                               & 0.795 (0.170)                                                & 0.846 (0.097)                                              \\
                    & wHMV                                                                  & 0.847 (0.179)                                               & 0.865 (0.167)                                               & 0.835 (0.123)                                              \\
                    & SMV                                                                    & 0.905 (0.130)                                                  & 0.795 (0.206)                                               & 0.831 (0.142)                                              \\
                    & wSMV                                                                   & 0.950 (0.112)                                                   & 0.750 (0.221)                                                & 0.821 (0.161)                                              \\
RBF SVC             & HMV                                                                   & 0.880 (0.108)                                                  & 0.880 (0.174)                                                & 0.867 (0.110)                                                 \\
                    & wHMV                                                                  & 0.888 (0.169)                                                  & 0.880 (0.174)                                                & 0.873 (0.156)                                                 \\
                    & SMV                                                                    & 0.913 (0.114)                                                  & 0.880 (0.174)                                                & 0.885 (0.124)                                                 \\
                    & wSMV                                                                   & 0.933 (0.110)                                                  & 0.835 (0.190)                                                & 0.868 (0.131)                                                 \\
Logistic Regression & HMV                                                                   & 0.863 (0.164)                                                  & 0.860 (0.170)                                                 & 0.849 (0.140)                                                 \\
                    & wHMV                                                                  & 0.888 (0.169)                                                  & 0.745 (0.242)                                               & 0.793 (0.188)                                                 \\
                    & SMV                                                                    & 0.830 (0.137)                                                  & 0.840 (0.163)                                                & 0.816 (0.088)                                                 \\
                    & wSMV                                                                   & 0.933 (0.110)                                                  & 0.750 (0.187)                                                & 0.814 (0.128)                                                 \\
Naive Bayes         & HMV                                                                   & 0.908 (0.165)                                                  & 0.790 (0.207)                                                & 0.835 (0.171)                                                 \\
                    & wHMV                                                                  & 0.950 (0.180)                                                  & 0.765 (0.226)                                               & 0.831 (0.176)                                                 \\
                    & SMV                                                                    & 0.950 (0.180)                                                  & 0.770 (0.193)                                                & 0.839 (0.176)                                                 \\
                    & wSMV                                                                   & 0.908 (0.195)                                                  & 0.790 (0.207)                                                & 0.835 (0.181)                                                 \\* \bottomrule
\end{longtable}

In the case of continuous scaling, there is a significant improvement in all three measures, with RF, DT and KNN still being the best performing ML algorithms. In fact, most of the combinations perform better than in the binary case. This is because when scaling using continuous functions, the models predict based on more granular correlations between biomarker values, i.e., the algorithms aim to predict based on a continuous space of biomarker representations rather than a discrete, fragmented space.

Thirdly, Table \ref{tab:PerformanceProbScalingMainText} lists the results obtain for all combinations of ML algorithms and ensemble schemes when probabilistic scaling (defined in Section \ref{subsubsec:prob}) was used to integrate the $KB$ statistical information.

\begin{longtable}[t]{@{}llrrr@{}}
\caption[tab]{Performance of \textit{M2--CRS} with \textbf{probabilistic scaling} across several combinations of ML algorithms and voting schemes.} \\   
\label{tab:PerformanceProbScalingMainText}\\
\toprule
Model                                & \begin{tabular}[c]{@{}l@{}}Majority Voting \\ strategy\end{tabular} & \multicolumn{1}{l}{\begin{tabular}[c]{@{}l@{}}Precision \\ mean (SD)\end{tabular}} & \multicolumn{1}{l}{\begin{tabular}[c]{@{}l@{}}Recall \\ mean (SD)\end{tabular}} & \multicolumn{1}{l}{\begin{tabular}[c]{@{}l@{}}F1-score \\ mean (SD)\end{tabular}} \\* \midrule
\endhead

\bottomrule
\endfoot
\endlastfoot
Random Forest       & HMV                                                                   & 0.938 (0.101)                                                                      & 0.880 (0.174)                                                                   & 0.896 (0.115)                                                                     \\
                                     & wHMV                                                                  & 0.858 (0.167)                                                                      & 0.925 (0.121)                                                                   & 0.888 (0.141)                                                                     \\
                                     & SMV                                                                    & 0.938 (0.101)                                                                      & 0.925 (0.121)                                                                   & \textbf{0.926 (0.088)}                                                                     \\
                                     & wSMV                                                                   & \textbf{0.958 (0.090)}                                                                      & 0.855 (0.172)                                                                   & 0.893 (0.116)                                                                     \\
Decision Tree       & HMV                                                                   & 0.878 (0.142)                                                                      & 0.930 (0.114)                                                                   & 0.899 (0.114)                                                                     \\
                                     & wHMV                                                                  & 0.842 (0.128)                                                                      & \textbf{0.950 (0.105)}                                                                   & 0.890 (0.109)                                                                     \\
                                     & SMV                                                                    & \textbf{0.958 (0.090)}                                                                      & 0.875 (0.177)                                                                   & 0.904 (0.120)                                                                     \\
                                     & wSMV                                                                   & \textbf{0.958 (0.090)}                                                                      & 0.855 (0.172)                                                                   & 0.893 (0.116)                                                                     \\
Nearest Neighbors   & HMV                                                                   & 0.938 (0.101)                                                                      & 0.880 (0.174)                                                                   & 0.896 (0.115)                                                                     \\
                                     & wHMV                                                                  & 0.886 (0.123)                                                                      & \textbf{0.950 (0.105)}                                                                   & 0.911 (0.087)                                                                     \\
                                     & SMV                                                                    & 0.938 (0.101)                                                                      & 0.905 (0.124)                                                                   & 0.915 (0.084)                                                                     \\
                                     & wSMV                                                                   & 0.886 (0.123)                                                                      & \textbf{0.950 (0.105)}                                                                   & 0.911 (0.087)                                                                     \\
Linear SVC          & HMV                                                                   & 0.803 (0.177)                                                                      & 0.790 (0.141)                                                                   & 0.787 (0.143)                                                                     \\
                                     & wHMV                                                                  & 0.922 (0.130)                                                                      & 0.725 (0.153)                                                                   & 0.803 (0.123)                                                                     \\
                                     & SMV                                                                    & 0.873 (0.146)                                                                      & 0.770 (0.153)                                                                   & 0.804 (0.109)                                                                     \\
                                     & wSMV                                                                   & 0.888 (0.167)                                                                      & 0.770 (0.153)                                                                   & 0.808 (0.119)                                                                     \\
RBF SVC            & HMV                                                                   & 0.943 (0.132)                                                                      & 0.855 (0.172)                                                                   & 0.885 (0.129)                                                                     \\
                                     & wHMV                                                                  & \textbf{0.958 (0.090)}                                                                      & 0.835 (0.190)                                                                   & 0.879 (0.124)                                                                     \\
                                     & SMV                                                                    & \textbf{0.958 (0.090)}                                                                      & 0.855 (0.172)                                                                   & 0.893 (0.116)                                                                     \\
                                     & wSMV                                                                   & 0.943 (0.132)                                                                      & 0.855 (0.172)                                                                   & 0.885 (0.129)                                                                     \\
Logistic Regression & HMV                                                                   & 0.817 (0.191)                                                                      & 0.855 (0.172)                                                                   & 0.822 (0.161)                                                                     \\
                                     & wHMV                                                                  & 0.918 (0.143)                                                                      & 0.770 (0.153)                                                                   & 0.823 (0.110)                                                                     \\
                                     & SMV                                                                    & 0.898 (0.144)                                                                      & 0.770 (0.153)                                                                   & 0.815 (0.108)                                                                     \\
                                     & wSMV                                                                   & 0.898 (0.144)                                                                      & 0.770 (0.153)                                                                   & 0.815 (0.108)                                                                     \\
Naive Bayes       & HMV                                                                   & 0.838 (0.165)                                                                      & 0.790 (0.207)                                                                   & 0.804 (0.171)                                                                     \\
                                     & wHMV                                                                  & 0.805 (0.194)                                                                      & 0.830 (0.204)                                                                   & 0.804 (0.175)                                                                     \\
                                     & SMV                                                                    & 0.805 (0.194)                                                                      & 0.830 (0.204)                                                                   & 0.804 (0.175)                                                                     \\
                                     & wSMV                                                                   & 0.800 (0.195)                                                                      & 0.810 (0.217)                                                                   & 0.791 (0.181)                                                                     \\* \bottomrule
\end{longtable}

The observations made in the previous case regarding the use of continuous scaling functions hold for the probabilistic approach. In fact, the differences from the continuous results are marginal, with the best performance still achieved by RF, DT, and KNN. We therefore choose RF as the ML algorithm choice to compare against a simple data--driven RF baseline, with no $KB$ integration. The results are in Table \ref{tab:PerformanceRandomForest}.

\begin{longtable}[c]{@{}lllllll@{}}
\caption[tab]{Performance of Random Forest for best combination of hyperparameters and thresholds using data integrated knowledge from KB performed for three different type of scaling of each biomarker value with respect to each study in KB} \\   
\label{tab:PerformanceRandomForest}\\
\toprule
 & \begin{tabular}[c]{@{}l@{}}Majority \\ Voting \\ strategy\end{tabular} & \begin{tabular}[c]{@{}l@{}}Precision \\ mean (SD)\end{tabular} & \begin{tabular}[c]{@{}l@{}}Recall \\ mean (SD)\end{tabular} & \begin{tabular}[c]{@{}l@{}}F1-score \\ mean (SD)\end{tabular} & \begin{tabular}[c]{@{}l@{}}AUC \\ mean (SD)\end{tabular} & \begin{tabular}[c]{@{}l@{}}Accuracy \\ mean (SD)\end{tabular} \\* \midrule
\endhead
\bottomrule
\endfoot
\endlastfoot
no $KB$--integration                   &                                                                          & \multicolumn{1}{r}{0.960 (0.084)}                                                               & \multicolumn{1}{r}{0.795 (0.297)}                           &    \multicolumn{1}{r}{0.830 (0.212)}                                                       & \multicolumn{1}{r}{0.881 (0.139)}                        & \multicolumn{1}{r}{0.893 (0.115)}                                                              \\
 \hline \\
prob. scaling        & HMV                                                                       & 0.938 (0.101)                                                  & 0.880 (0.174)                                                & 0.896 (0.115)                                                 & 0.915 (0.09)                                            & 0.921 (0.079)                                                 \\
                                       & wHMV                                                                      & 0.858 (0.167)                                                  & \textbf{0.925 (0.121)}                                                 & 0.888 (0.141)                                                 & 0.904 (0.123)                                            & 0.901 (0.125)                                                 \\
                                       & SMV                                                                        & 0.938 (0.101)                                                  & \textbf{0.925 (0.121)}                                               & \textbf{0.926 (0.088)}                                                 & \textbf{0.938 (0.074)}                                            & \textbf{0.941 (0.069)}                                                 \\
                                       & wSMV                                                                       & 0.958 (0.090)                                                  & 0.855 (0.172)                                                & 0.893 (0.116)                                                 & 0.911 (0.091)                                            & 0.921 (0.079)                                                 \\
binary scaling        & HMV                                                                       & 0.832 (0.178)                                                  & 0.830 (0.204)                                                & 0.820 (0.172)                                                 & 0.845 (0.142)                                            & 0.851 (0.136)                                                 \\
                                       & wHMV                                                                      & 0.810 (0.163)                                                  & 0.900 (0.175)                                                 & 0.845 (0.145)                                                 & 0.863 (0.119)                                            & 0.862 (0.116)                                                 \\
                                       & SMV                                                                        & 0.857 (0.183)                                                  & 0.855 (0.172)                                               & 0.850 (0.161)                                                 & 0.867 (0.136)                                            & 0.871 (0.134)                                                 \\
                                       & wSMV                                                                       & 0.810 (0.163)                                                  & 0.900 (0.175)                                                 & 0.845 (0.145)                                                 & 0.863 (0.119)                                            & 0.862 (0.116)                                                 \\
cont. scaling    & HMV                                                                       & \textbf{0.963 (0.078)}                                                  & 0.905 (0.123)                                               & \textbf{0.926 (0.066)}                                                 & 0.936 (0.057)                                            & \textbf{0.941 (0.051)}                                                 \\
                                       & wHMV                                                                      & 0.938 (0.101)                                                  & 0.905 (0.123)                                               & 0.915 (0.084)                                                 & 0.928 (0.070)                                             & 0.931 (0.067)                                                 \\
                                       & SMV                                                                        & \textbf{0.963 (0.078)}                                                  & 0.905 (0.123)                                               & \textbf{0.926 (0.066)}                                                 & 0.936 (0.057)                                            & \textbf{0.941 (0.051)}                                                 \\
                                       & wSMV                                                                       & \textbf{0.963 (0.078)}                                                  & 0.905 (0.123)                                               & \textbf{0.926 (0.066)}                                                 & 0.936 (0.057)                                            & \textbf{0.941 (0.051)}                                                 \\* \bottomrule
\end{longtable}

\textit{M2--CRS} using RF outperformed the baseline achieving AUC=0.938. Furthermore, the metareview--informed approach with probabilistic or continuous scaling proved to perform better regardless the the chosen voting strategy (i.e., AUC$>$0.904).

\textbf{Explainability analysis}

\begin{figure}[t]
\centering
\caption{SHAP visualizations}
\begin{subfigure}{.45\textwidth}
    \centering
    \includegraphics[width=\linewidth]{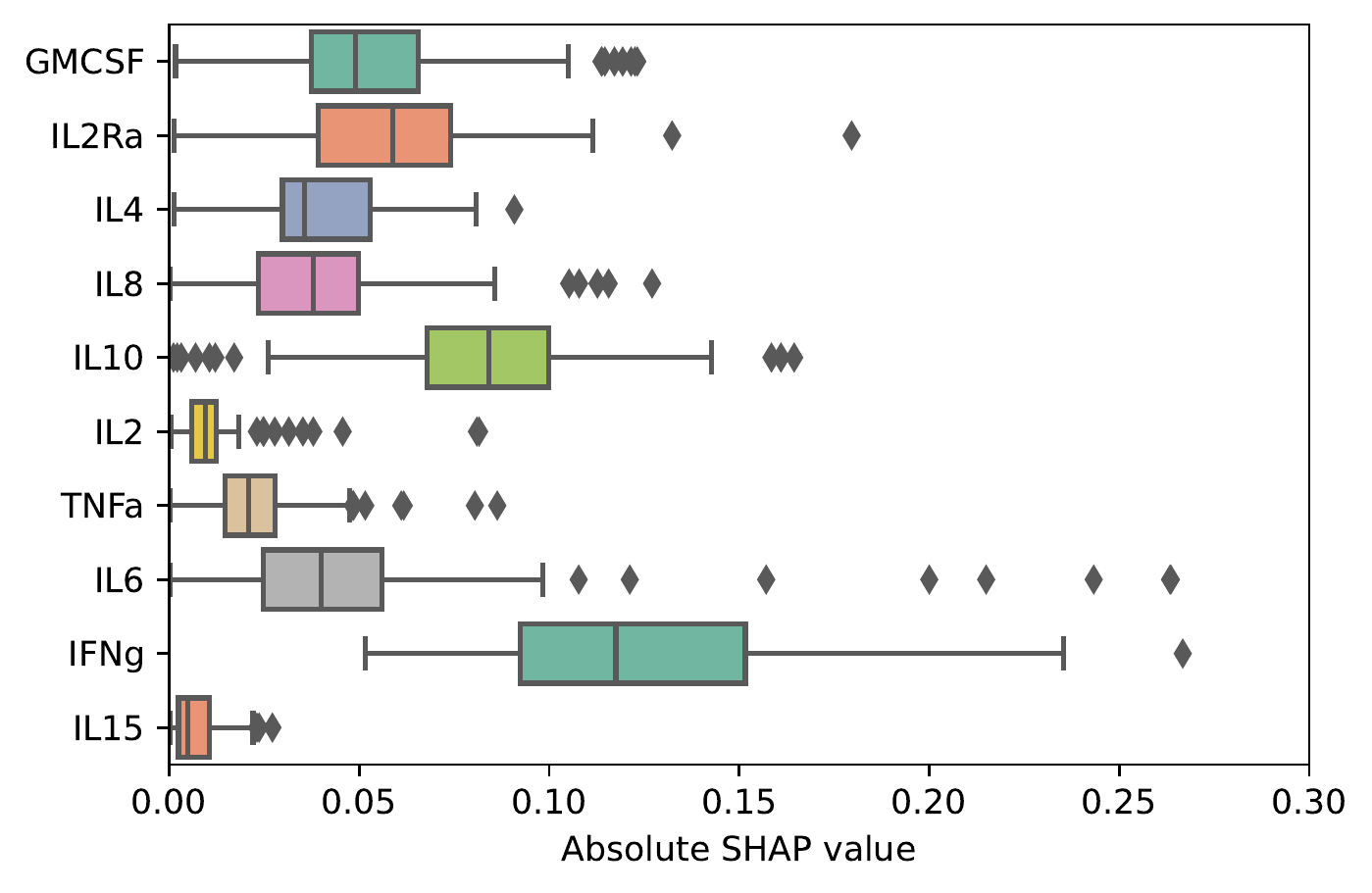}
    \caption{SHAP explanation for baseline ML model.}
    \label{fig:boxplots_shap_no_scaling}
\end{subfigure}
\begin{subfigure}{.45\textwidth}
    \centering
    \includegraphics[width=\linewidth]{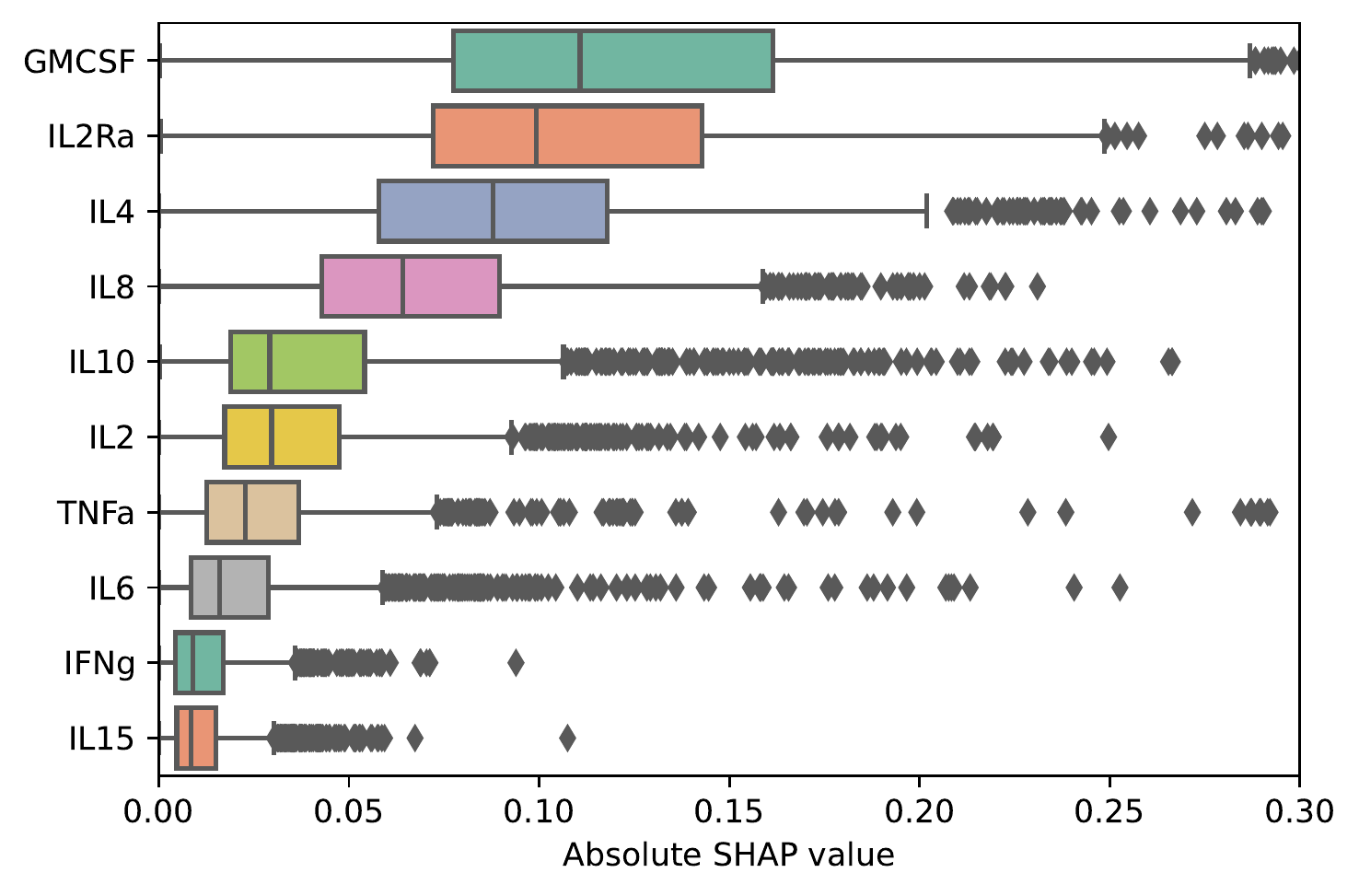}
    \caption{SHAP explanation for RF with probabilistic scaling.}
    \label{fig:boxplots_shap_prob}
\end{subfigure}
\label{fig:shap}
\end{figure}

In addition to the ML--specific evaluation presented in the tables above, we also conduct an additional analysis with respect to the explainability characteristics of \textit{M2--CRS}. To this end, we employ the SHAPley Additive exPlanations (SHAP) method as a visualization tool that can be used for making a machine learning model more explainable by visualizing its output. More specifically, we use this tool for explaining the prediction of our baseline and probability scaled ML models by computing the contribution of each feature to the prediction. This contribution is quantified in the form of Shapely values \cite{lundberg17SHAP} that denote the importance weight each feature had on the prediction. The results are shown in Figure \ref{fig:shap}. We observe that \textit{M2--CRS}'s top 3 predictive biomarkers with the strongest predictive power are \textit{GMCSF, IL2R$\alpha$} and \textit{IL4}. This is regardless of the $KB$ integration strategy or voting scheme used (see Suppl. Figs \ref{fig:boxplots_shap_minmax} and\ref{fig:boxplots_shap_binary} for additional details). Interestingly, for our baseline model without $KB$ integration, IFN--$\gamma$ and IL10 have been the most important biomarkers. This, together with the fact that \textit{M2--CRS} performs significantly better than when $KB$ information is not used, support our hypothesis from Section \ref{subsec:kb_integ} that the knowledge informed model can leverage correlations between biomarker values that would otherwise be missed.

The above SHAP method has been proposed as a general explainability approach for interpreting the predictions ML methods output. In this paper, we contribute an extension to this interpretability method in the form of a first--order logic abductive inference approach that links the predictions of our ML methods to the relevant literature. Collectively, the SHAP--based feature importance, exemplified in Figure \ref{fig:shap}, and our abductive reasoning process, exemplified in Section \ref{subsec:expl}, have the potential to motivate a given prediction with clear references from the domain literature.

\section{Discussion}

In this paper we set out to pioneer an ML--based method that could assist clinicians in diagnosing the onset of CRS in patients under CAR--T therapy. To this end, our governing desiderata has been to propose a method that is:

\begin{itemize}
    \item[(i)] \textit{functional in scenarios characterized by clinical data scarcity},
    \item[(ii)] \textit{interpretable}, and
    \item[(iii)] \textit{safe to apply in practice}.
\end{itemize}
In achieving these desiderata, we have started from the observations that, often, when diagnosing the occurrence of CRS, clinicians rely on symptoms, specific biomarker measurement values and past knowledge about known combinations of these and how they can signal CRS. Therefore, our ultimate goal has been to  use the predictive power of known biomarkers to extend the reach human reasoning can achieve with respect to the space of correlations relevant for such diagnosing process. As such, we have chosen ten cytokines as predictive biomarkers and developed a method for incorporating statistical literature knowledge about these into ML algorithms. This offered a potential path to overcoming data scarcity problems, as described in Section \ref{sec:method}, and allowed us to address (i) from above. Furthermore, by integrating statistical information from clinical studies into our ML methods, we can further track back relevant studies that support or refute specific ML predictions, as described in Section \ref{subsec:expl}. We therefore addressed (ii) from the above as well.

With respect to desideratum (iii), the safety of the method is ultimately decided by the clinician. The aim of our proposals in this paper is to safely assist clinicians. In this context, the experimental evaluation from Section \ref{sec:eval} has shown that \textit{M2--CRS} can systematically leverage statistical information extracted by meta--reviewing relevant studies to achieve significantly better results in detecting CRS than when external information is not used. Furthermore, multiple of our design choices (e.g., ML algorithm + scaling method + majority voting strategy) consistently achieved practical values of precision/recall/f1--score/accuracy above $90\%$, as hypothesized by $H_1$ at the beginning of Section \ref{sec:eval}. We have shown that this level of performance is a consequence of $KB$ integration, and that a meta--review informed approach is superior to a purely data--driven alternative, as hypothesized by $H_2$. In addition, to directly contribute to prediction safety and trust, \textit{M2--CRS} extends classical ML explanation methods with a more granular approach that leverages the same $KB$ to offer inference to the best available evidence in the form of relevant literature studies, as hypothesized by $H_3$.

Finally, we mention two main limitations of our approach introduced in this paper. These limitations are subject to current and future research:

\begin{itemize}
    \item \textit{M2--CRS} is sensible to the $KB$ size and similarity. In other words, the data scarcity problem can be overcome as long as there are sufficient and similar studies in the knowledge base that could extend the current data. In addition, the range of biomarkers available in the knowledge base defines the feature space that governs the algorithm's predictions. Although, we do note that \textit{M2--CRS} is easily extensible to include additional biomarkers from the ones used in this paper. 
    \item \textit{M2--CRS} is \textit{time--independent}. In other words, CRS is predicted based on biomarker measurements at a given time point, when the condition may already be in effect. However, a \textit{time--depended} analysis that would predict a future onset of the condition would require more data points than the $KB$ integration can offer.
\end{itemize}

\printbibliography %Prints bibliography

\newpage
\setcounter{secnumdepth}{0}
\section{Supplementary Information}
\setcounter{secnumdepth}{0}
\label{sec:SuppMat}

\setcounter{table}{0}
\renewcommand\thetable{S.\arabic{table}}
\setcounter{figure}{0}
\renewcommand\thefigure{S.\arabic{figure}}

\subsection{Supplementary Methods} \label{subsec:SuppleMethods}

\textbf{Methods for SP-X ELISA assays}

Human cytokine 10-plex panel catalog number 85-0002 was purchased from Quanterix (formerly Aushon). The following analytes are in the panel (hIFN$\gamma$, hIL1a, hIL1b, hIL2, hIL4, hIL6, hIL8, hIL10, hIL12p70, hTNF$\alpha$). Human cytokine 2-plex panel catalog number 100-0447 was purchased from Quanterix (formerly Aushon) for hIL15, hGMCSF. Human IL-2R$\alpha$ 1-plex assay (catalog number 100-0083) and human IL-5 1-plex assay (catalog number 100-0442) were also purchased from Quanterix (formerly Aushon) (Table \ref{tab:ELISAassays}). Serum samples were analyzed according to the manufacturers’ protocols.

% \usepackage{booktabs}
% \usepackage{multirow}
% \usepackage{longtable}
% Note: It may be necessary to compile the document several times to get a multi-page table to line up properly
%\fontsize{7}{11}\selectfont{
\fontsize{7}{11}\selectfont{
\begin{longtable}[c]{@{}llllll@{}}
\caption[tab]{Technical information for the used assays to gather cytokine data} \\   
\label{tab:ELISAassays}\\
\toprule
\textbf{Manufacturer}                                                                   & \textbf{Technology}         & \textbf{Manufacturer Assay Name} & \textbf{Manufacturer Product No.} & \textbf{Analytes*}                                                                                                 & \textbf{Internal name} \\* \midrule
\endhead
\bottomrule
\endfoot
\endlastfoot
\begin{tabular}[c]{@{}l@{}}Quanterix \\ (formerly Aushon)\end{tabular} & SP-X ELISA & Human Cytokine 2 10-Plex Array   & 85-0002                              & \begin{tabular}[c]{@{}l@{}}\textbf{hIFN$\gamma$}, hIL1a, hIL1b, \\ \textbf{hIL2}, \textbf{hIL4}, \textbf{hIL6}, \textbf{hIL8}, \\ \textbf{hIL10}, hIL12p70, \textbf{hTNF$\alpha$}\end{tabular} & IMMUN\_01                                \\
                                                                                        &                             & Human 2-Plex Array               & 100-0447                             & \textbf{hIL15}, \textbf{hGMCSF}                                                                                                     & IMMUN\_02                                \\
                                                                                        &                             & Human IL-2R$\alpha$ 1-Plex Assay        & 100-0083                             & \textbf{hIL2R}$\alpha$ & IMMUN\_03                                \\
                                        &                             & Human IL-5 1-Plex Assay          & 100-0442                             & hIL5                                                                                                              & IMMUN\_04                                \\* \bottomrule
\end{longtable}
*The data included in the model are bold.
}
\normalsize
%}

\textbf{Detailed description of final model development}

Following description aims to explain in details the modeling approach depicted in Fig. \ref{fig:model_derivation_diagram}.

\textbf{\textit{Pre-processing}}

Pre-processing was performed on 34 patients and comprised:
\begin{itemize}
    \item Excluding cases without information about CRS
    \item Excluding cases where one or more numeric value were missing from the following features: IL2, IL4, IL6, IL8, IL10, IL15, IL2R$\alpha$, TNF--$\alpha$, IFN--$\gamma$ or GM-CSF. These features were identified as the most significant in CRS based on previous works.
\end{itemize}
As a result of the pre-processing stage, 9 patients (102 timepoints together) were selected for further analysis. 

\textbf{\textit{Model derivation}}

%In the derivation phase, all results refer to mean values and standard deviations (std) obtained from calculations performed on a dataset of 102 timepoints (from 9 patients), scaled by 17 papers from KB. This dataset was derived via the following steps:

\begin{figure}[h!]
\centering
\begin{subfigure}{0.35\textwidth}
  \centering
  \includegraphics[width= \textwidth]{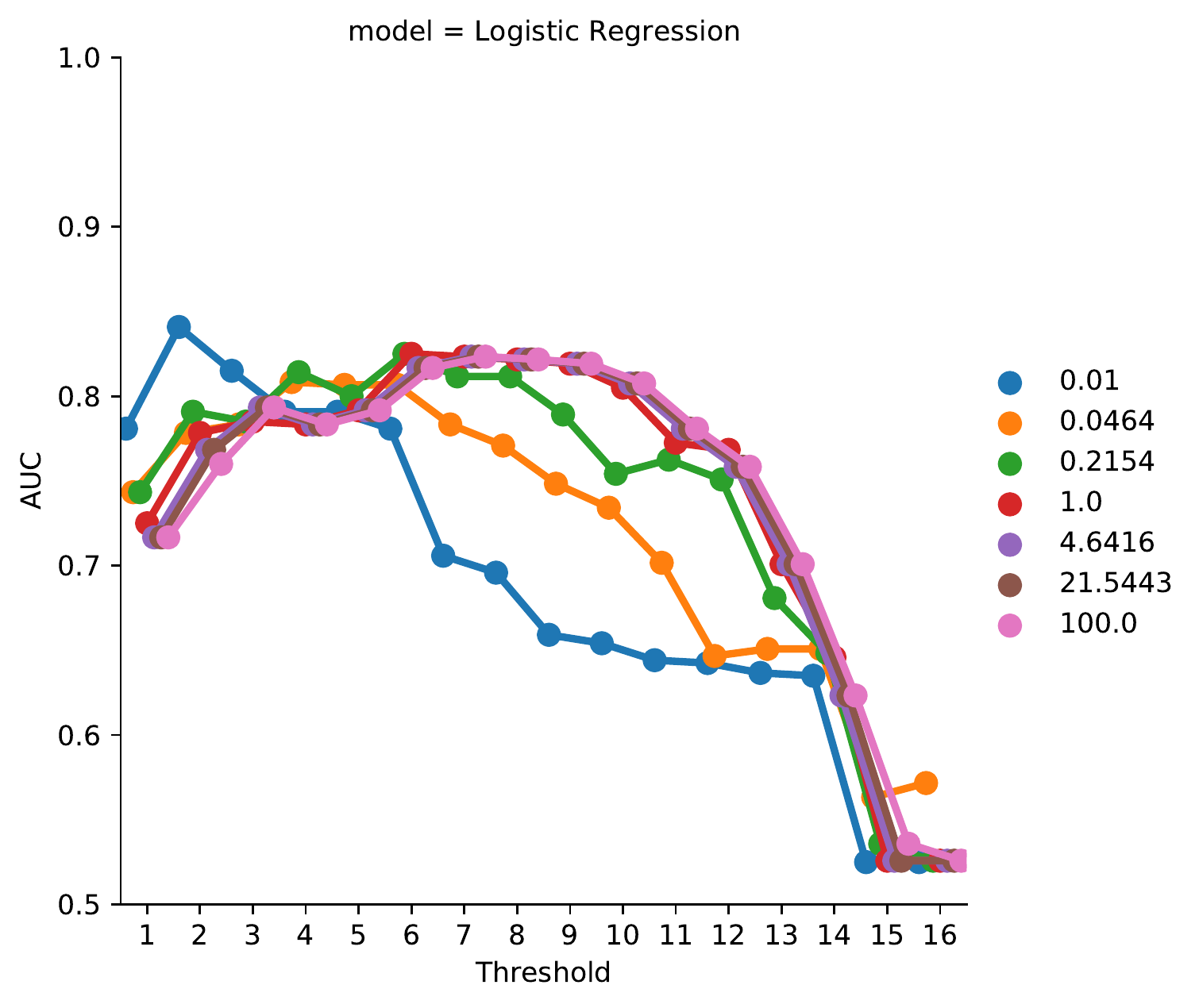}
\caption{}
\label{fig:ProbScalingMV-LR}
\end{subfigure}%
\begin{subfigure}{0.6\textwidth}
  \centering
  \includegraphics[width= \textwidth]{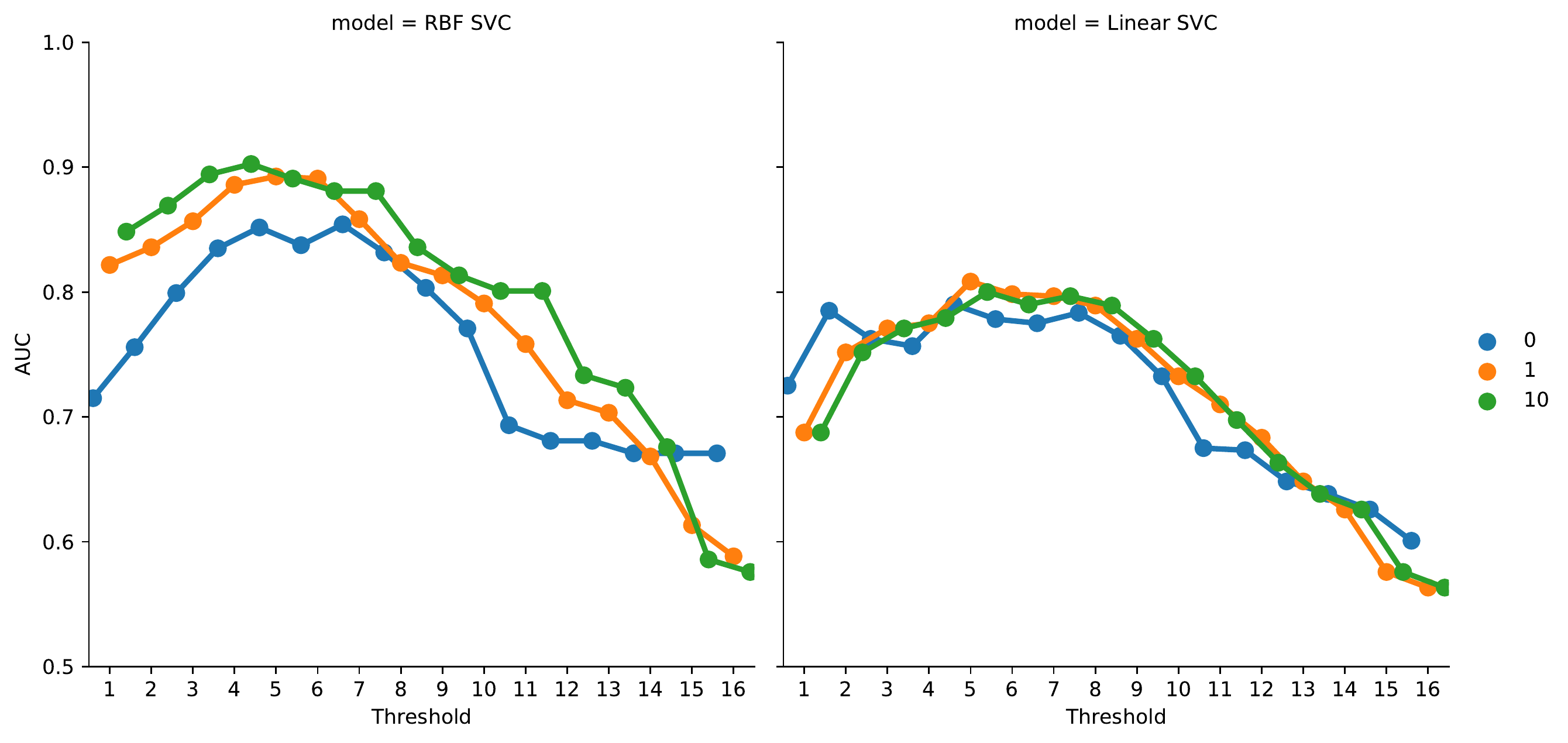}
\caption{}
\label{fig:ProbScalingMV-SVC}
\end{subfigure}%
\\
\begin{subfigure}{.99\textwidth}
  \centering
  \includegraphics[width= \textwidth]{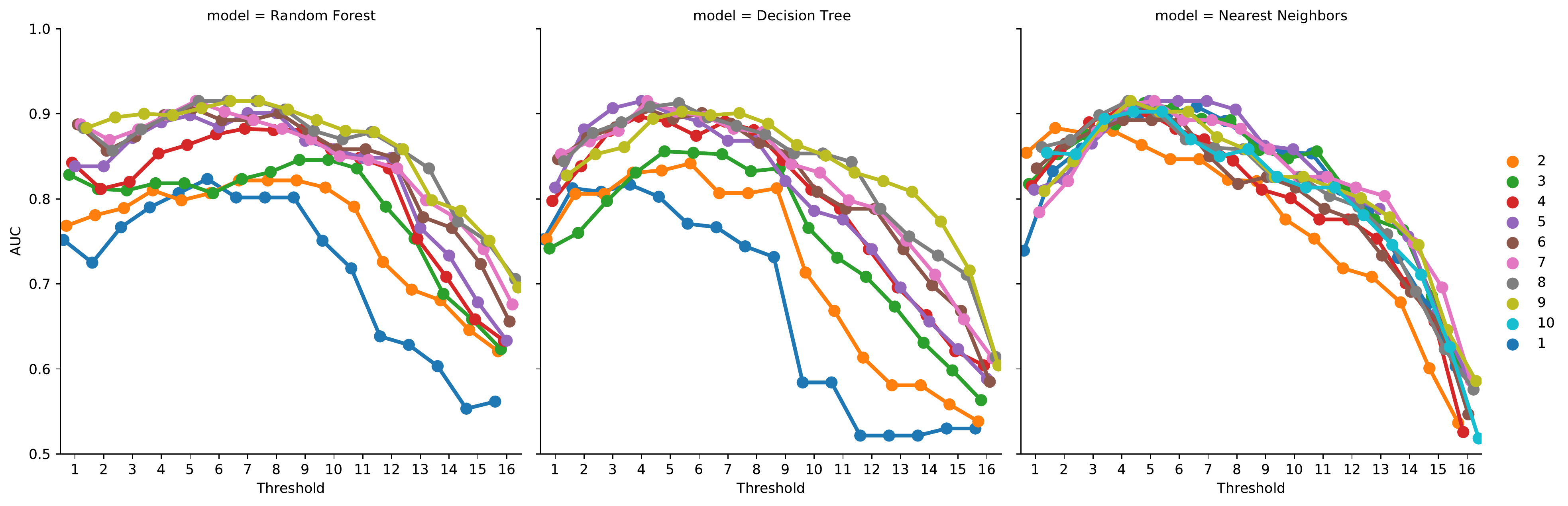}
\caption{}
\label{fig:ProbScalingMV-RF}
\end{subfigure}%
\caption{AUC for models with the probabilistic scaling variant of KB integration and HMV strategy as multi-perspective CRS prediction strategies, stratified by the hyperparameter values a) Logistic Regression model; b) RBF SVC and Linear SVC models; c) Random Forest, Decision Tree and Nearest Neighbors models}
\label{fig:ProbScalingHMV}
\end{figure}

\begin{figure}[h!]
\centering
\begin{subfigure}{0.35\textwidth}
  \centering
  \includegraphics[width= \textwidth]{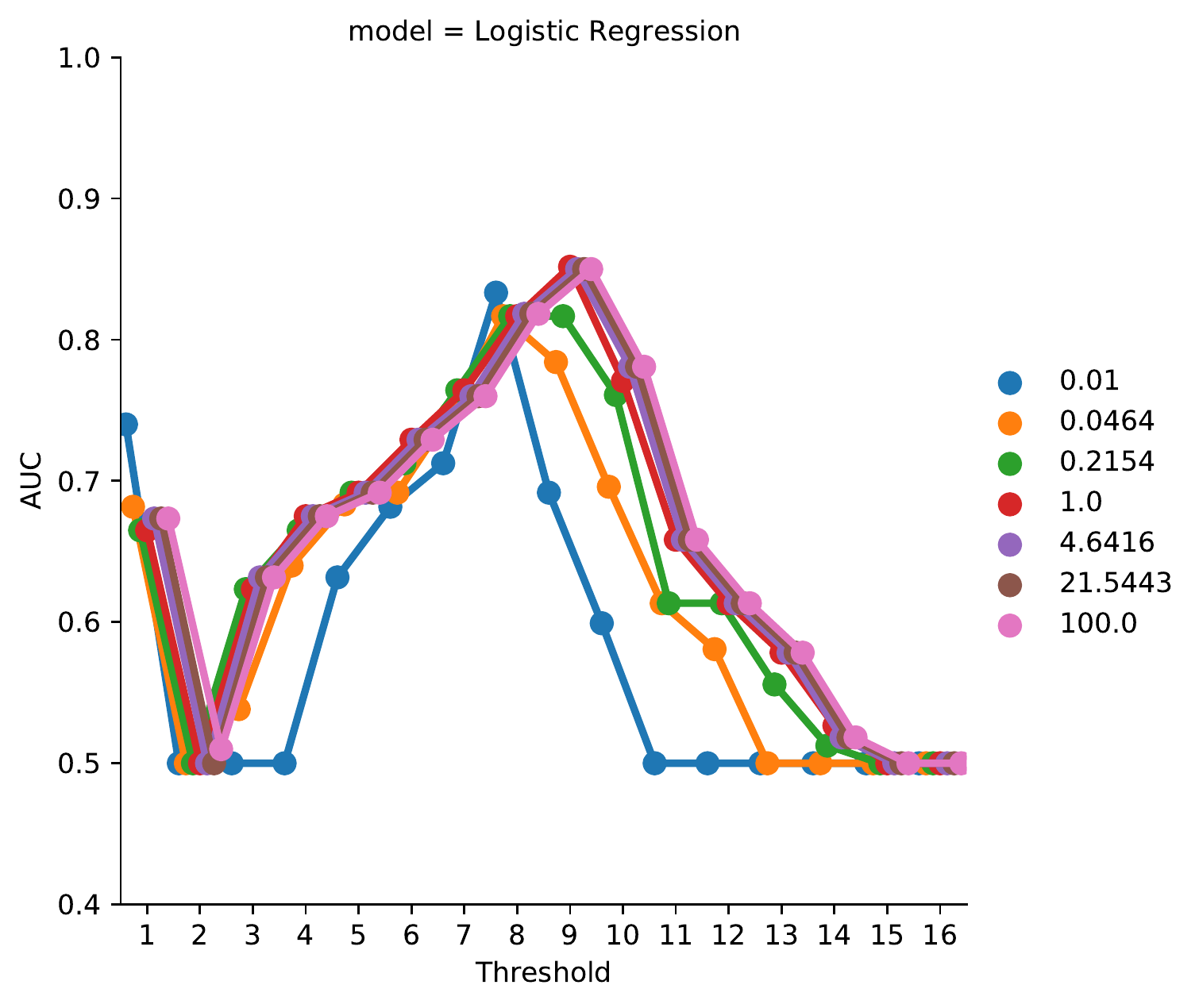}
\caption{}
\label{fig:ProbScalingMV-LR}
\end{subfigure}%
\begin{subfigure}{0.6\textwidth}
  \centering
  \includegraphics[width= \textwidth]{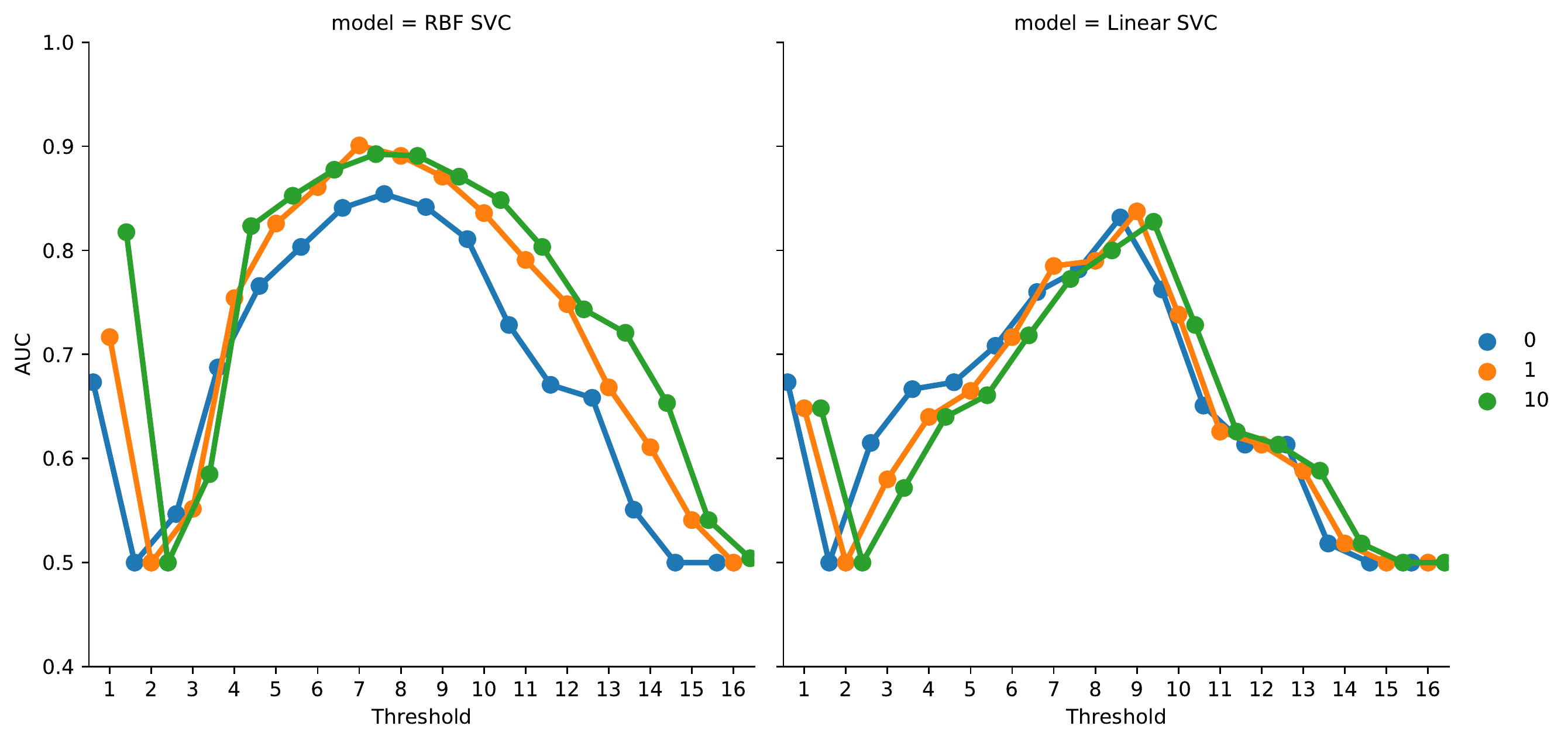}
\caption{}
\label{fig:ProbScalingMV-SVC}
\end{subfigure}%
\\
\begin{subfigure}{.99\textwidth}
  \centering
  \includegraphics[width= \textwidth]{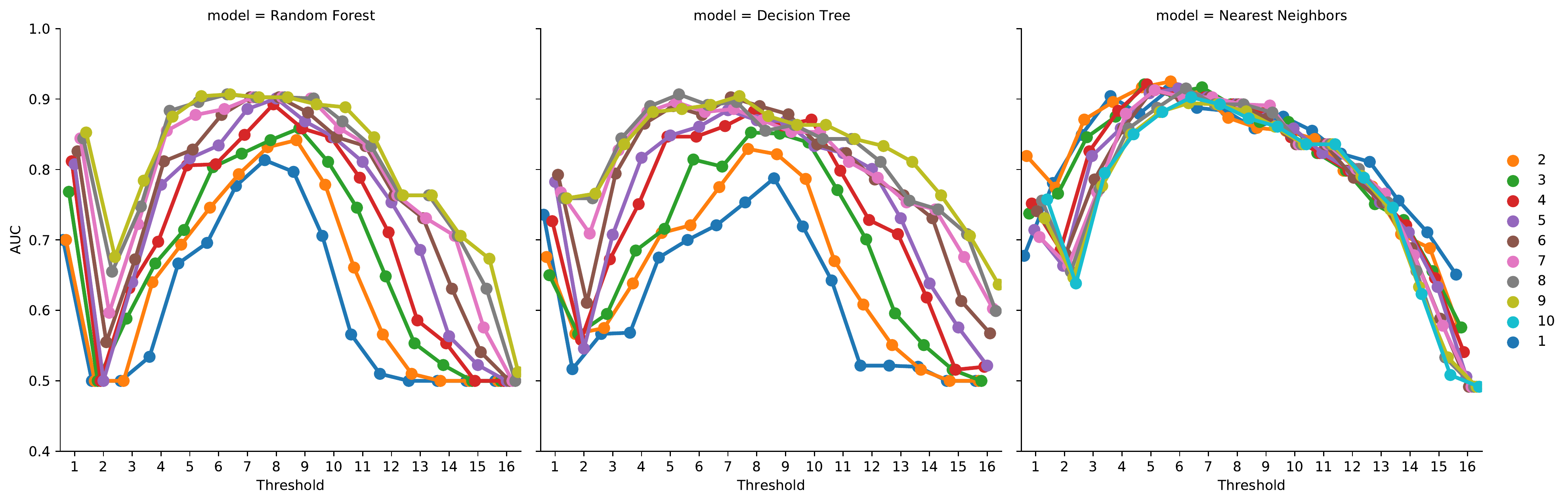}
\caption{}
\label{fig:ProbScalingMV-RF}
\end{subfigure}%
\caption{AUC for models with the probabilistic scaling variant of KB integration and wHMV strategy as multi-perspective CRS prediction strategies, stratified by the hyperparameter values a) Logistic Regression model; b) RBF SVC and Linear SVC models; c) Random Forest, Decision Tree and Nearest Neighbors models}
\label{fig:ProbScalingwHMV}
\end{figure}

\begin{figure}[h!]
\centering
\begin{subfigure}{0.35\textwidth}
  \centering
  \includegraphics[width= \textwidth]{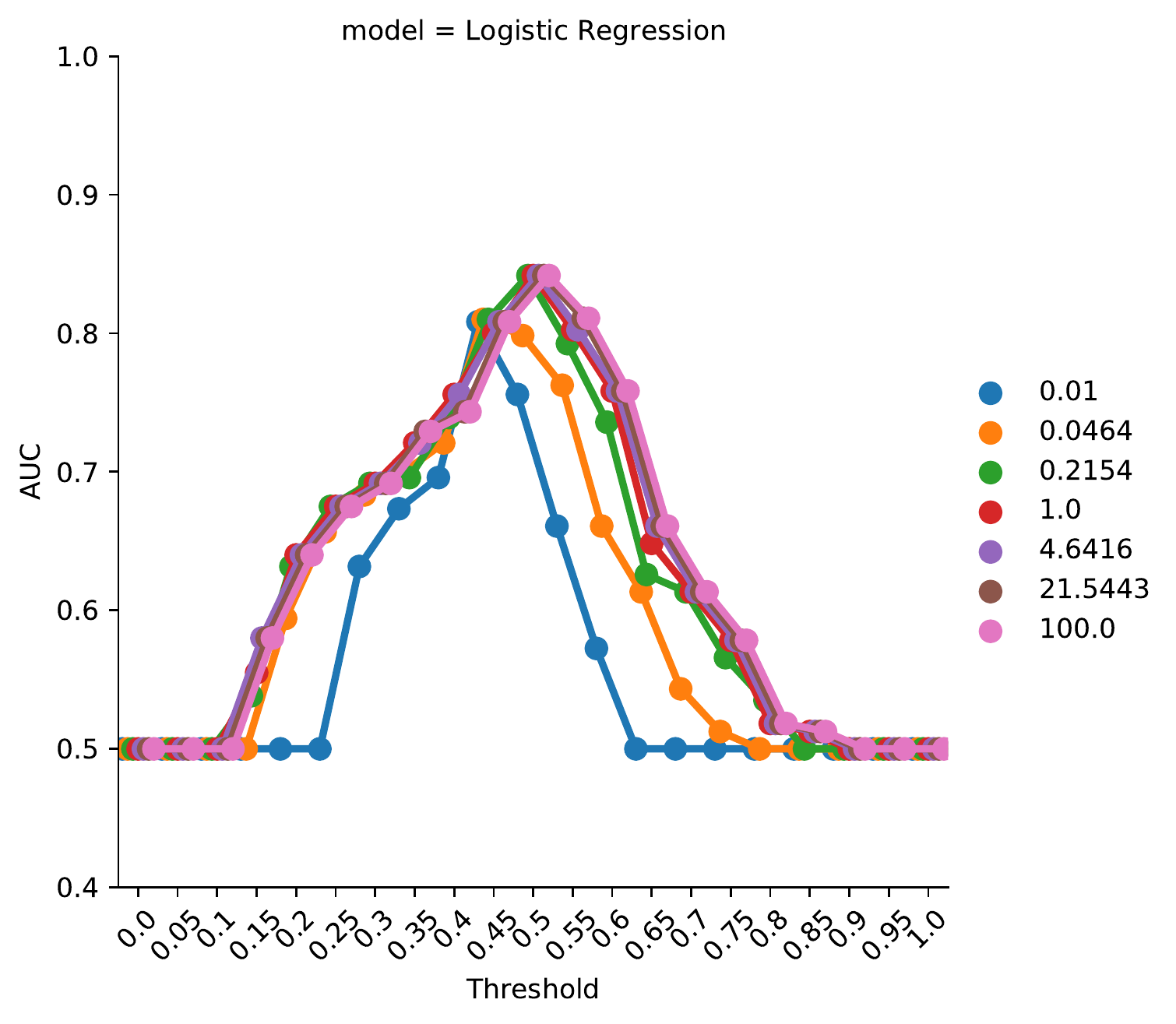}
\caption{}
\label{fig:ProbScalingMV-LR}
\end{subfigure}%
\begin{subfigure}{0.6\textwidth}
  \centering
  \includegraphics[width= \textwidth]{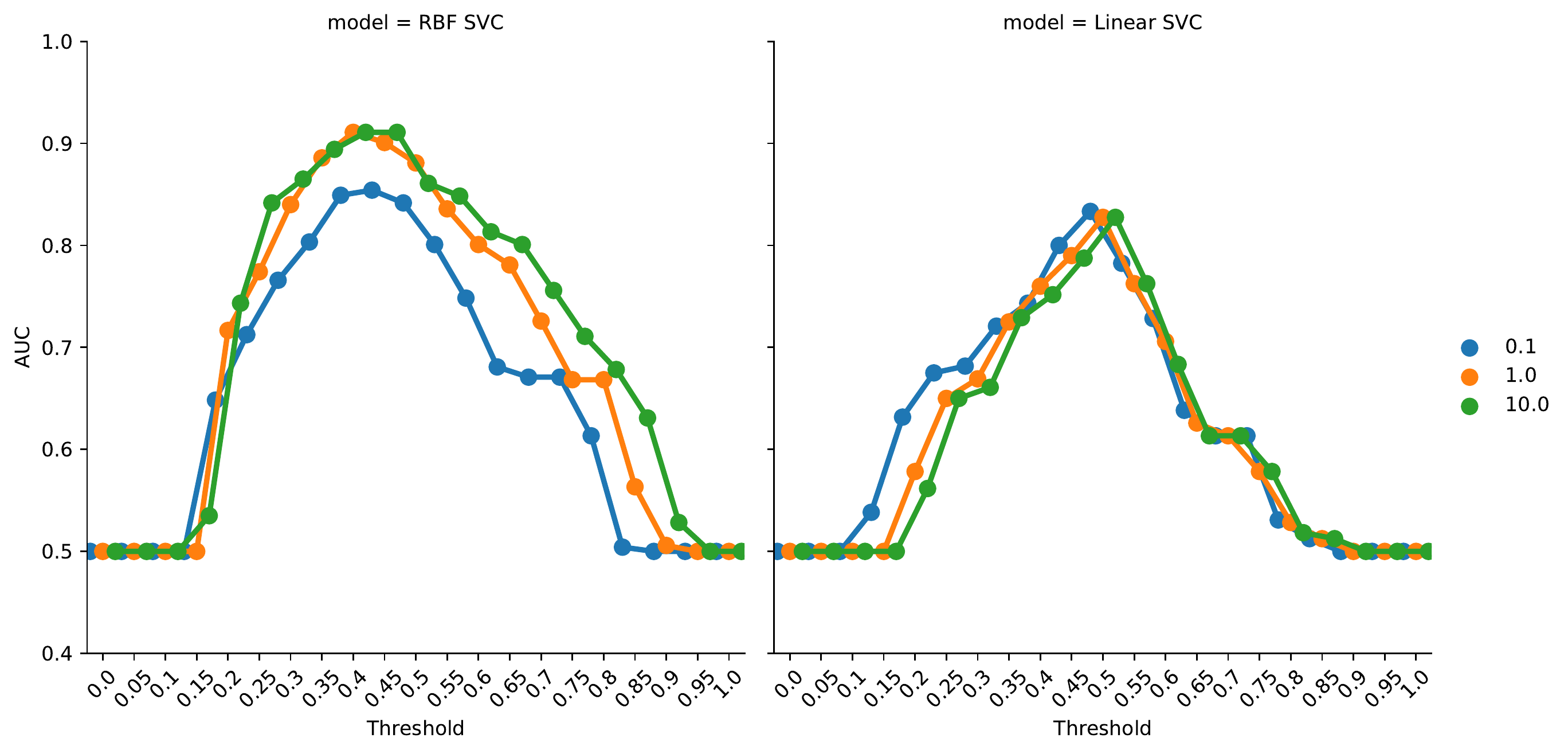}
\caption{}
\label{fig:ProbScalingMV-SVC}
\end{subfigure}%
\\
\begin{subfigure}{.99\textwidth}
  \centering
  \includegraphics[width= \textwidth]{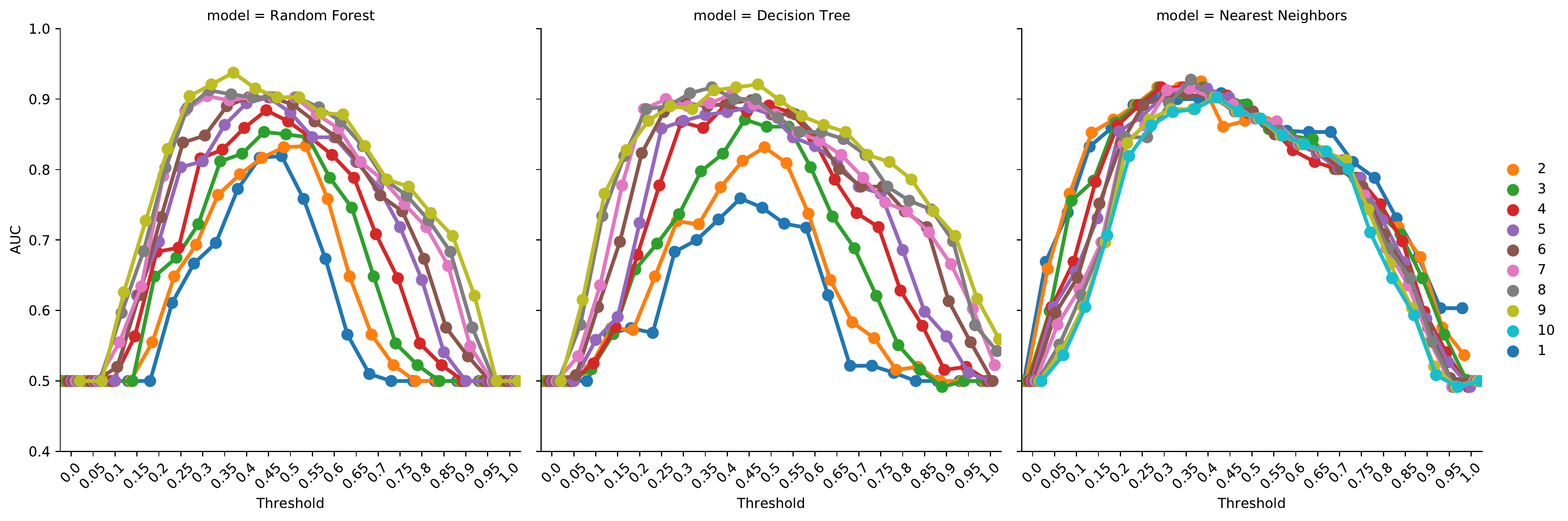}
\caption{}
\label{fig:ProbScalingMV-RF}
\end{subfigure}%
\caption{AUC for models with the probabilistic scaling variant of KB integration and SMV strategy as multi-perspective CRS prediction strategies, stratified by the hyperparameter values a) Logistic Regression model; b) RBF SVC and Linear SVC models; c) Random Forest, Decision Tree and Nearest Neighbors models}
\label{fig:ProbScalingSMV}
\end{figure}

\begin{figure}[h!]
\centering
\begin{subfigure}{0.35\textwidth}
  \centering
  \includegraphics[width= \textwidth]{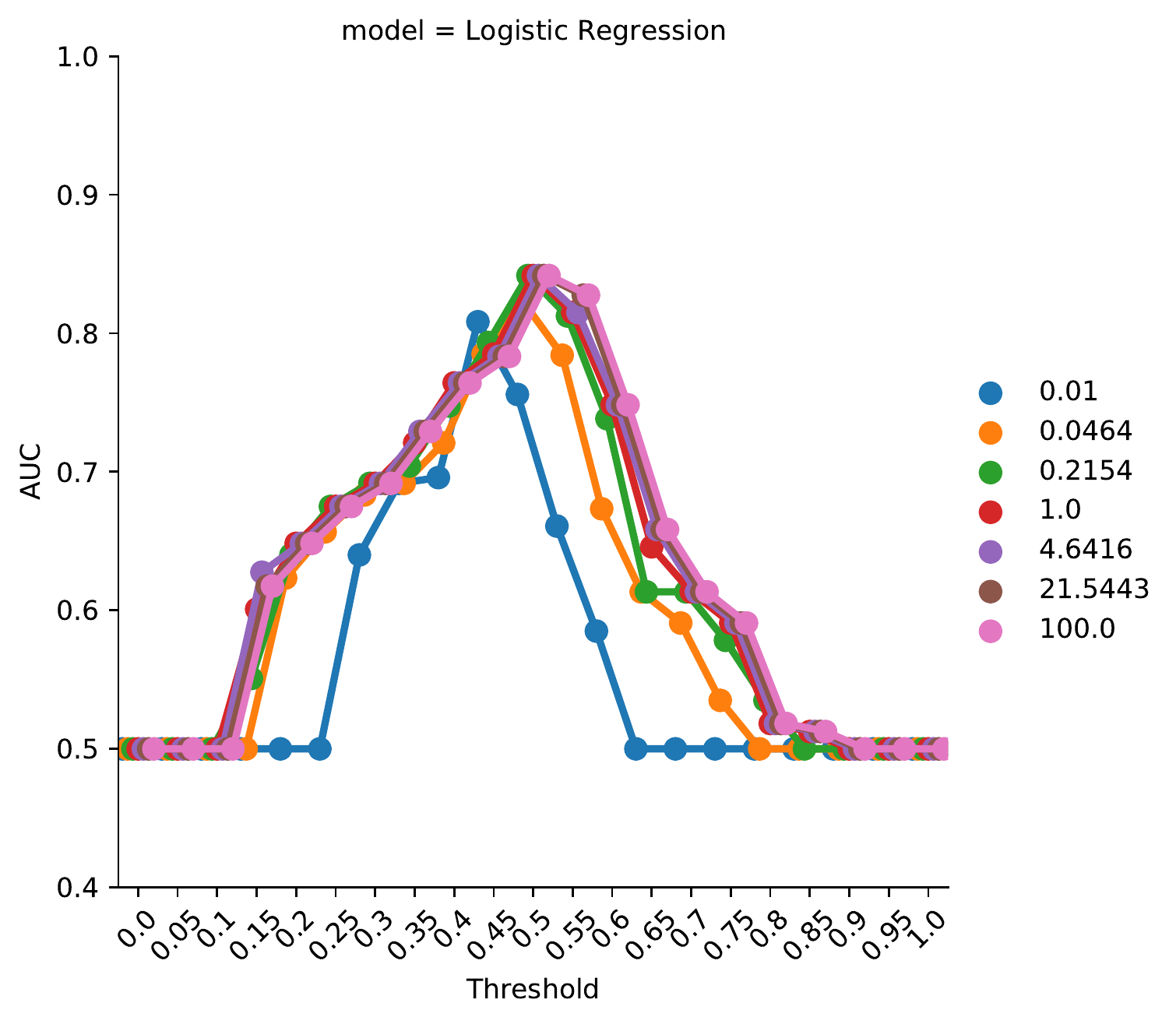}
\caption{}
\label{fig:ProbScalingMV-LR}
\end{subfigure}%
\begin{subfigure}{0.6\textwidth}
  \centering
  \includegraphics[width= \textwidth]{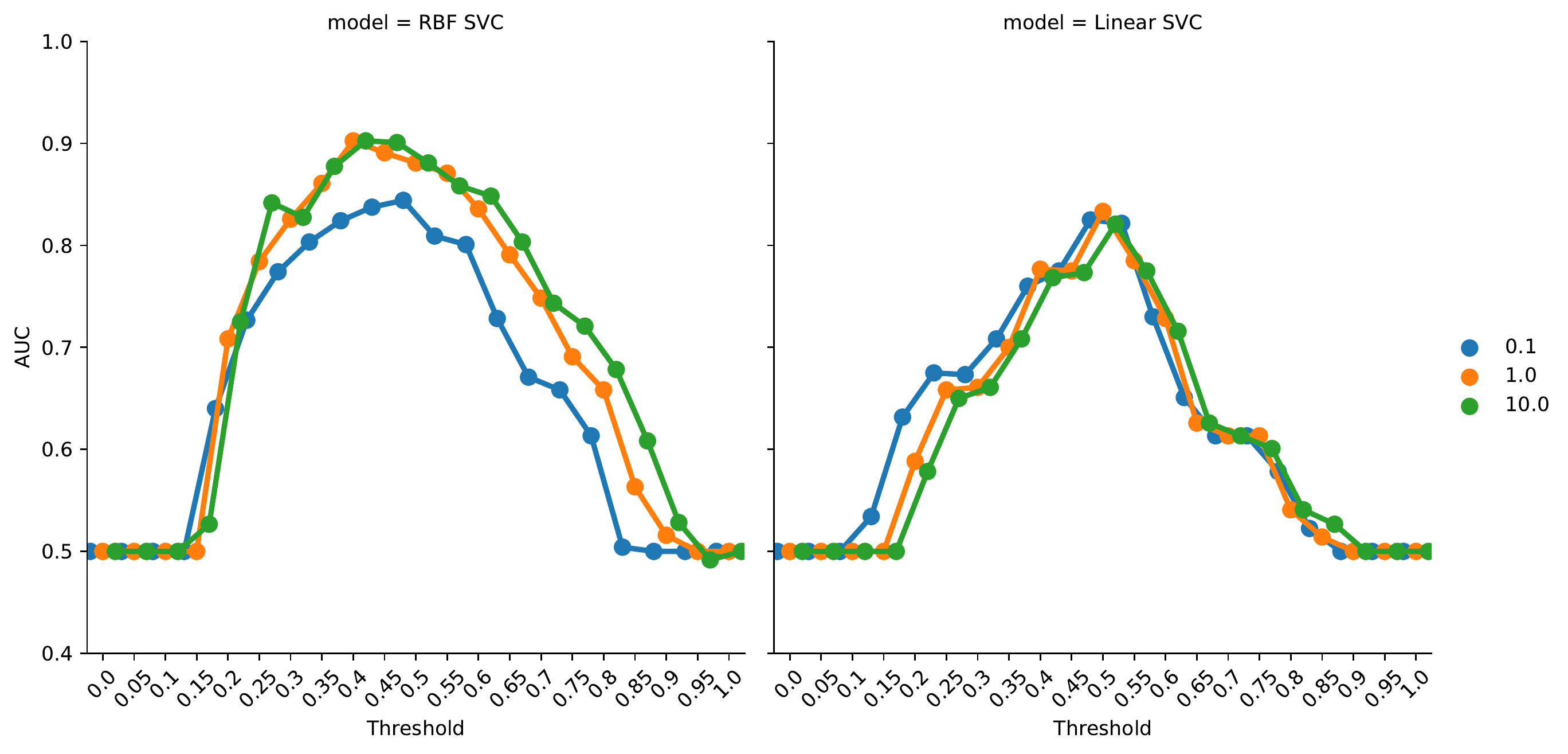}
\caption{}
\label{fig:ProbScalingMV-SVC}
\end{subfigure}%
\\
\begin{subfigure}{.99\textwidth}
  \centering
  \includegraphics[width= \textwidth]{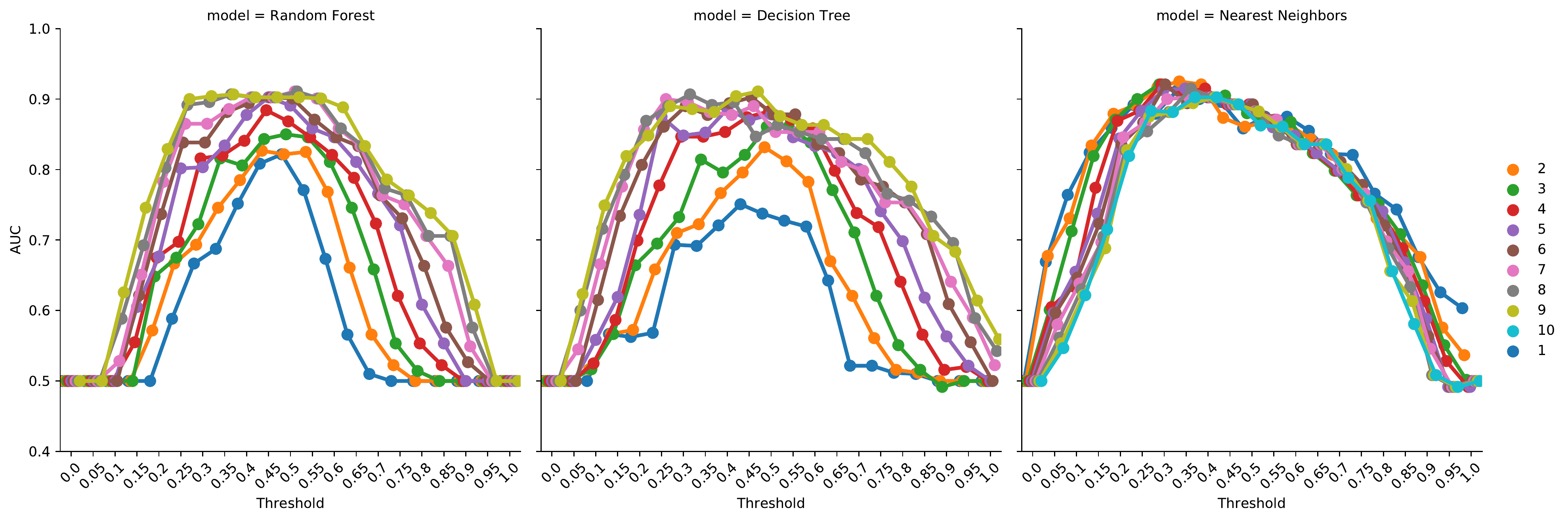}
\caption{}
\label{fig:ProbScalingMV-RF}
\end{subfigure}%
\caption{AUC for models with the probabilistic scaling variant of KB integration and wSMV strategy as multi-perspective CRS prediction strategies, stratified by the hyperparameter values a) Logistic Regression model; b) RBF SVC and Linear SVC models; c) Random Forest, Decision Tree and Nearest Neighbors models}
\label{fig:ProbScalingwSMV}
\end{figure}

\begin{figure}[h!]
\centering
\begin{subfigure}{0.35\textwidth}
  \centering
  \includegraphics[width= \textwidth]{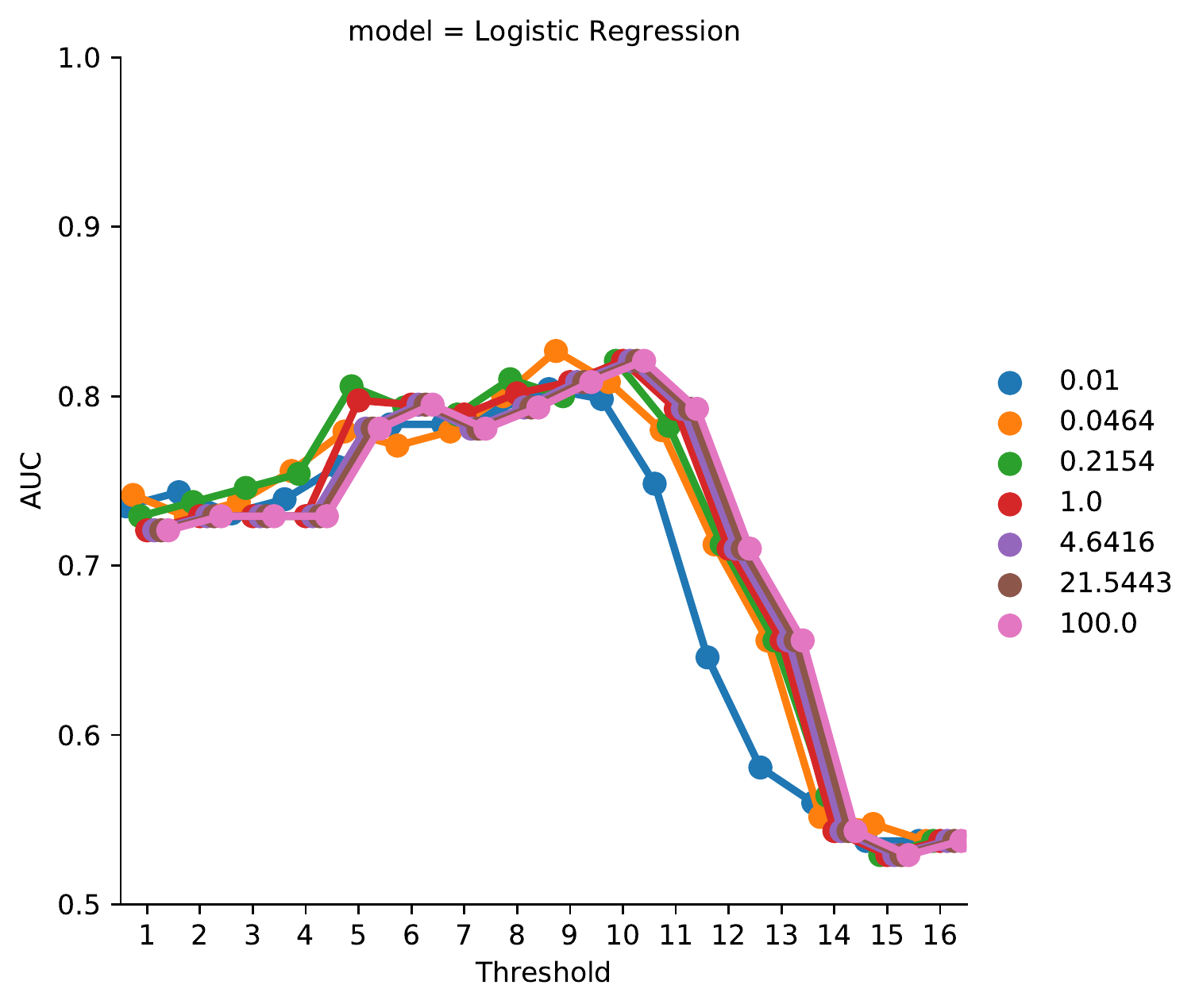}
\caption{}
\label{fig:ProbScalingMV-LR}
\end{subfigure}%
\begin{subfigure}{0.6\textwidth}
  \centering
  \includegraphics[width= \textwidth]{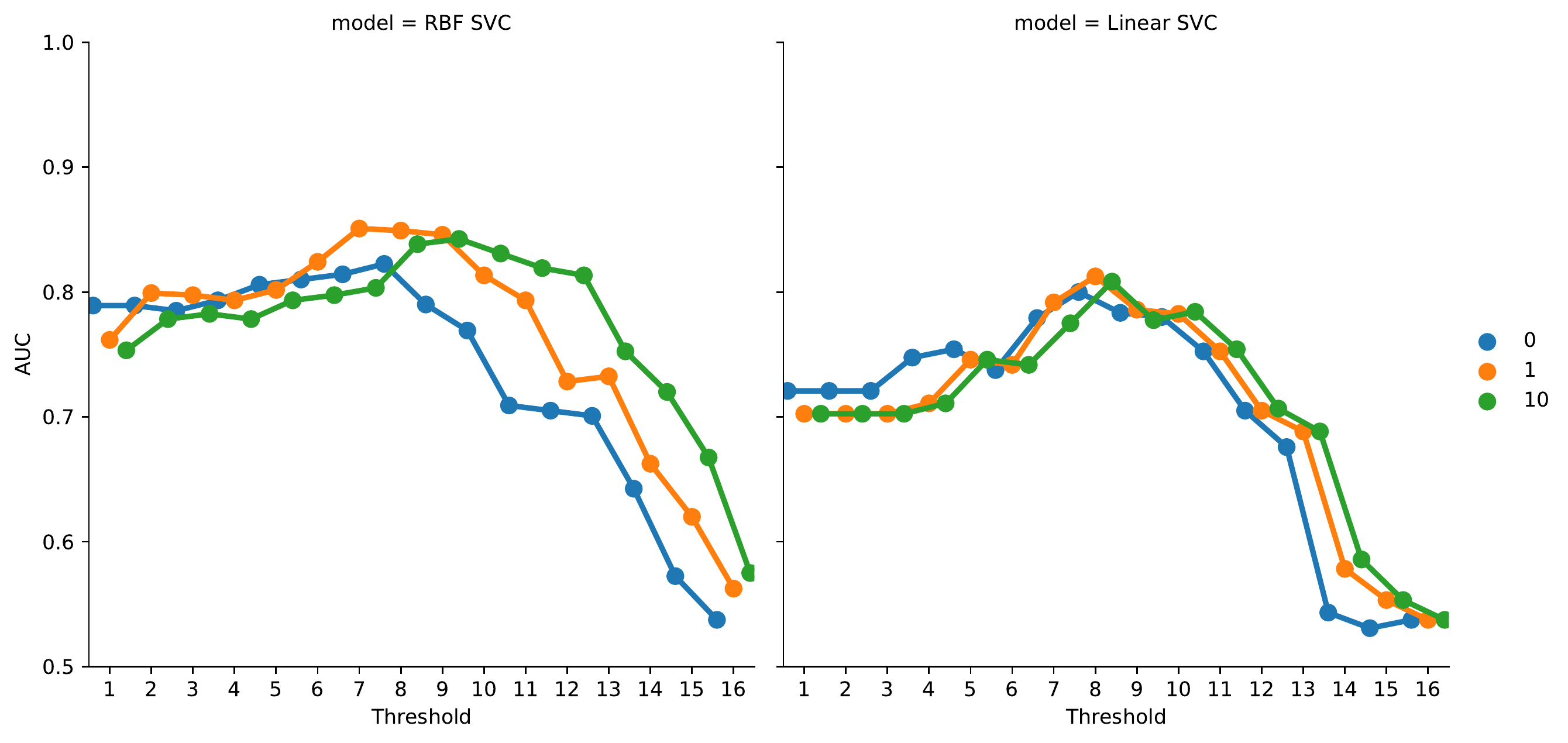}
\caption{}
\label{fig:ProbScalingMV-SVC}
\end{subfigure}%
\\
\begin{subfigure}{.99\textwidth}
  \centering
  \includegraphics[width= \textwidth]{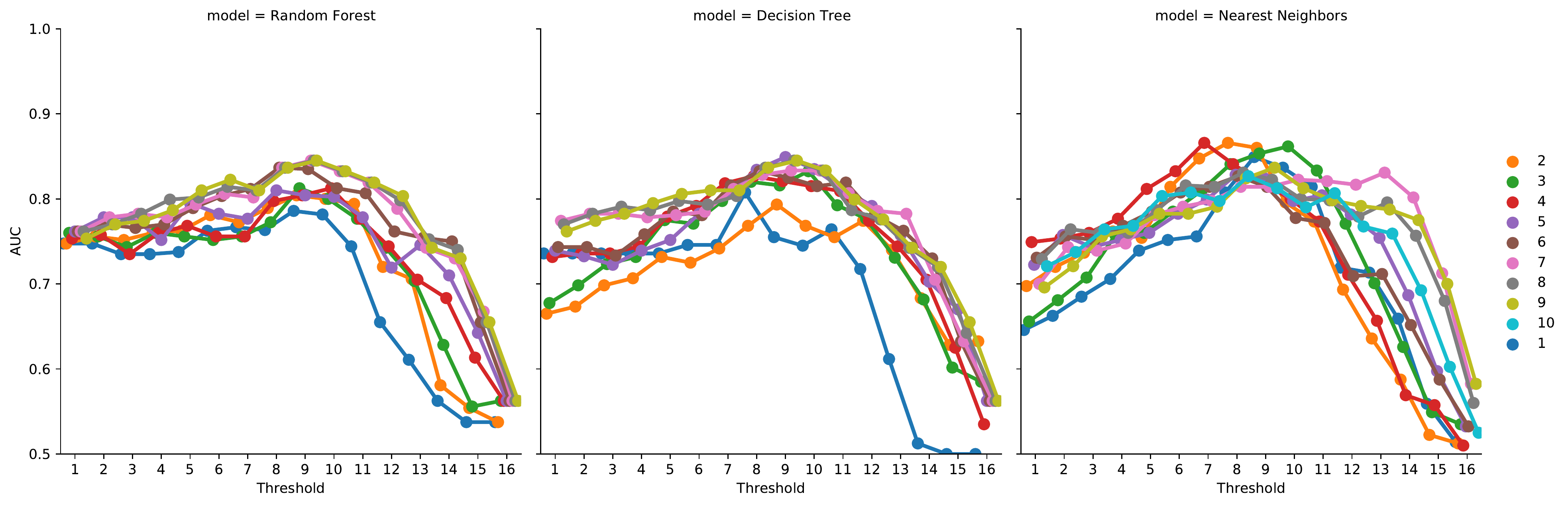}
\caption{}
\label{fig:ProbScalingMV-RF}
\end{subfigure}%
\caption{AUC for models with the binary scaling variant of KB integration and HMV strategy as multi-perspective CRS prediction strategies, stratified by the hyperparameter values a) Logistic Regression model; b) RBF SVC and Linear SVC models; c) Random Forest, Decision Tree and Nearest Neighbors models}
\label{fig:BinaryScalingHMV}
\end{figure}

\begin{figure}[h!]
\centering
\begin{subfigure}{0.35\textwidth}
  \centering
  \includegraphics[width= \textwidth]{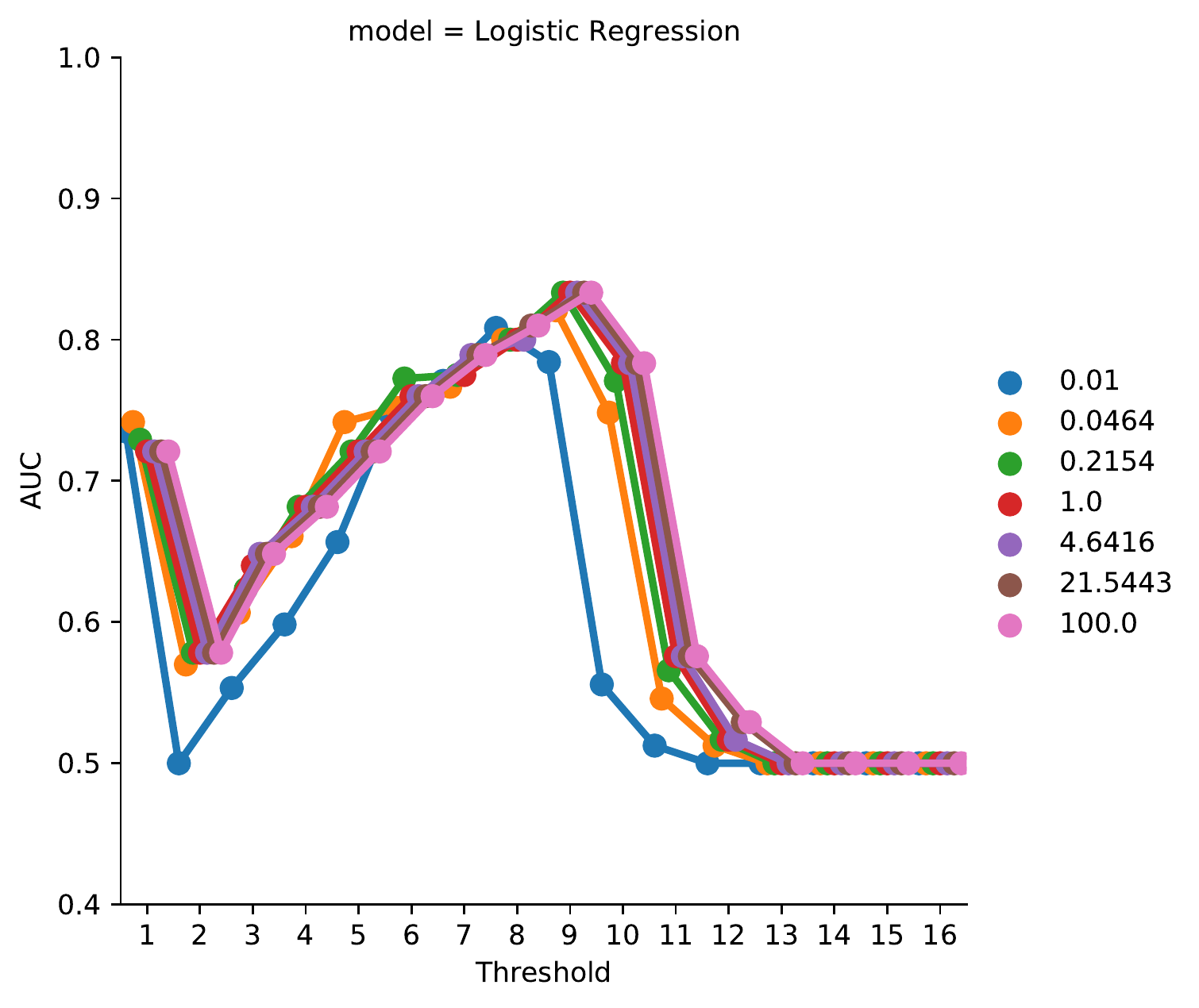}
\caption{}
\label{fig:ProbScalingMV-LR}
\end{subfigure}%
\begin{subfigure}{0.6\textwidth}
  \centering
  \includegraphics[width= \textwidth]{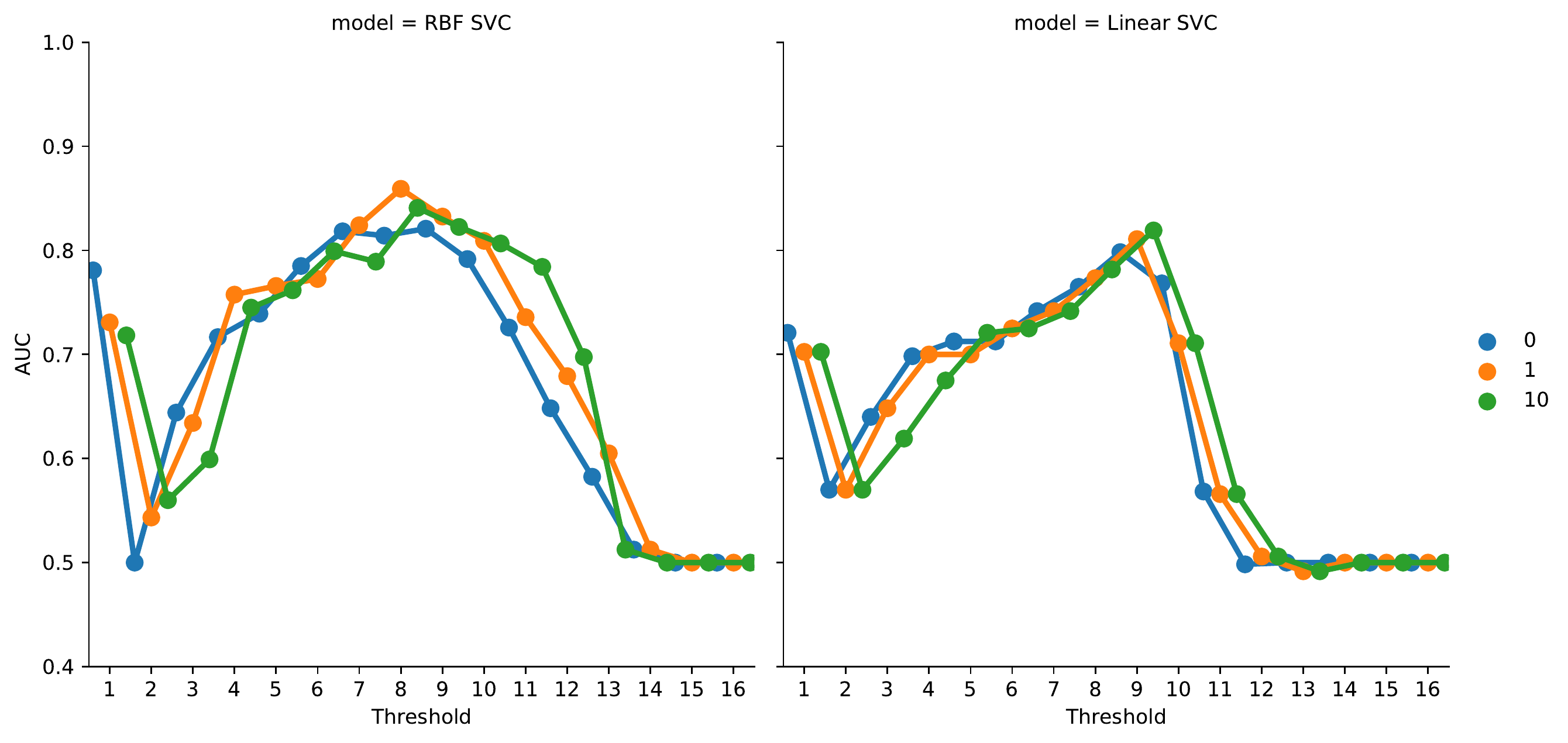}
\caption{}
\label{fig:ProbScalingMV-SVC}
\end{subfigure}%
\\
\begin{subfigure}{.99\textwidth}
  \centering
  \includegraphics[width= \textwidth]{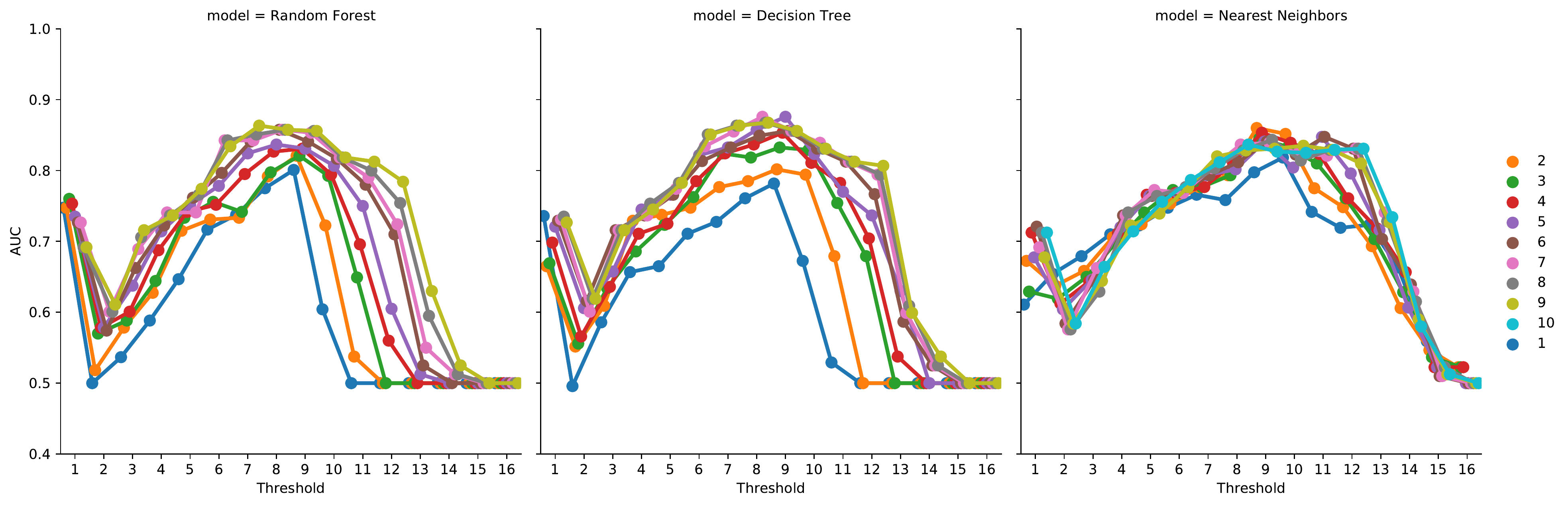}
\caption{}
\label{fig:ProbScalingMV-RF}
\end{subfigure}%
\caption{AUC for models with the binary scaling variant of KB integration and wHMV strategy as multi-perspective CRS prediction strategies, stratified by the hyperparameter values a) Logistic Regression model; b) RBF SVC and Linear SVC models; c) Random Forest, Decision Tree and Nearest Neighbors models}
\label{fig:BinaryScalingwHMV}
\end{figure}

\begin{figure}[h!]
\centering
\begin{subfigure}{0.35\textwidth}
  \centering
  \includegraphics[width= \textwidth]{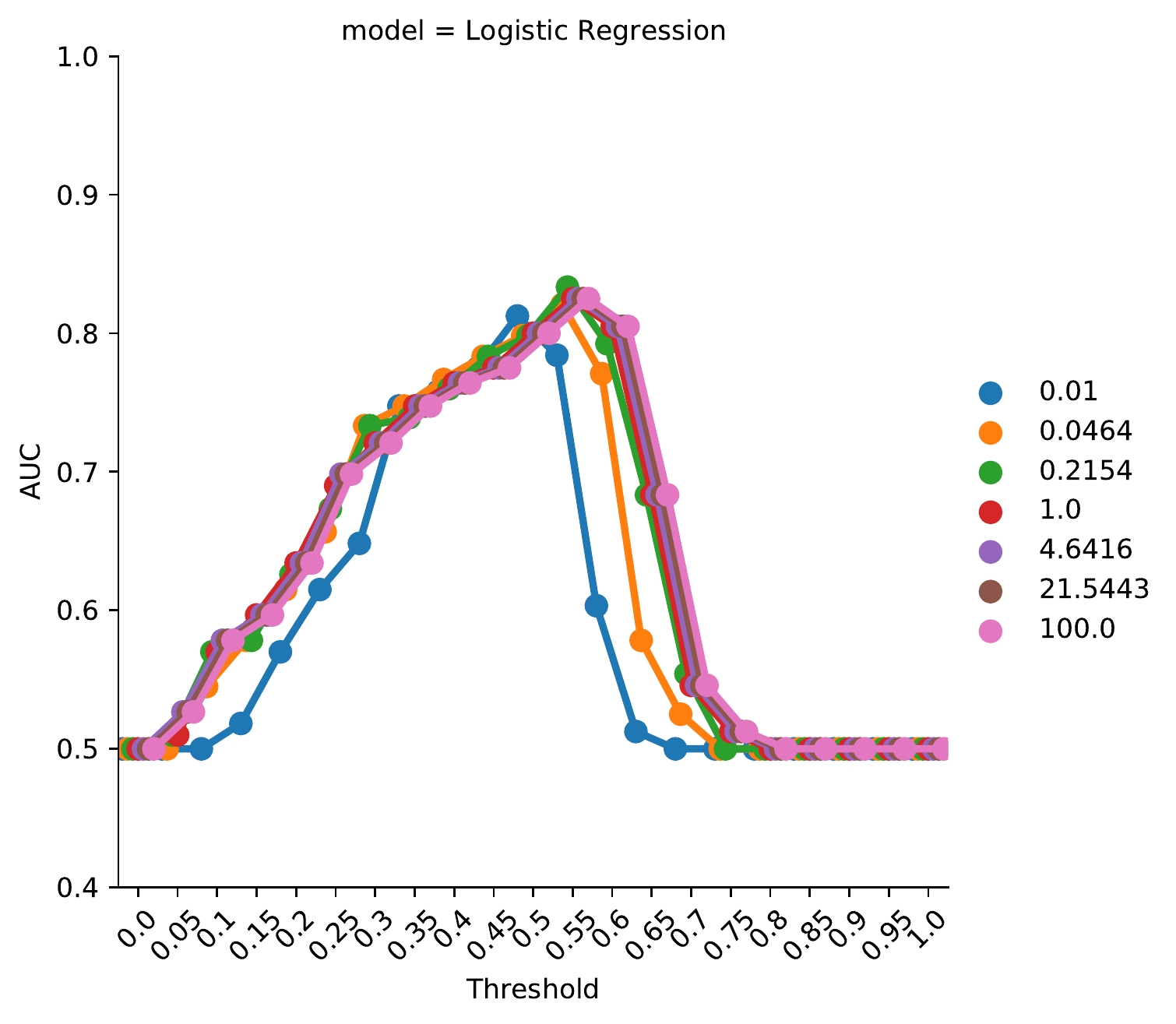}
\caption{}
\label{fig:ProbScalingMV-LR}
\end{subfigure}%
\begin{subfigure}{0.6\textwidth}
  \centering
  \includegraphics[width= \textwidth]{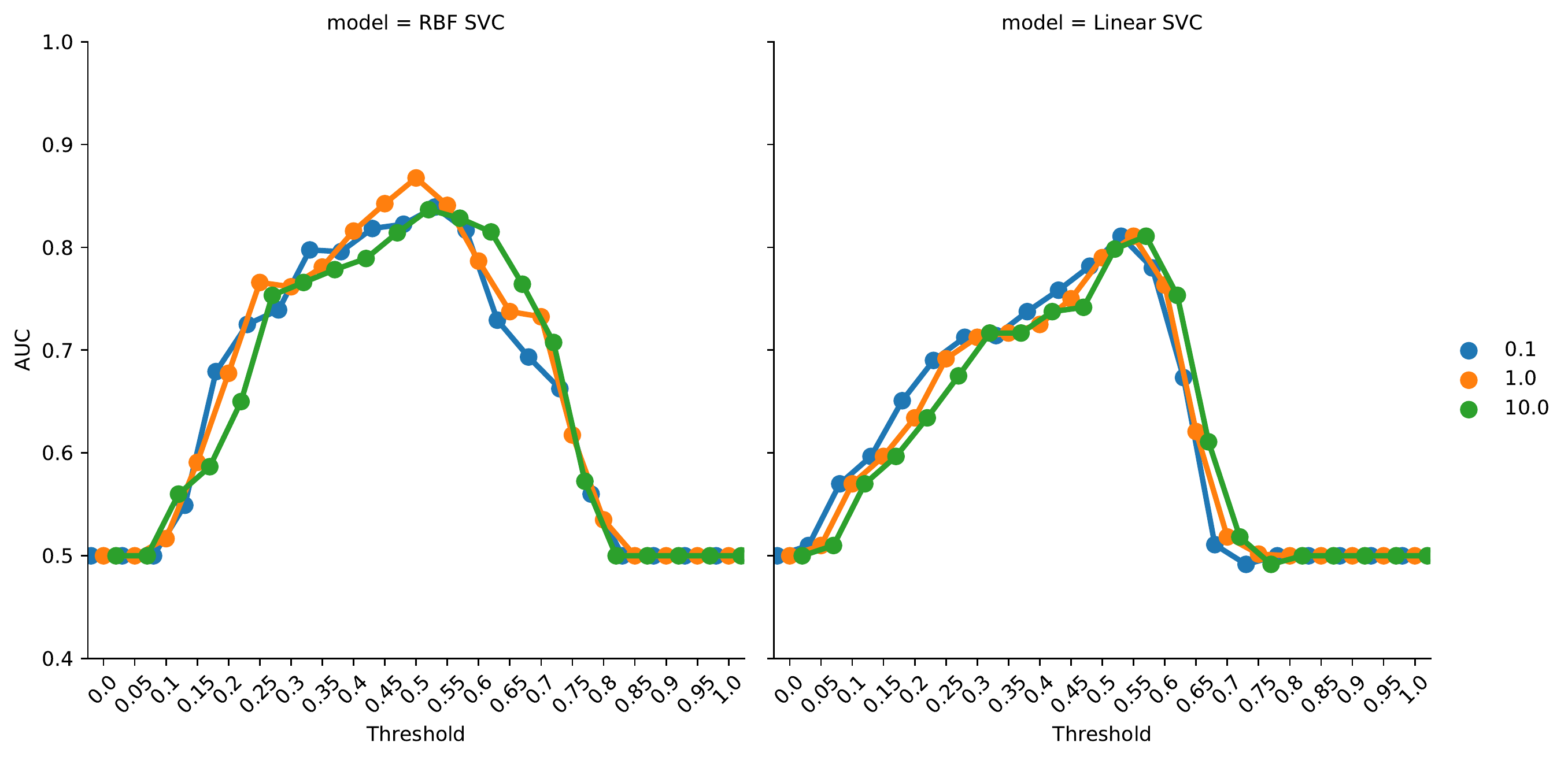}
\caption{}
\label{fig:ProbScalingMV-SVC}
\end{subfigure}%
\\
\begin{subfigure}{.99\textwidth}
  \centering
  \includegraphics[width= \textwidth]{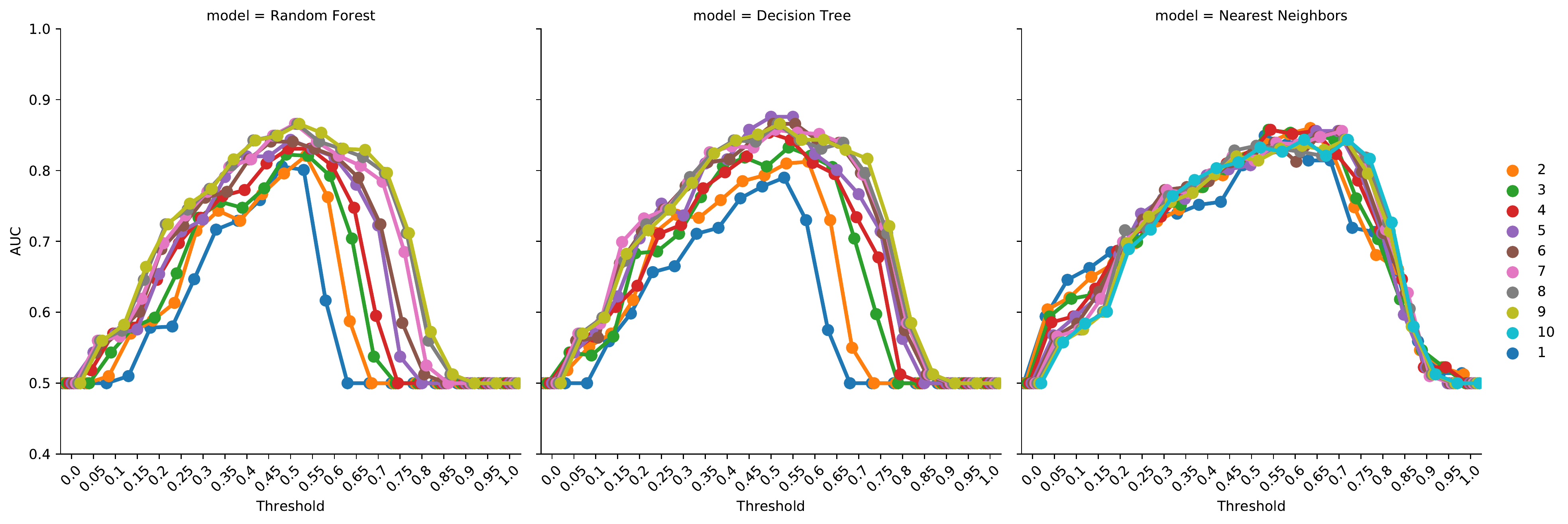}
\caption{}
\label{fig:ProbScalingMV-RF}
\end{subfigure}%
\caption{AUC for models with the binary scaling variant of KB integration and SMV strategy as multi-perspective CRS prediction strategies, stratified by the hyperparameter values a) Logistic Regression model; b) RBF SVC and Linear SVC models; c) Random Forest, Decision Tree and Nearest Neighbors models}
\label{fig:BinaryScalingSMV}
\end{figure}

\begin{figure}[h!]
\centering
\begin{subfigure}{0.35\textwidth}
  \centering
  \includegraphics[width= \textwidth]{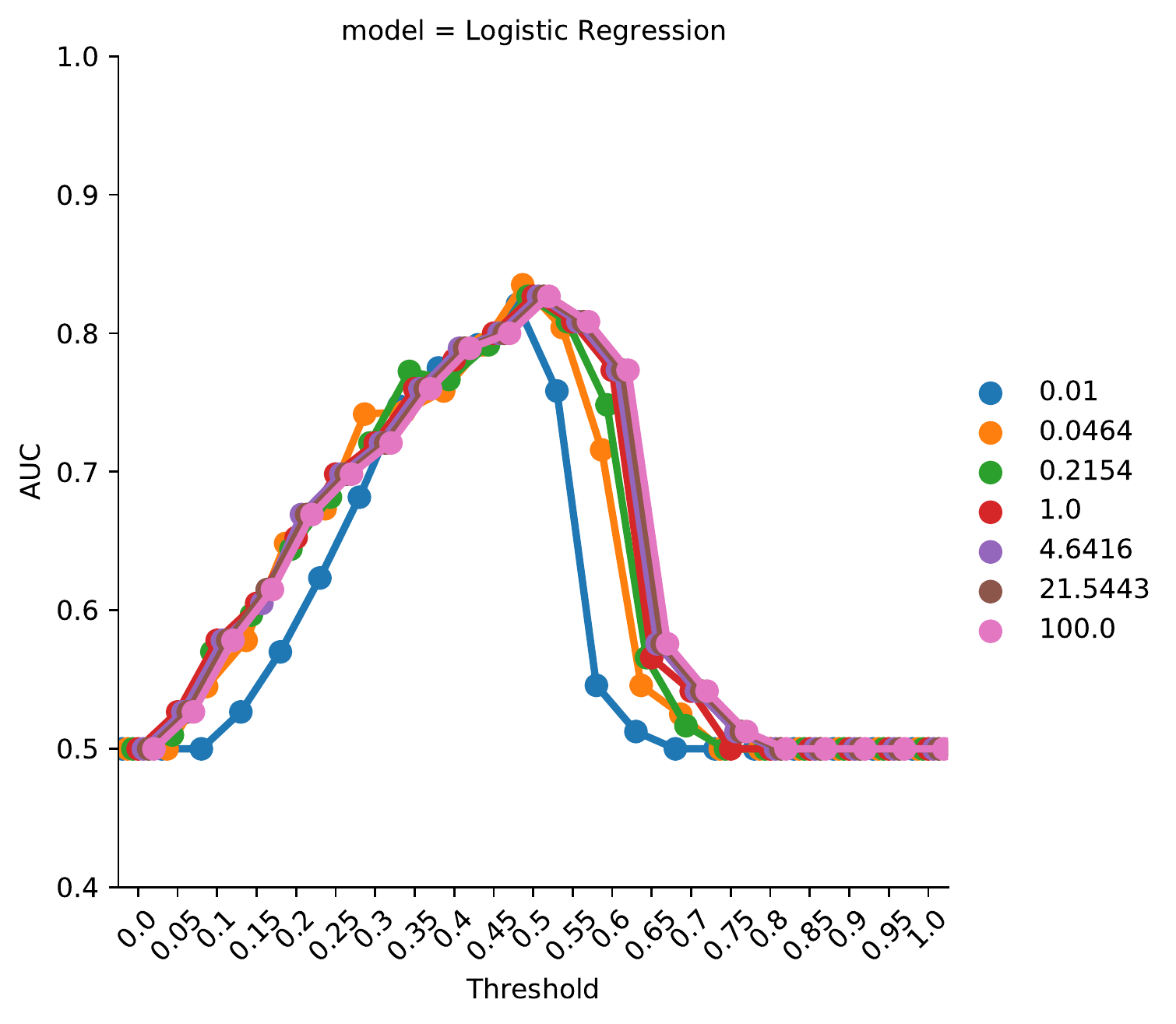}
\caption{}
\label{fig:ProbScalingMV-LR}
\end{subfigure}%
\begin{subfigure}{0.6\textwidth}
  \centering
  \includegraphics[width= \textwidth]{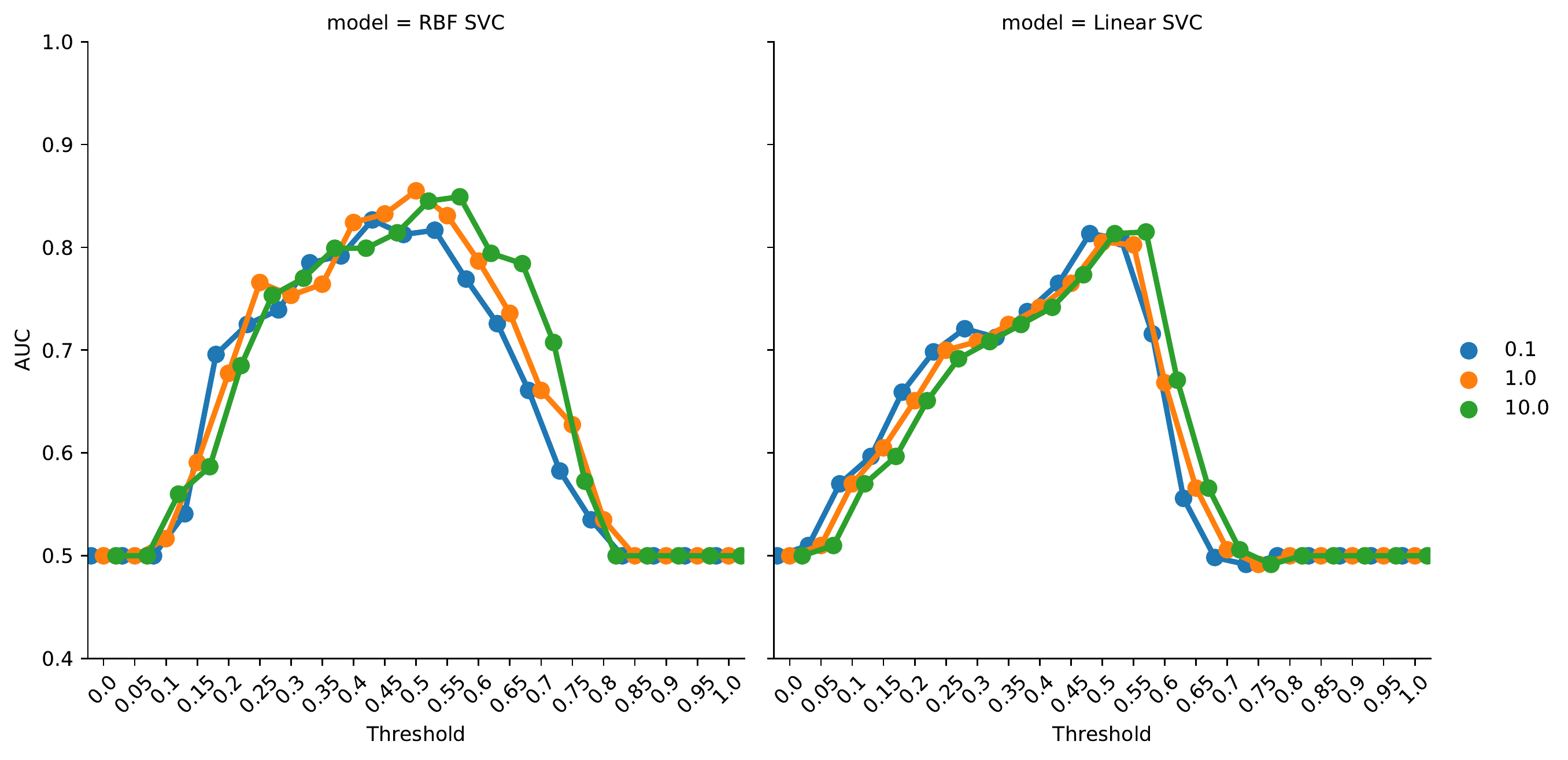}
\caption{}
\label{fig:ProbScalingMV-SVC}
\end{subfigure}%
\\
\begin{subfigure}{.99\textwidth}
  \centering
  \includegraphics[width= \textwidth]{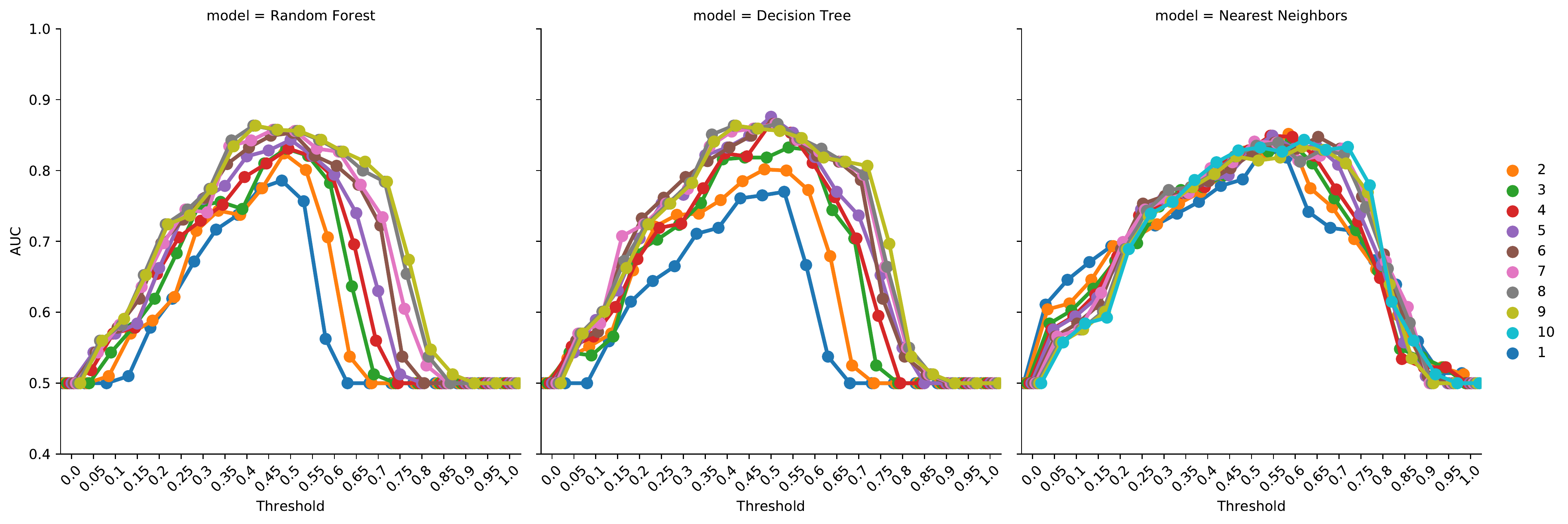}
\caption{}
\label{fig:ProbScalingMV-RF}
\end{subfigure}%
\caption{AUC for models with the binary scaling variant of KB integration and wSMV strategy as multi-perspective CRS prediction strategies, stratified by the hyperparameter values a) Logistic Regression model; b) RBF SVC and Linear SVC models; c) Random Forest, Decision Tree and Nearest Neighbors models}
\label{fig:BinaryScalingwSMV}
\end{figure}

\begin{figure}[h!]
\centering
\begin{subfigure}{0.35\textwidth}
  \centering
  \includegraphics[width= \textwidth]{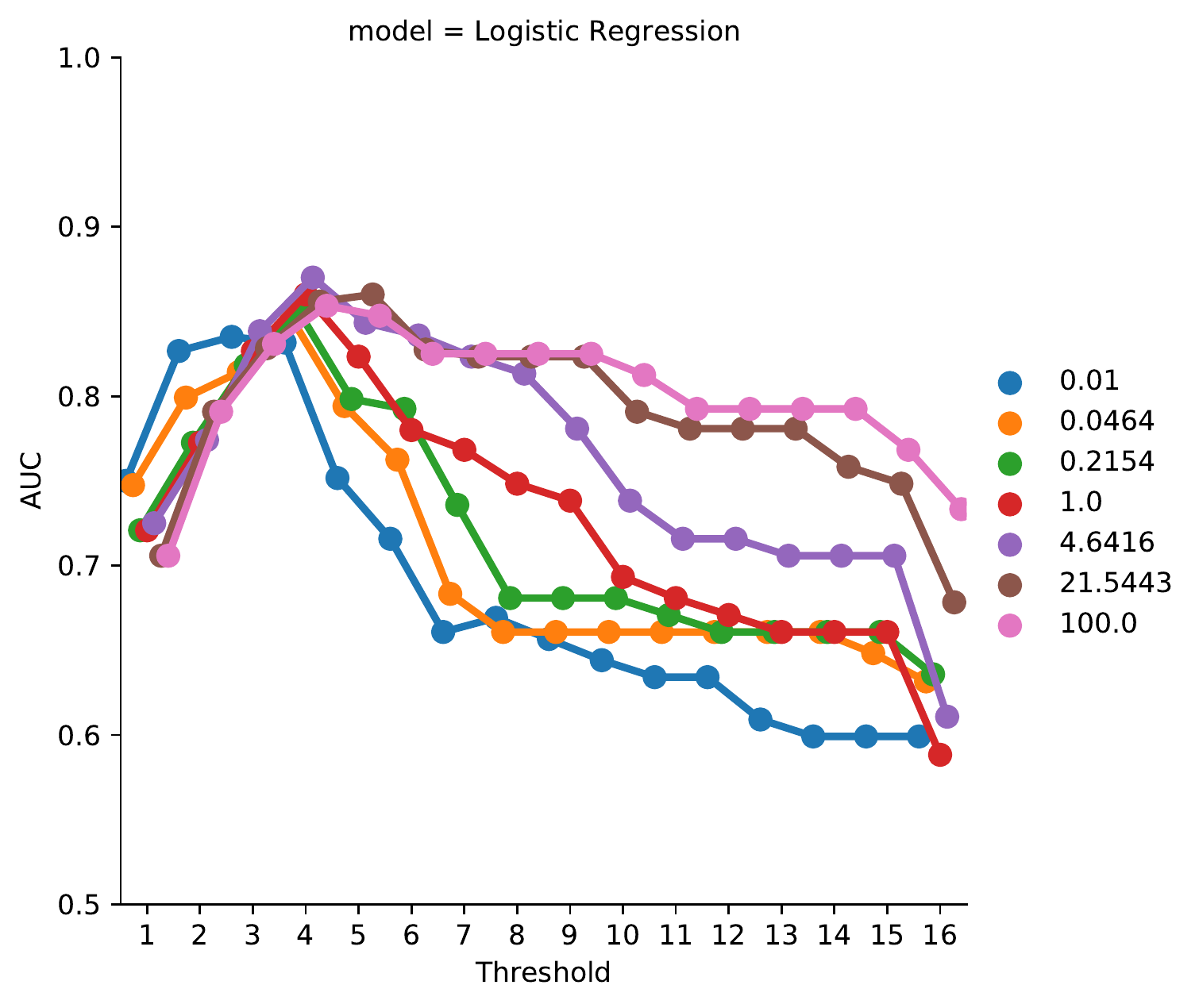}
\caption{}
\label{fig:ProbScalingMV-LR}
\end{subfigure}%
\begin{subfigure}{0.6\textwidth}
  \centering
  \includegraphics[width= \textwidth]{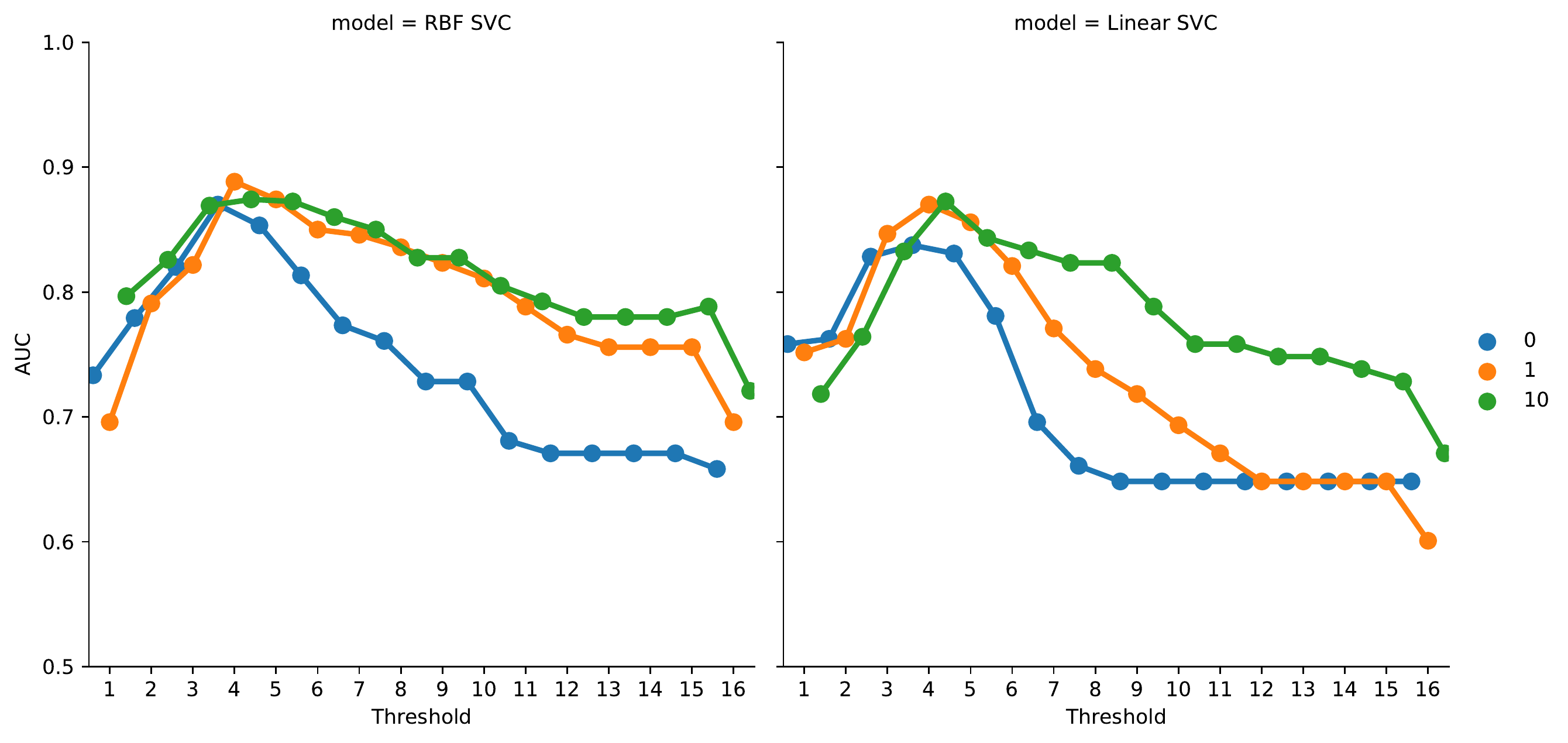}
\caption{}
\label{fig:ProbScalingMV-SVC}
\end{subfigure}%
\\
\begin{subfigure}{.99\textwidth}
  \centering
  \includegraphics[width= \textwidth]{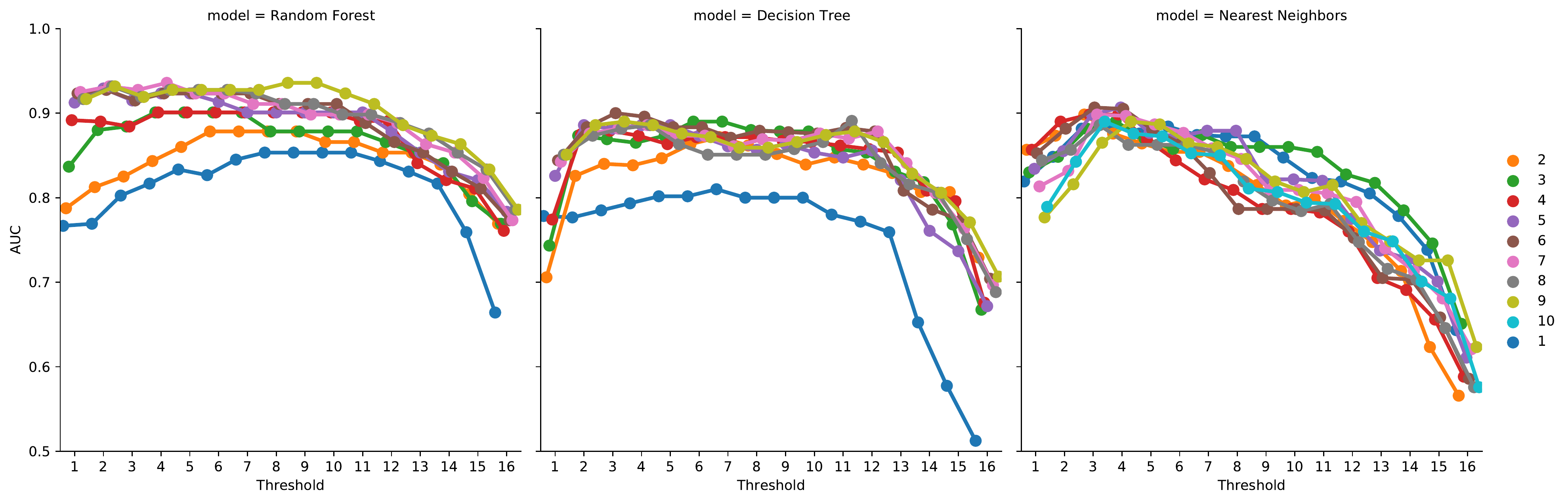}
\caption{}
\label{fig:ProbScalingMV-RF}
\end{subfigure}%
\caption{AUC for models with the continuous scaling variant of KB integration and HMV strategy as multi-perspective CRS prediction strategies, stratified by the hyperparameter values a) Logistic Regression model; b) RBF SVC and Linear SVC models; c) Random Forest, Decision Tree and Nearest Neighbors models}
\label{fig:ContinuousScalingHMV}
\end{figure}

\begin{figure}[h!]
\centering
\begin{subfigure}{0.35\textwidth}
  \centering
  \includegraphics[width= \textwidth]{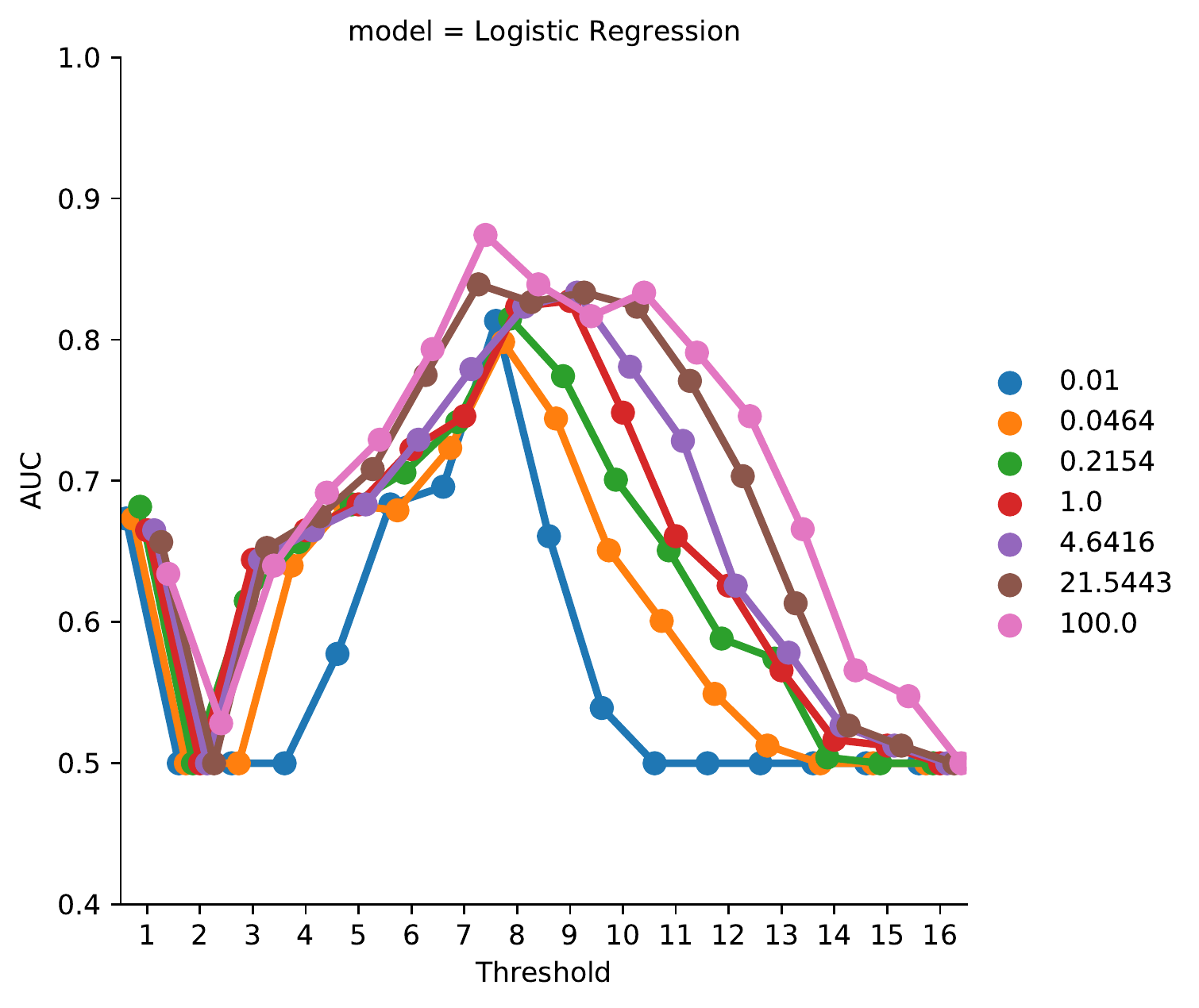}
\caption{}
\label{fig:ProbScalingMV-LR}
\end{subfigure}%
\begin{subfigure}{0.6\textwidth}
  \centering
  \includegraphics[width= \textwidth]{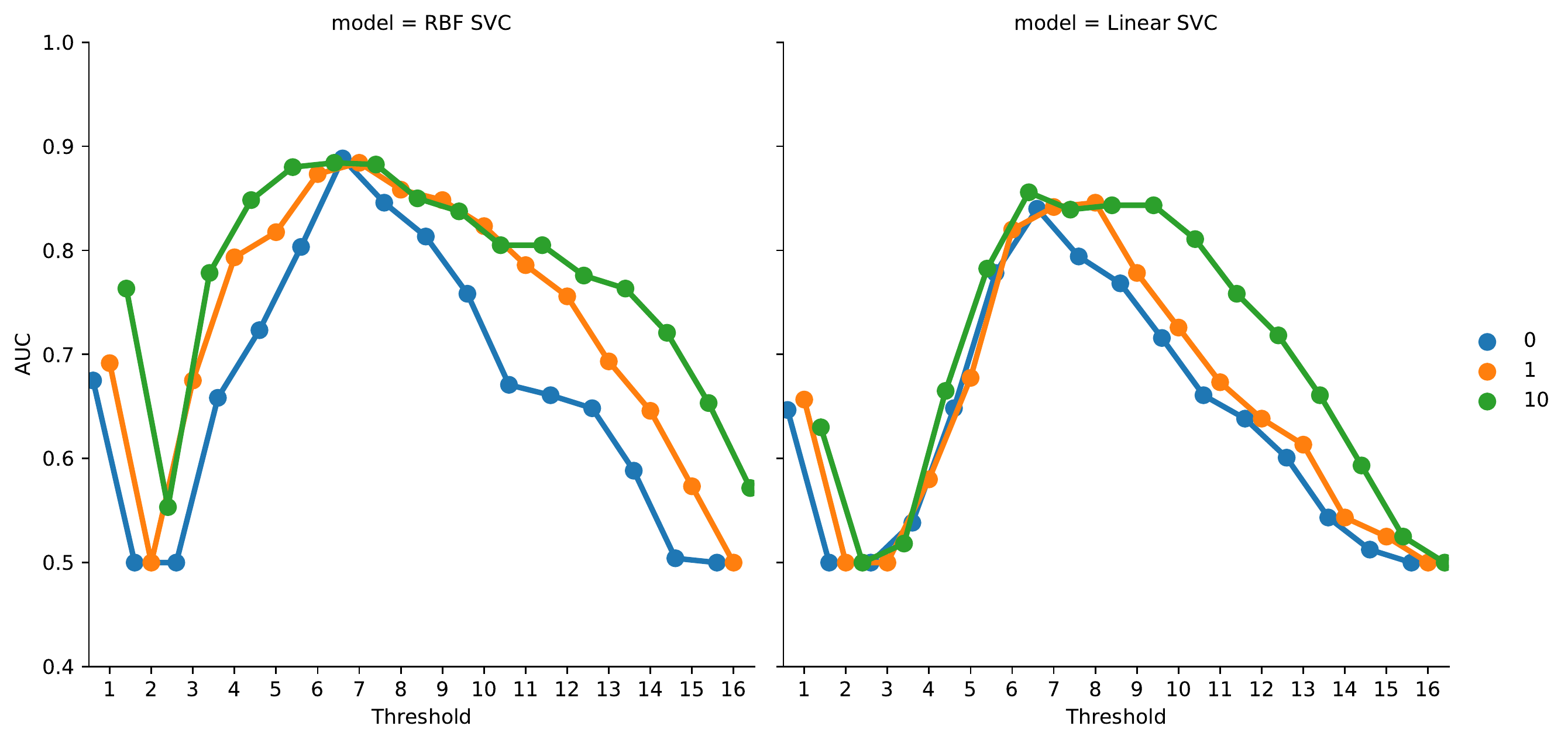}
\caption{}
\label{fig:ProbScalingMV-SVC}
\end{subfigure}%
\\
\begin{subfigure}{.99\textwidth}
  \centering
  \includegraphics[width= \textwidth]{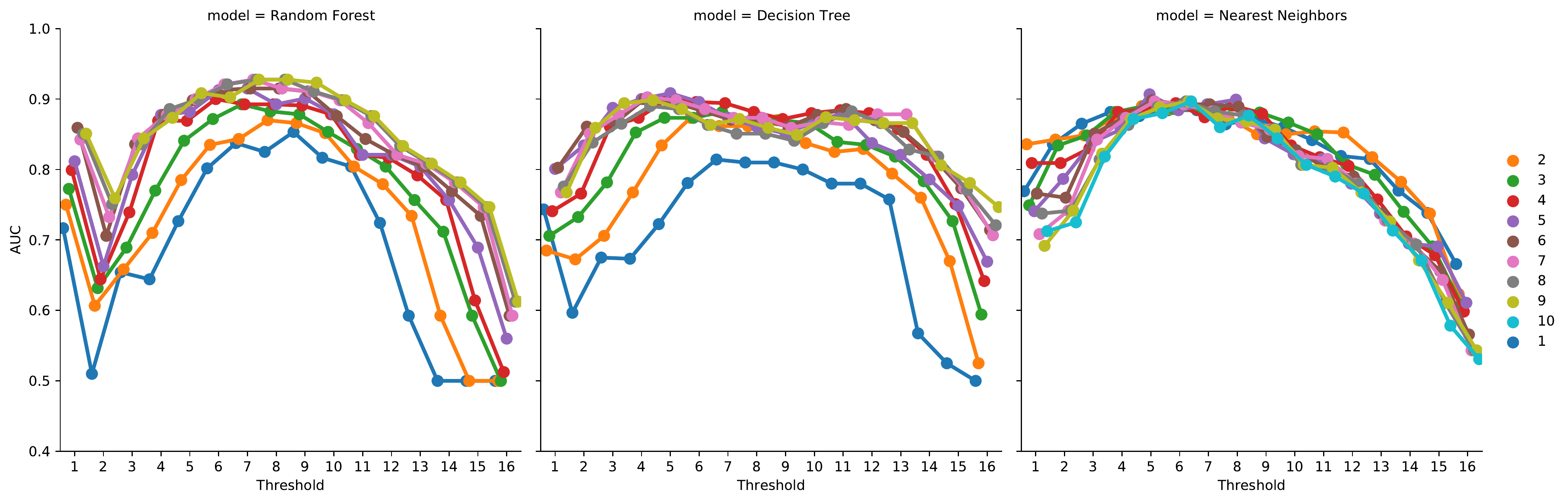}
\caption{}
\label{fig:ProbScalingMV-RF}
\end{subfigure}%
\caption{AUC for models with the continuous scaling variant of KB integration and wHMV strategy as multi-perspective CRS prediction strategies, stratified by the hyperparameter values a) Logistic Regression model; b) RBF SVC and Linear SVC models; c) Random Forest, Decision Tree and Nearest Neighbors models}
\label{fig:ContinuousScalingwHMV}
\end{figure}

\begin{figure}[h!]
\centering
\begin{subfigure}{0.35\textwidth}
  \centering
  \includegraphics[width= \textwidth]{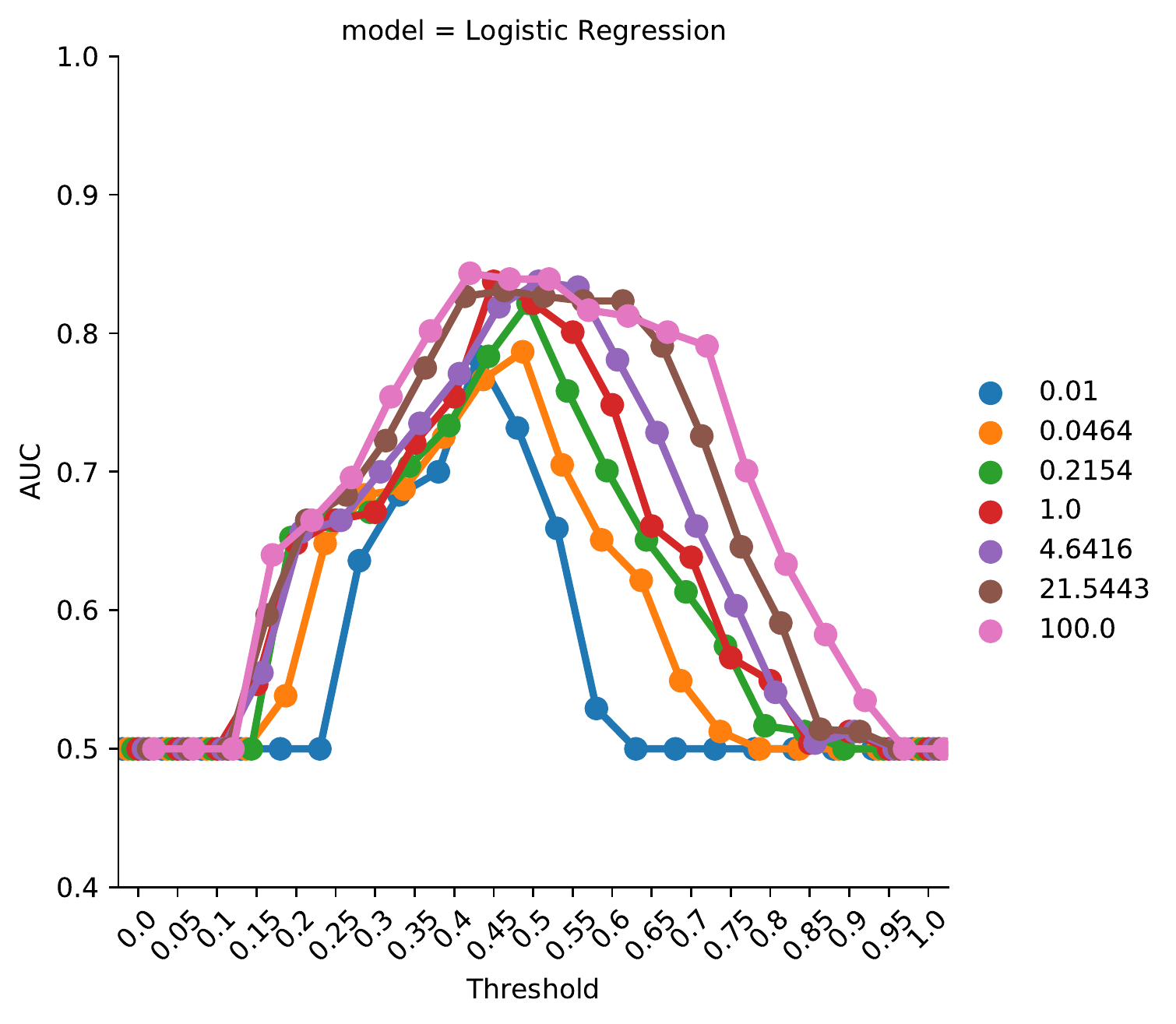}
\caption{}
\label{fig:ProbScalingMV-LR}
\end{subfigure}%
\begin{subfigure}{0.6\textwidth}
  \centering
  \includegraphics[width= \textwidth]{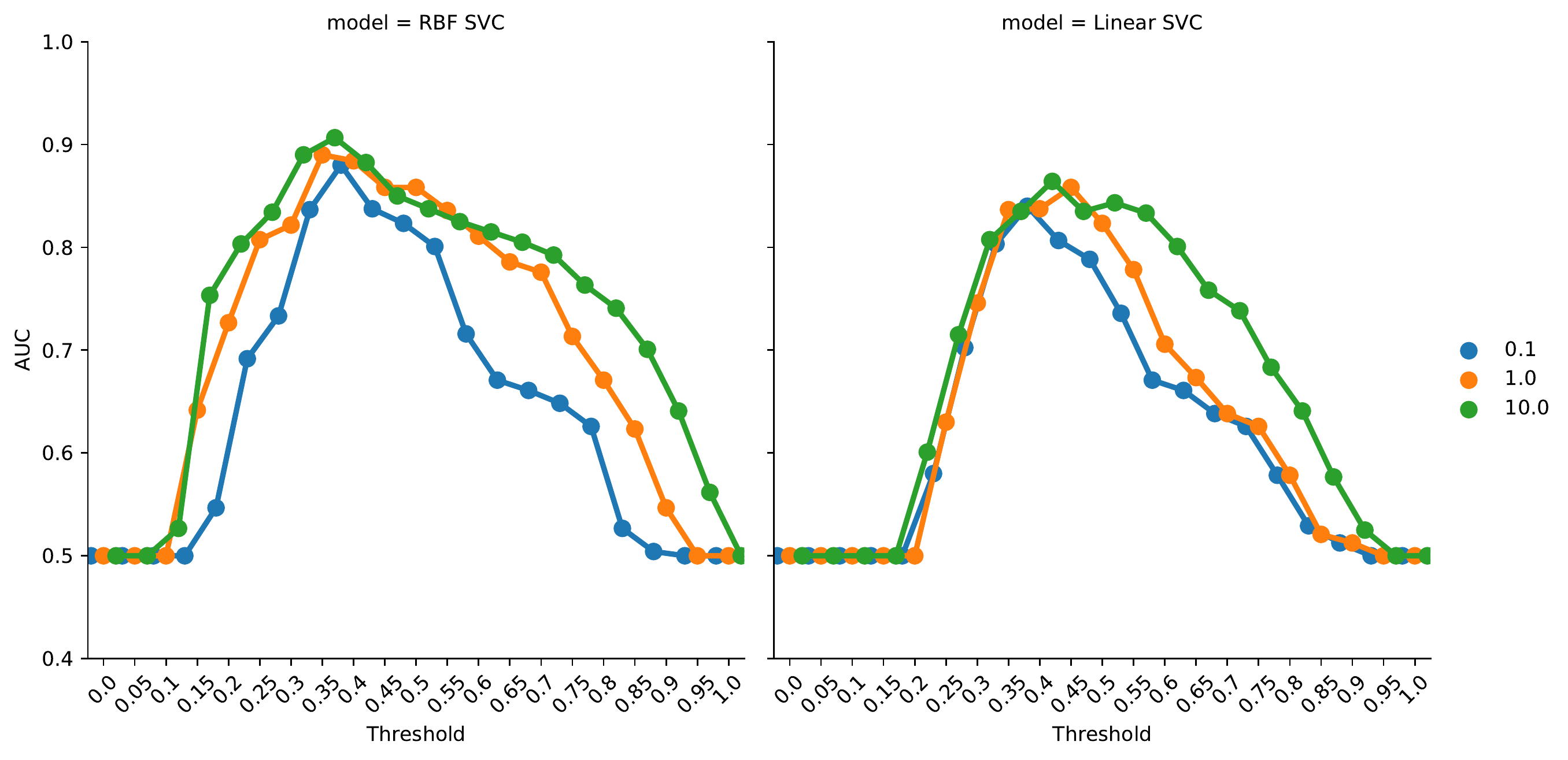}
\caption{}
\label{fig:ProbScalingMV-SVC}
\end{subfigure}%
\\
\begin{subfigure}{.99\textwidth}
  \centering
  \includegraphics[width= \textwidth]{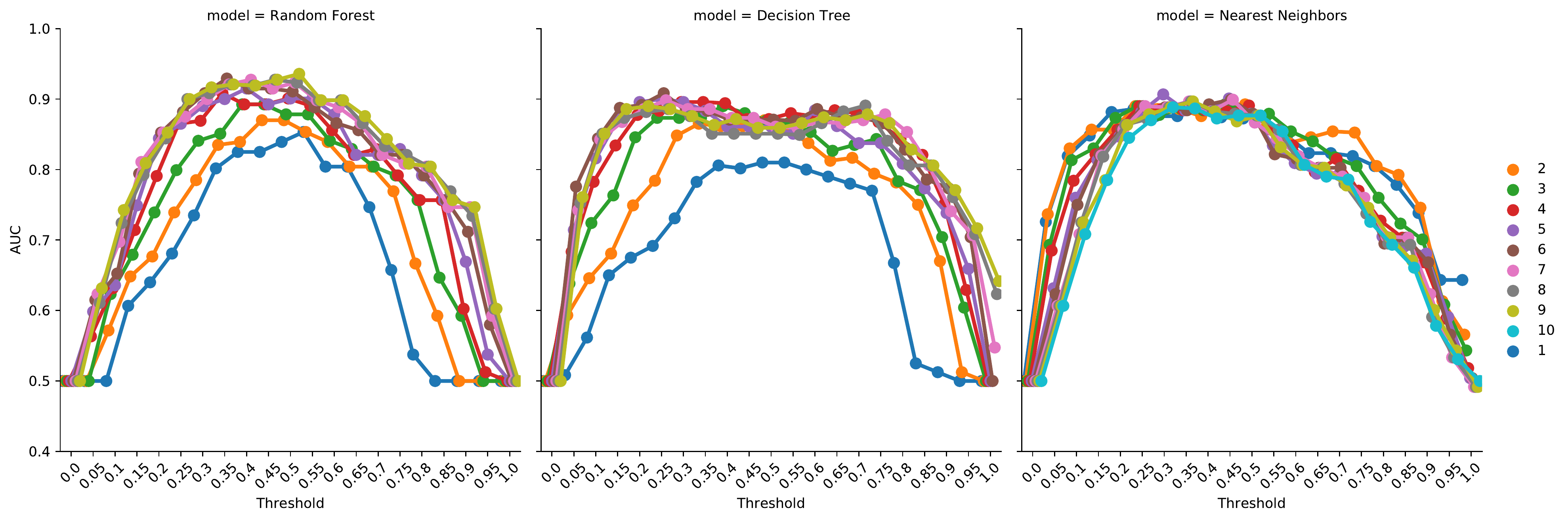}
\caption{}
\label{fig:ProbScalingMV-RF}
\end{subfigure}%
\caption{AUC for models with the continuous scaling variant of KB integration and SMV strategy as multi-perspective CRS prediction strategies, stratified by the hyperparameter values a) Logistic Regression model; b) RBF SVC and Linear SVC models; c) Random Forest, Decision Tree and Nearest Neighbors models}
\label{fig:ContinuousScalingSMV}
\end{figure}

\begin{figure}[h!]
\centering
\begin{subfigure}{0.35\textwidth}
  \centering
  \includegraphics[width= \textwidth]{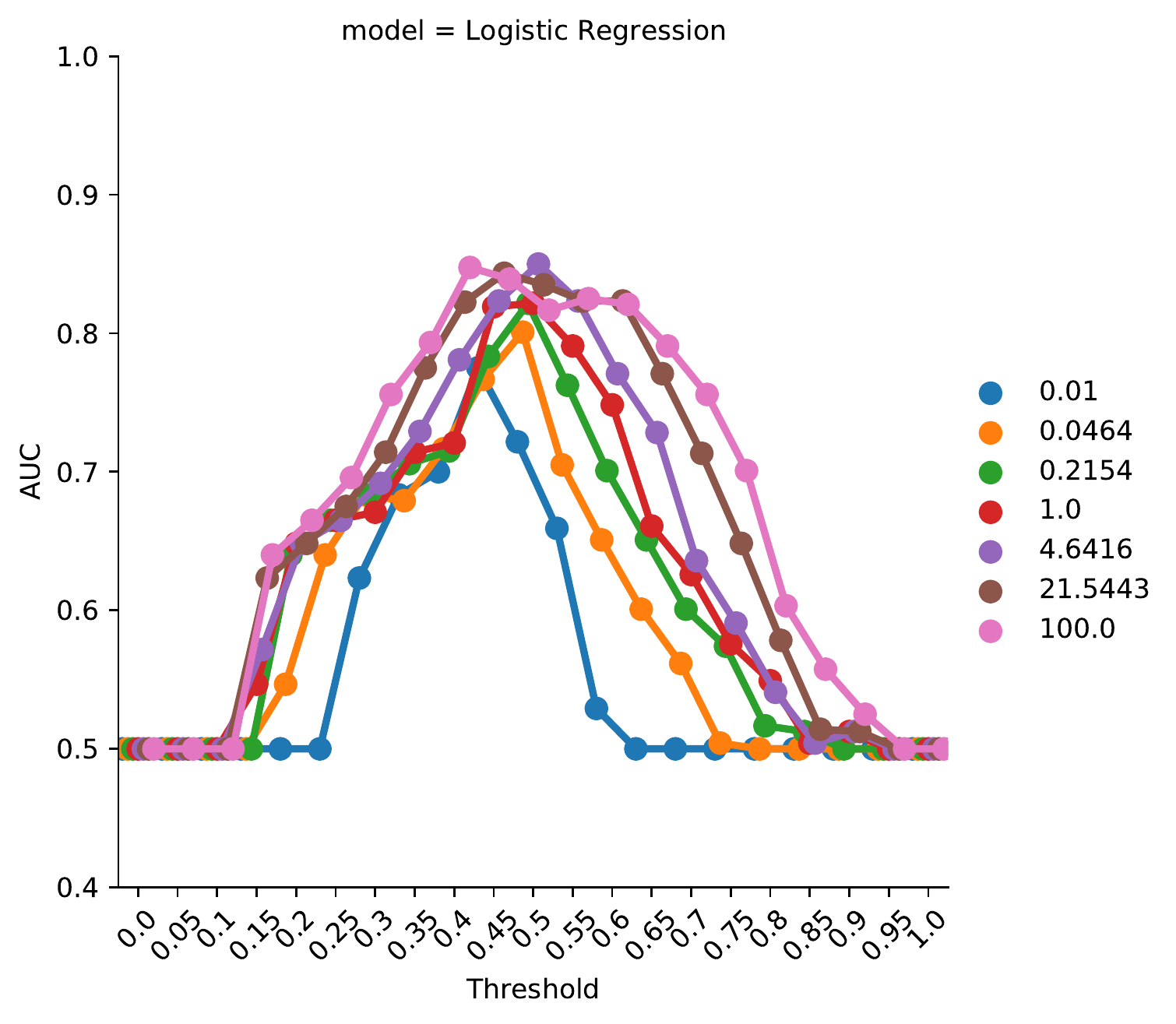}
\caption{}
\label{fig:ProbScalingMV-LR}
\end{subfigure}%
\begin{subfigure}{0.6\textwidth}
  \centering
  \includegraphics[width= \textwidth]{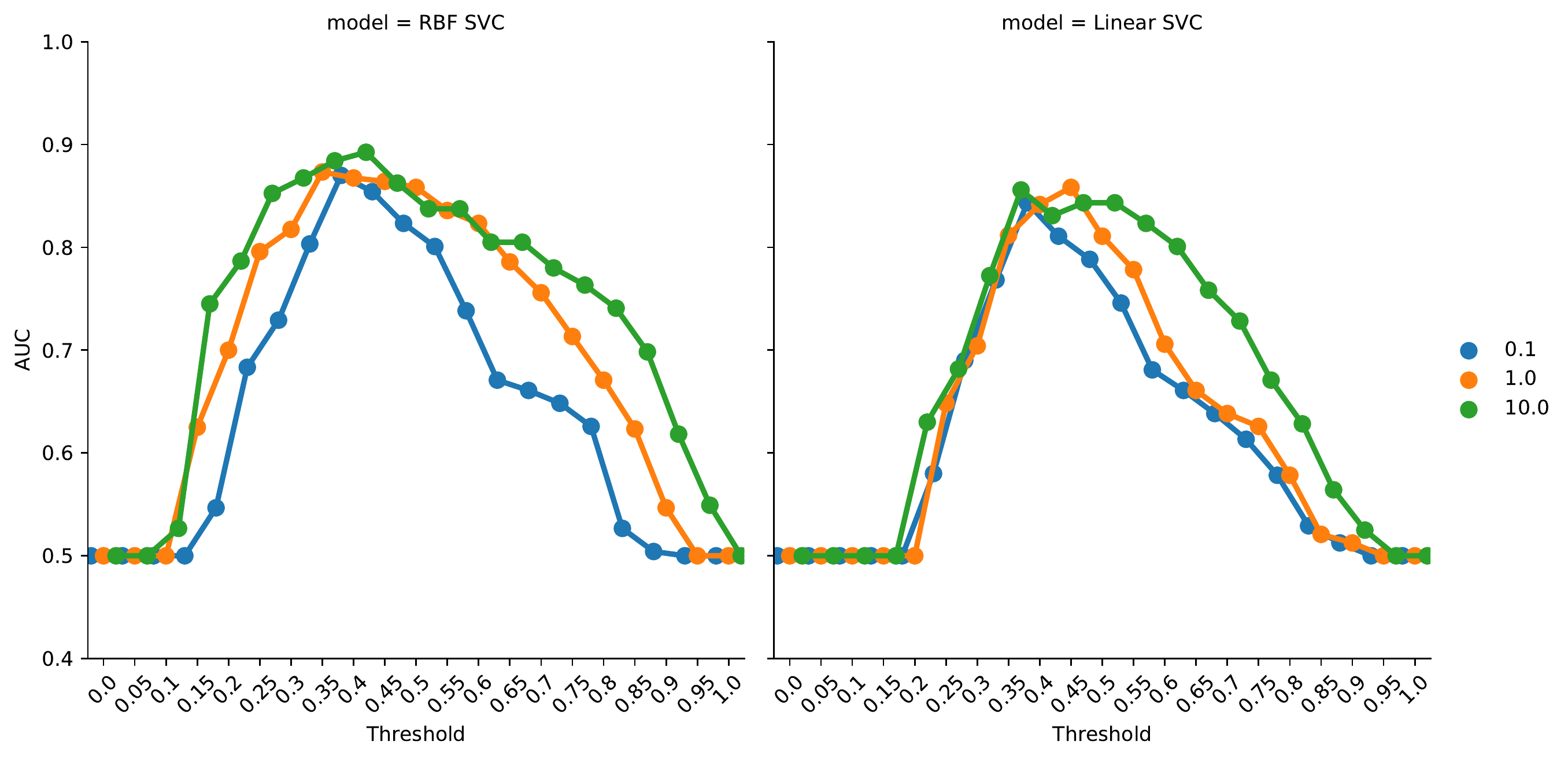}
\caption{}
\label{fig:ProbScalingMV-SVC}
\end{subfigure}%
\\
\begin{subfigure}{.99\textwidth}
  \centering
  \includegraphics[width= \textwidth]{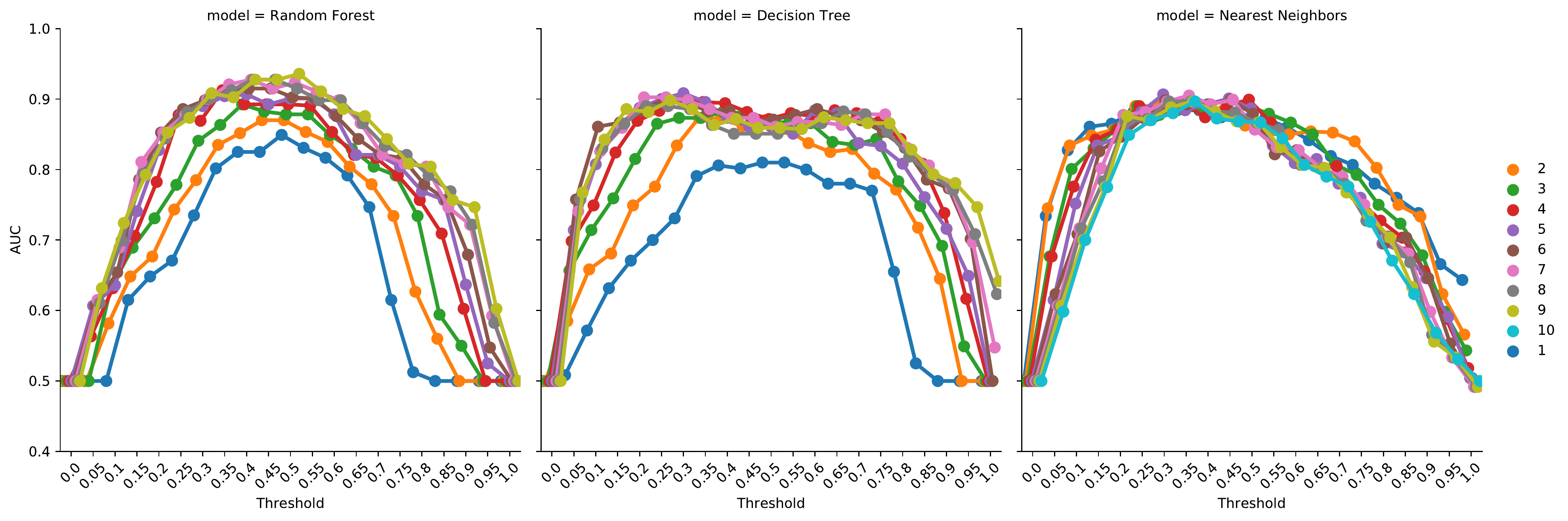}
\caption{}
\label{fig:ProbScalingMV-RF}
\end{subfigure}%
\caption{AUC for models with the continuous scaling variant of KB integration and wSMV strategy as multi-perspective CRS prediction strategies, stratified by the hyperparameter values a) Logistic Regression model; b) RBF SVC and Linear SVC models; c) Random Forest, Decision Tree and Nearest Neighbors models}
\label{fig:ContinuousScalingwSMV}
\end{figure}

Hyperparameters and thresholds considered in model derivation are summarized in Suppl. Tables \ref{tab:ClassifierComparisonNoScaling}, \ref{tab:ClassifierComparisonProbScaling}, \ref{tab:ClassifierComparisonBinaryScaling}, \ref{tab:ClassifierComparisonContinuousScaling} and Suppl. Figures \ref{fig:ProbScalingHMV}-\ref{fig:ContinuousScalingwSMV}.

\textbf{\textit{Literature-informed ML model} 'M2--CRS-CRS' }

Using the 102 timepoints, we have integrated this data with the knowledge base data by scaling the points across 17 papers available in KB, which resulted in 1734 datapoints.Using these data points, we performed Grouped Stratified 10-Folds cross-validation (train/test split ratio 90/10). 102 groups were assigned corresponding to 102 timepoints.
This ensures that each 17 datapoints coming from a single datapoint are not split between train and test sets. 
As a result of this process, 10 datasets were obtained with a training size of 1564 and test size 170.

%Next, we investigated seven candidate algorithms:  Random Forest (RF), logistic regression (LR), Support Vector Classifier (SVC) (RBF  and linear kernels), Decision Tree (DT), K-Nearest Neighbour (KNN) and Naive Bayes (NB). For each algorithm, we tuned hyperparameters: maximum depth for RF and DT, \textit{k} neighbours for KNN, \textit{C} for Linear SVC, RBF SVC and LR. Furthermore, as the prediction for a single timepoint is an aggregate of 17 predictions we investigated 4 aggregation strategies: majority voting (MV) and weighted majority voting (WMV) which use 17 binary predictions for voting; probabilities (P) and weighted probabilities (WP) which use 17 probabilities output from the classifier. 

%Using the predicted score as a determinant of a binary outcome, i.e. CRS vs. no CRS, we calculated Area Under the Receiver Operating Characteristics (AUC) and sensitivity (recall). Performance and 

In our experiments we used cohort size $N$ in a study as the study weight, following the formula:
\begin{equation} \label{eq:study_size_weight}
w_j = \frac{log_{10}(N_j)}{\sum_{j=1,e} (w_j)}
\end{equation}
where $w_j$ - weight of prediction in multi-view voting, $N_j$ - cohort size in study $j$.

\textbf{\textit{Knowledge integration and model derivation}}

Tuning hyperparameters based on AUC, we determined max depth for RF and DT, n neighbours for NN, C for Linear SVC, RBF SVC and LR; for all scaling variants and aggregated decision strategies (Suppl. Fig. \ref{fig:ProbScalingHMV}-\ref{fig:ContinuousScalingwSMV}). %For all models, the learning curves flattened at ≈50\% of the training set suggesting that the current size of the dataset provides sufficient model accuracy. 
The model using only exemplar data, without integration of knowledge from $KB$, is treated as the baseline. The RF, DT and NN achieved the highest AUC for all scaling variants compared to other models (Suppl. Table \ref{tab:ClassifierComparisonProbScaling} \ref{tab:ClassifierComparisonBinaryScaling}
\ref{tab:ClassifierComparisonContinuousScaling}).

\textbf{\textit{Threshold derivation and establishment of the final model}}

Therefore the metareview--informed RF model after probabilistic scaling of data and soft majority voting strategy was selected to proceed to final model development (Table \ref{tab:PerformanceRandomForest}). The increasing importance of features used in the final model are shown in Fig. \ref{fig:shap}, in which <> were considered as contributing the most to CRS prediction. 

%\textbf{\textit{Final model}}

%To evaluate the models, we performed Stratified K-Folds cross-validation on the dataset with imputed values from the model validation phase. In cross-validation, the model was trained X times, each time using a different observation as the testing set, and the remaining as the training set. As a result, we obtained the result/not aggregate output for each observation in the dataset, predicted by a model trained on the training set size.
%Based on the AUC, RF was selected for further development. 
%SHAP values from the final model were investigated via a feature contribution summary plot and dependency plots with detailed contributions of each training point.
%Although we defined thresholds in this decision-support tool, the continuous model score will be presented to the user online as the main model output, followed by a recommendation derived from the threshold. Continuous predictions will thus allow for more refined decision-making at an individual patient level and combined with the calibration plot, will deliver more information about the model’s performance and limitations to the user of the tool (i.e. the healthcare professional).

\newpage
\subsection{Supplementary Tables} \label{subsec:SuppleTables}

% -----------------------------------------------------------------------------------------------------------------------------------------
% Characteristics of the included studies

% \usepackage{booktabs}
% \usepackage{lscape}
% \usepackage{longtable}
% Note: It may be necessary to compile the document several times to get a multi-page table to line up properly
\fontsize{7}{11}\selectfont{
\begin{landscape}
\begin{longtable}[c]{@{}rLrLLLLLLLM@{}}
\caption[tab]{Characteristics of the included studies.} \\   
\label{tab:StudiesCharacteristics}\\
\toprule
\multicolumn{1}{l}{\textbf{}} & \textbf{Reference}                          & \textbf{Year} & \textbf{Source}        & \textbf{Study}         & \textbf{Trial Phase} & \textbf{Patients evaluated, N*} & \textbf{Cancer type(s)}                       & \textbf{CAR antigen target}    & \textbf{Co-stim domain} & \textbf{CRS grading scale}                                     \\* \midrule
\endhead
\bottomrule
\endfoot
\endlastfoot
1                               & Jacobson et al. \cite{jacobsonAxicabtageneCiloleucelRelapsed2022} & 2022          & Lancet Oncol           & Clinical Trial: ZUMA-5 & II        & 148                           & r/r indolent NHL                       & CD19                       & ND                            & Lee et al. 2014 criteria                                       \\
2                               & Hong et al. \cite{hongClinicalCharacterizationRisk2021}     & 2021          & Bone Marrow Transplant & Chinese Clinical Trial & R**          & 41                            & ALL, MM, NHL (COVID-19 - not included) & ND                         & ND                            & CTCAE v 5.0                                                    \\
3                               & Yan et al. \cite{yanCharacteristicsRiskFactors2021}      & 2021          & Front. Immunol.        & Clinical Trial         &           & 142                           & r/r ALL, Lymphoma, MM                  & CD19, CD19+BCMA, CD19+CD20 & ND                            & Lee et al. 2014 criteria                                       \\
4                               & Topp et al. \cite{toppEarlierCorticosteroidUse2021}     & 2021          & BJHaem                 & Clinical Trial: ZUMA-4 &          & 41                            & DLBCL, PMBCL, TFL, HGBCL               & CD19                       & ND                            & Modified criteria of Lee and colleagues                        \\
5                               & Shah et al. \cite{shahKTEX19RelapsedRefractory2021}     & 2021          & Lancet                 & Clinical Trial: ZUMA-3 & II        & 55                            & r/r B-ALL                              & CD19                       & ND                            & Lee et al. 2014 criteria                                       \\
6                               & Liu et al. \cite{liuNovelDominantnegativePD12021}      & 2021          & Translational Oncology & Chinese Clinical Trial & I        & 9                             & DLBCL, TFL, FL                         & CD19                       & ND                            & Lee et al. 2014 criteria                                       \\
7                               & Sang et al. \cite{sangPhaseIITrial2020}     & 2020          & Cancer Med.            & Clinical Trial         & II        & 21                            & r/r DLBCL                              & CD19, CD20                 & 4-1BB/CD3                   & Lee et al. 2014 criteria                                       \\
8                               & Yan et al. \cite{yanCombinationHumanisedAntiCD192019}      & 2019          & Lancet Haematol        & Clinical Trial         & II        & 21                            & MM                                     & CD19, BCMA                 & 4-BB, 4-1BB                   & Modified criteria of Lee and colleagues and NCI CTCAE v4.03    \\
9                               & Zhao et al. \cite{zhaoPhaseOpenlabelStudy2018}     & 2018          & J Hematol Oncol        & Clinical Trial         & I          & 57                            & MM                                     & BCMA                       & 4-1BB                         & Modified criteria of Lee and colleagues and NCI CTCAE v4.03    \\
10                              & Neelapu et al. \cite{neelapuAxicabtageneCiloleucelCAR2017}   & 2017          & N Eng J Med            & Clinical Trial: ZUMA-1 & II          & 111                           & DLBCL, PMBCL, TLF                      & CD19                       & CD28                          & Lee et al. 2014 criteria‡                                      \\
11                              & Hay et al. \cite{hayKineticsBiomarkersSevere2017}      & 2017          & Blood                  & Clinical Trial         & I/II          & 133                           & r/r B-ALL, CLL, NHL                    & CD19                       & 4-1BB                         & Lee et al. 2014 criteria                                       \\
12                              & Turtle et al. \cite{turtleDurableMolecularRemissions2017}   & 2017          & J Clin Oncol           & Clinical Trial         & I/II          & 24                            & CLL                                    & CD19                       & ND                            & Modified criteria of Lee and colleagues and NCI CTCAE v4.03    \\
13                              & Hu et al. \cite{huPotentAntileukemiaActivities2017}       & 2017          & Clin Cancer Res        & Chinese Clinical Trial         &           & 15                            & r/r ALL                                & CD19                       & 4-1BB/CD3                   & Modified criteria of Lee and colleagues and NCI CTCAE v4.03    \\
14                              & Teachey et al. \cite{teacheyIdentificationPredictiveBiomarkers2016}  & 2016          & Cancer Discov          & Clinical Trial         &           & 51                   & r/r ALL                                & CD19                       & 4-1BB                         & Custom CRS grading scale (Modified criteria of Lee and Davila) \\
15                              & Porter et al.  \cite{porterChimericAntigenReceptor2015}   & 2015          & Sci Transl Med         & Pilot Clinical Trial   &           & 14                            & r/r CLL                                & CD19                       & 4-1BB/CD3                   & Penn Grading System                                            \\
16                              & Davila et al. \cite{davilaEfficacyToxicityManagement2014}   & 2014          & Sci Transl Med         & Clinical Trial         & I          & 16                            & r/r B-ALL                              & CD19                       & ND                            & Davila et al. criteria                                         \\
17                              & Kalos et al. \cite{kalosCellsChimericAntigen2011}    & 2011          & Leukemia               & Pilot Clinical Trial         &           & 3                             & CLL                                    & CD19                       & 4-1BB                         & ND                                                             \\* \bottomrule
\end{longtable}
*Population included in the clinical trials (Teachey et al. included children in the cohort)

Abbreviations: B-ALL: B-cell acute lymphoblastic leukemia, DLBCL: diffuse large B-cell lymphoma, CLL: chronic lymphocytic leukemia, NHL: non-hodgkin lymphoma. R/R: relapsed/refractory, ALL: acute lymphocytic leukemia, Multiple cytokine elevation, three or more cytokines elevated with levels 10 folds higher than the baseline level, MSKCC, Memorial Sloan-Kettering Cancer Center

**Retrospectively registered
\end{landscape}
}

% -----------------------------------------------------------------------------------------------------------------------------------------

% Patients specific parameters
% \usepackage{booktabs}
% \usepackage{lscape}
% \usepackage{longtable}
% Note: It may be necessary to compile the document several times to get a multi-page table to line up properly
%\fontsize{7}{11}\selectfont{
%\begin{landscape}
\begin{longtable}[c]{@{}rlLLMLN@{}}
\caption[tab]{Patients specific parameters} \\   
\label{tab:PatientsParameters}\\
\toprule
\multicolumn{1}{l}{\textbf{}} & \textbf{Study}                          & \textbf{Patient Cohort Size, N} & \textbf{Patient Cohort Male, n (\%)} & \textbf{Patient Cohort Age (A), year median (range/IQR*)} & \textbf{CRS, n (\%)} & \textbf{Severe or grade 3+ CRS**, n (\%)} \\* \midrule
\endhead
\bottomrule
\endfoot
\endlastfoot
1                               & Jacobson et al. \cite{jacobsonAxicabtageneCiloleucelRelapsed2022} & 148                             & 84 (57)                              & 61 (53-68)*                                                & 121 (82)                         & 10 (8)                                             \\
2                               & Hong et al. \cite{hongClinicalCharacterizationRisk2021}     & 41                              & 21 (51)                              & 51 (32.5 - 62.5)*                                          & 41 (100)                        & 41 (100)**                                           \\
3                               & Yan et al. \cite{yanCharacteristicsRiskFactors2021}      & 142                             & 87 (61)                              & 45 (24 - 59)*                                              & 123 (87)                         & 30 (24)                                            \\
4                               & Topp et al. \cite{toppEarlierCorticosteroidUse2021}     & 41                              & 28 (68)                              & 61 (19 - 77)                                               & 38 (93)                         & 1 (2.6)                                           \\
5                               & Shah et al. \cite{shahKTEX19RelapsedRefractory2021}     & 55                              & 33 (60)                              & 40 (28 - 52)*                                              & 49 (89)                         & 13 (26.5)                                          \\
6                               & Liu et al. \cite{liuNovelDominantnegativePD12021}      & 9                               & 5 (56)                               & 51 (22 - 62)                                               & 9 (100)                        & 1 (11)                                            \\
\bottomrule                                        
\end{longtable}
**The present study only included patients with sCRS.
%\end{landscape}
%}

% -----------------------------------------------------------------------

% \usepackage{booktabs}
% \usepackage{longtable}
% Note: It may be necessary to compile the document several times to get a multi-page table to line up properly
\begin{longtable}[c]{@{}rlcccccccccc@{}}
\caption[tab]{Characteristic of missing values (ranges) in the $KB$.} \\   
\label{tab:KB-MissingValues}\\
\toprule
\multicolumn{1}{l}{} & Study                                    & IL2 & IL4 & IL6 & IL8 & IL10 & IL15 & IL2R$\alpha$ & TNF--$\alpha$ & IFN--$\gamma$ & GM-CSF \\* \midrule
\endhead
\bottomrule
\endfoot
\endlastfoot
1                    & Jacobson et al. \cite{jacobsonAxicabtageneCiloleucelRelapsed2022} & R   & R   & R   & R   & R    & R    & R     & R   & R   & R       \\
2                    & Hong et al. \cite{hongClinicalCharacterizationRisk2021}     & R   & MV  & R   & MV  & R    & MV   & MV    & R   & R   & MV      \\
3                    & Yan et al. \cite{yanCharacteristicsRiskFactors2021}      & MV  & MV  & R   & MV  & MV   & MV   & MV    & MV  & MV  & MV      \\
4                    & Topp et al. \cite{toppEarlierCorticosteroidUse2021}     & R   & R   & R   & R   & R    & R    & R     & R   & R   & R       \\
5                    & Shah et al. \cite{shahKTEX19RelapsedRefractory2021}     & MV  & MV  & R   & R   & R    & R    & R     & R   & R   & R       \\
6                    & Liu et al. \cite{liuCytokineReleaseSyndrome2018}      & R   & R   & R   & MV  & R    & MV   & MV    & R   & R   & MV      \\
7                    & Sang et al. \cite{sangPhaseIITrial2020}     & MV  & MV  & R   & MV  & MV   & MV   & MV    & MV  & R   & MV      \\
8                    & Yan et al. \cite{yanCombinationHumanisedAntiCD192019}      & MV  & MV  & R   & MV  & MV   & MV   & MV    & MV  & MV  & MV      \\
\bottomrule
\end{longtable}
Abbreviations: R - reported, MV - missing value

% \usepackage{booktabs}
% \usepackage{longtable}
% Note: It may be necessary to compile the document several times to get a multi-page table to line up properly
\fontsize{7}{11}\selectfont{
\begin{longtable}[c]{@{}lrrrrrr@{}}
\caption[tab]{Comparison of performance of data-driven ML models} \\   
\label{tab:ClassifierComparisonNoScaling}\\
\toprule
Model               & \multicolumn{1}{l}{\begin{tabular}[c]{@{}l@{}}Selected \\ hyperparameter\end{tabular}} & \multicolumn{1}{l}{\begin{tabular}[c]{@{}l@{}}Precision \\ mean (SD)\end{tabular}} & \multicolumn{1}{l}{\begin{tabular}[c]{@{}l@{}}Recall \\ mean (SD)\end{tabular}} & \multicolumn{1}{l}{\begin{tabular}[c]{@{}l@{}}F1-score \\ mean (SD)\end{tabular}} & \multicolumn{1}{l}{\begin{tabular}[c]{@{}l@{}}AUC\\ mean (SD)\end{tabular}} & \multicolumn{1}{l}{\begin{tabular}[c]{@{}l@{}}Accuracy \\ mean (SD)\end{tabular}} \\* \midrule
\endhead
\bottomrule
\endfoot
\endlastfoot
Random Forest       & max\_depth = 4                                                                         & \textbf{0.960 (0.084)}                                                                      & 0.795 (0.297)                                                                   & \textbf{0.830 (0.212)}                                                                     & \textbf{0.881 (0.139)}                                                               & \textbf{0.893 (0.115)}                                                                     \\
Decision Tree       & max\_depth = 2                                                                         & 0.798 (0.311)                                                                      & 0.725 (0.346)                                                                   & 0.729 (0.291)                                                                     & 0.813 (0.153)                                                               & 0.825 (0.129)                                                                     \\
Nearest Neighbors   & n\_neighbors = 6                                                                       & 0.838 (0.311)                                                                      & 0.675 (0.382)                                                                   & 0.703 (0.339)                                                                     & 0.811 (0.176)                                                               & 0.834 (0.144)                                                                     \\
\bottomrule
\end{longtable}
}
% \usepackage{booktabs}
% \usepackage{multirow}
% \usepackage{longtable}
% Note: It may be necessary to compile the document several times to get a multi-page table to line up properly
\fontsize{7}{11}\selectfont{
{\setlength{\tabcolsep}{1.4pt}
\begin{longtable}[c]{@{}Lllrrrrrrr@{}}
\caption[tab]{Performance comparison of six algorithms using knowledge from KB with probabilistic scaling} \\   
\label{tab:ClassifierComparisonProbScaling}\\
\toprule
Model                                & Configuration                                                                                                           & \begin{tabular}[c]{@{}l@{}}Majority Voting \\ strategy\end{tabular} & \multicolumn{1}{l}{\begin{tabular}[c]{@{}l@{}}Selected \\ hyperparameter\end{tabular}} & \multicolumn{1}{l}{\begin{tabular}[c]{@{}l@{}}Best \\ threshold\end{tabular}} & \multicolumn{1}{l}{\begin{tabular}[c]{@{}l@{}}Precision \\ mean (SD)\end{tabular}} & \multicolumn{1}{l}{\begin{tabular}[c]{@{}l@{}}Recall \\ mean (SD)\end{tabular}} & \multicolumn{1}{l}{\begin{tabular}[c]{@{}l@{}}F1-score \\ mean (SD)\end{tabular}} & \multicolumn{1}{l}{\begin{tabular}[c]{@{}l@{}}AUC \\ mean (SD)\end{tabular}} & \multicolumn{1}{l}{\begin{tabular}[c]{@{}l@{}}Accuracy \\ mean (SD)\end{tabular}} \\* \midrule
\endhead
\bottomrule
\endfoot
\endlastfoot
Random Forest      & \begin{tabular}[c]{@{}l@{}}max\_depth = 2-10, step = 1 \\ threshold = 1-16, step = 1\end{tabular}                       & HMV                                                                   & max\_depth = 8                                                                         & 5                                                                             & 0.938 (0.101)                                                                      & 0.880 (0.174)                                                                   & 0.896 (0.115)                                                                     & 0.915 (0.09)                                                                 & 0.921 (0.079)                                                                     \\
                                     &                                                                                                                         & wHMV                                                                  & max\_depth = 10                                                                        & 5                                                                             & 0.858 (0.167)                                                                      & 0.925 (0.121)                                                                   & 0.888 (0.141)                                                                     & 0.904 (0.123)                                                                & 0.901 (0.125)                                                                     \\
\bottomrule
\end{longtable}
Majority Voting strategy: HMV - hard majority voting; wHMV - weighted hard majority voting; SMV - soft majority voting; wSMV - weighted soft majority voting
}
}
\normalsize

% \usepackage{booktabs}
% \usepackage{longtable}
% Note: It may be necessary to compile the document several times to get a multi-page table to line up properly
\fontsize{7}{11}\selectfont{
{\setlength{\tabcolsep}{1.4pt}
\begin{longtable}[c]{@{}Llllllllll@{}}
\caption[tab]{Performance copmarison of six algorithms using knowledge from KB with binary scaling} \\   
\label{tab:ClassifierComparisonBinaryScaling}\\
\toprule
Model               & Configuration                                                                                                           & \begin{tabular}[c]{@{}l@{}}Majority Voting \\ strategy\end{tabular} & \begin{tabular}[c]{@{}l@{}}Selected \\ hyperparameter\end{tabular} & \begin{tabular}[c]{@{}l@{}}Best \\ threshold\end{tabular} & \begin{tabular}[c]{@{}l@{}}Precision \\ mean (SD)\end{tabular} & \begin{tabular}[c]{@{}l@{}}Recall\\ mean (SD)\end{tabular} & \begin{tabular}[c]{@{}l@{}}F1-score \\ mean (SD)\end{tabular} & \begin{tabular}[c]{@{}l@{}}AUC \\ mean (SD)\end{tabular} & \begin{tabular}[c]{@{}l@{}}Accuracy \\ mean (SD)\end{tabular} \\* \midrule
\endhead
\bottomrule
\endfoot
\endlastfoot
Random Forest       & \begin{tabular}[c]{@{}l@{}}max\_depth = 2-10, step = 1 \\ threshold = 1-16, step = 1\end{tabular}                       & HMV                                                                   & max\_depth = 8                                                     & \multicolumn{1}{r}{9}                                     & 0.832 (0.178)                                                  & 0.83 (0.204)                                               & 0.820 (0.172)                                                 & 0.845 (0.142)                                            & 0.851 (0.136)                                                 \\
                    &                                                                                                                         & wHMV                                                                  & max\_depth = 10                                                    & \multicolumn{1}{r}{7}                                     & 0.810 (0.163)                                                  & \textbf{0.900 (0.175)}                                                & 0.845 (0.145)                                                 & 0.863 (0.119)                                            & 0.862 (0.116)                                                 \\
\bottomrule
\end{longtable}
}
}
\normalsize

% \usepackage{booktabs}
% \usepackage{longtable}
% Note: It may be necessary to compile the document several times to get a multi-page table to line up properly
\fontsize{7}{11}\selectfont{
{\setlength{\tabcolsep}{1.4pt}
\begin{longtable}[c]{@{}Llllllllll@{}}
\caption[tab]{Performance comparison of six algorithms using knowledge from KB with continuous scaling} \\   
\label{tab:ClassifierComparisonContinuousScaling}\\
\toprule
Model               & Configuration                                                                                                           & \begin{tabular}[c]{@{}l@{}}Majority \\Voting\end{tabular} & \begin{tabular}[c]{@{}l@{}}Selected \\ hyperparameter\end{tabular} & \begin{tabular}[c]{@{}l@{}}Best \\ threshold\end{tabular} & \begin{tabular}[c]{@{}l@{}}Precision \\ mean (SD)\end{tabular} & \begin{tabular}[c]{@{}l@{}}Recall \\ mean (SD)\end{tabular} & \begin{tabular}[c]{@{}l@{}}F1-score \\ mean (SD)\end{tabular} & \begin{tabular}[c]{@{}l@{}}AUC \\ mean (SD)\end{tabular} & \begin{tabular}[c]{@{}l@{}}Accuracy \\ mean (SD)\end{tabular} \\* \midrule
\endhead
\bottomrule
\endfoot
\endlastfoot
Random Forest       & \begin{tabular}[c]{@{}l@{}}max\_depth = 2-10, step = 1 \\ threshold = 1-16, step = 1\end{tabular}                       & HMV                                                                   & max\_depth = 10                                                    & \multicolumn{1}{r}{8}                                     & \textbf{0.963 (0.078)}                                                  & 0.905 (0.123)                                               & \textbf{0.926 (0.066)}                                                 & \textbf{0.936 (0.057)}                                            & \textbf{0.941 (0.051)}                                                 \\
                    &                                                                                                                         & wHMV                                                                  & max\_depth = 8                                                     & \multicolumn{1}{r}{7}                                     & 0.938 (0.101)                                                  & 0.905 (0.123)                                               & 0.915 (0.084)                                                 & 0.928 (0.07)                                             & 0.931 (0.067)                                                 \\
\bottomrule
\end{longtable}
}
}
\normalsize

\newpage
\subsection{Supplementary Figures} \label{subsec:SuppleFigures}

%%%%%%%%%%%%%%%%%%%%%%% probabilistic scaling
%% MV

%% WMV

%% P

%% WP

%%%%%%%%%%%%%%%%%%%%%%% binary scaling
%% MV

%% WMV

%% P

%% WP

%%%%%%%%%%%%%%%%%%%%%%% continuous scaling
%% MV

%% WMV

%% P

%% WP

%%%%%%%%%%%%%%%%%%%%%%% boxplots data exploration
\begin{figure}[htb!]
\centering
\includegraphics[width= .8\textwidth]{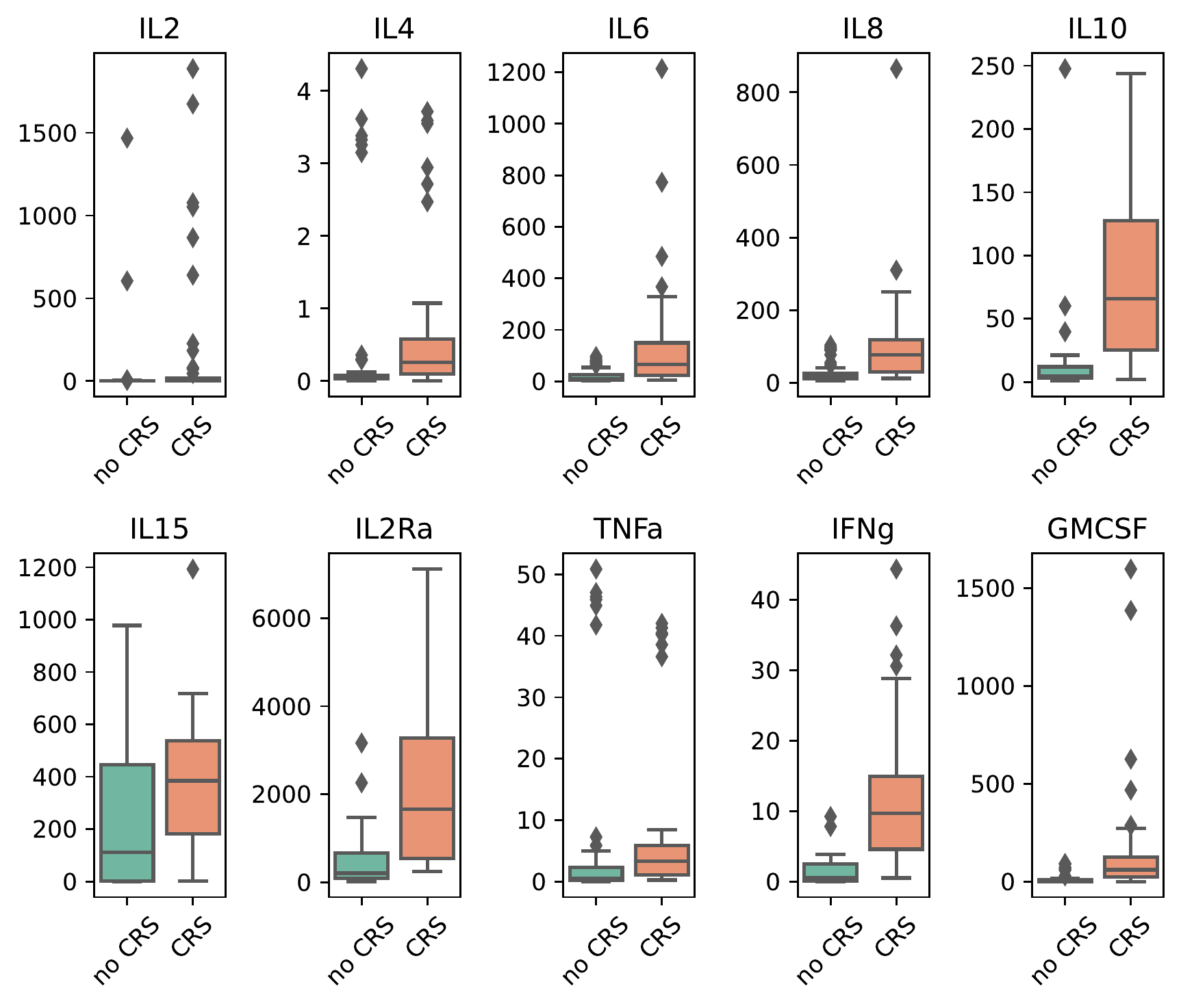}
\caption{Biomarkers for 102 datapoints stratified by no--CRS and CRS}
\label{fig:boxplots_exploration}
\end{figure}

%%%%%%%%%%%%%%%%%%%%%%% voting strategy example
%\begin{figure}[htb!]
%\centering
%\includegraphics[width= %.8\textwidth]{figures/voting_strategies_example.pdf}
%\caption{n - no. studies in KB}
%\label{fig:voting_strategies_example}
%\end{figure}

%%%%%%%%%%%%%%%%%%%%%%% SHAP abs values

\begin{figure}
\centering
\includegraphics[width= .8\textwidth]{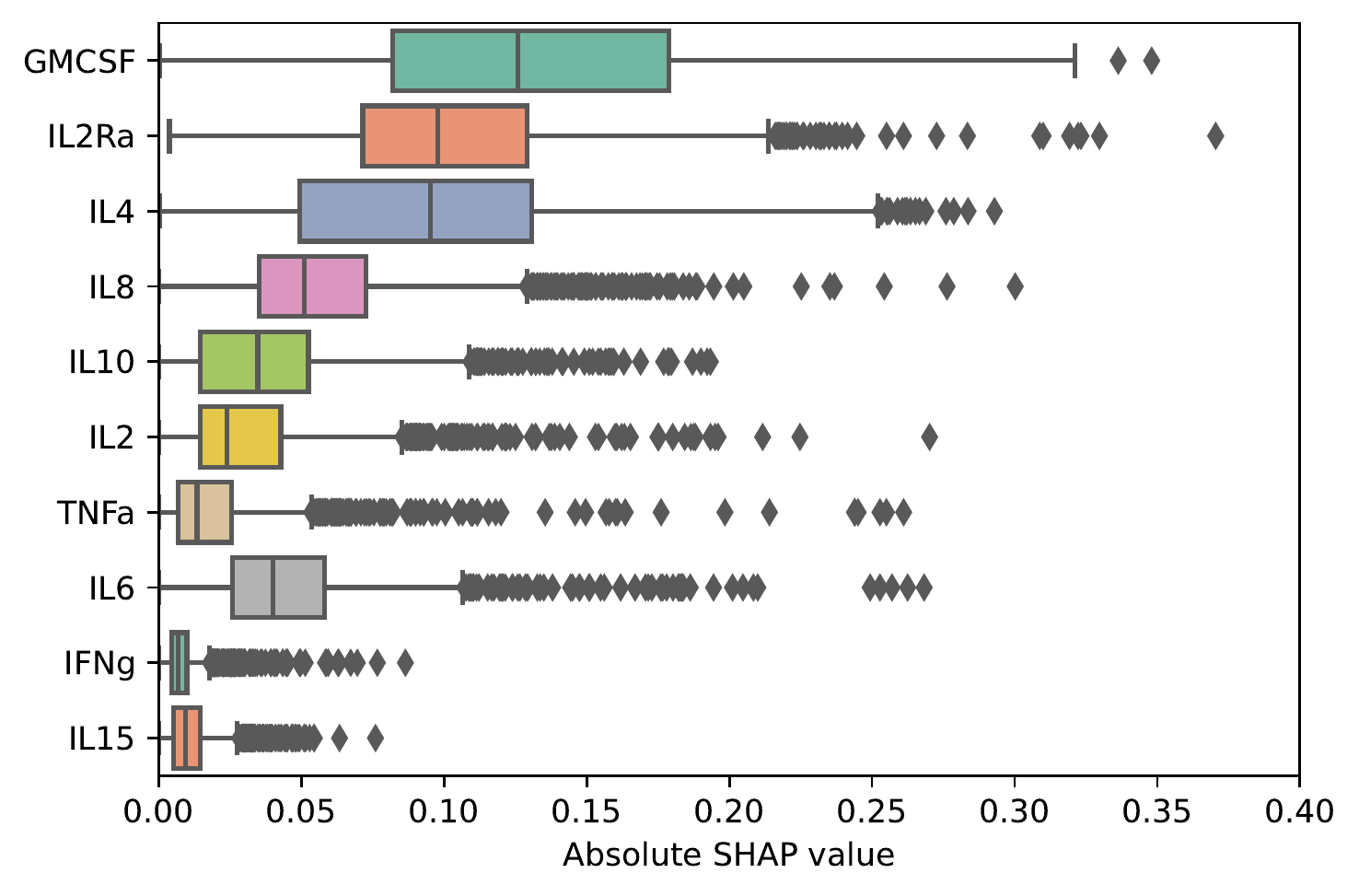}
\caption{SHAP explanation for RF model with continuous scaling variant of KB integration}
\label{fig:boxplots_shap_minmax}
\end{figure}

\begin{figure}
\centering
\includegraphics[width= .8\textwidth]{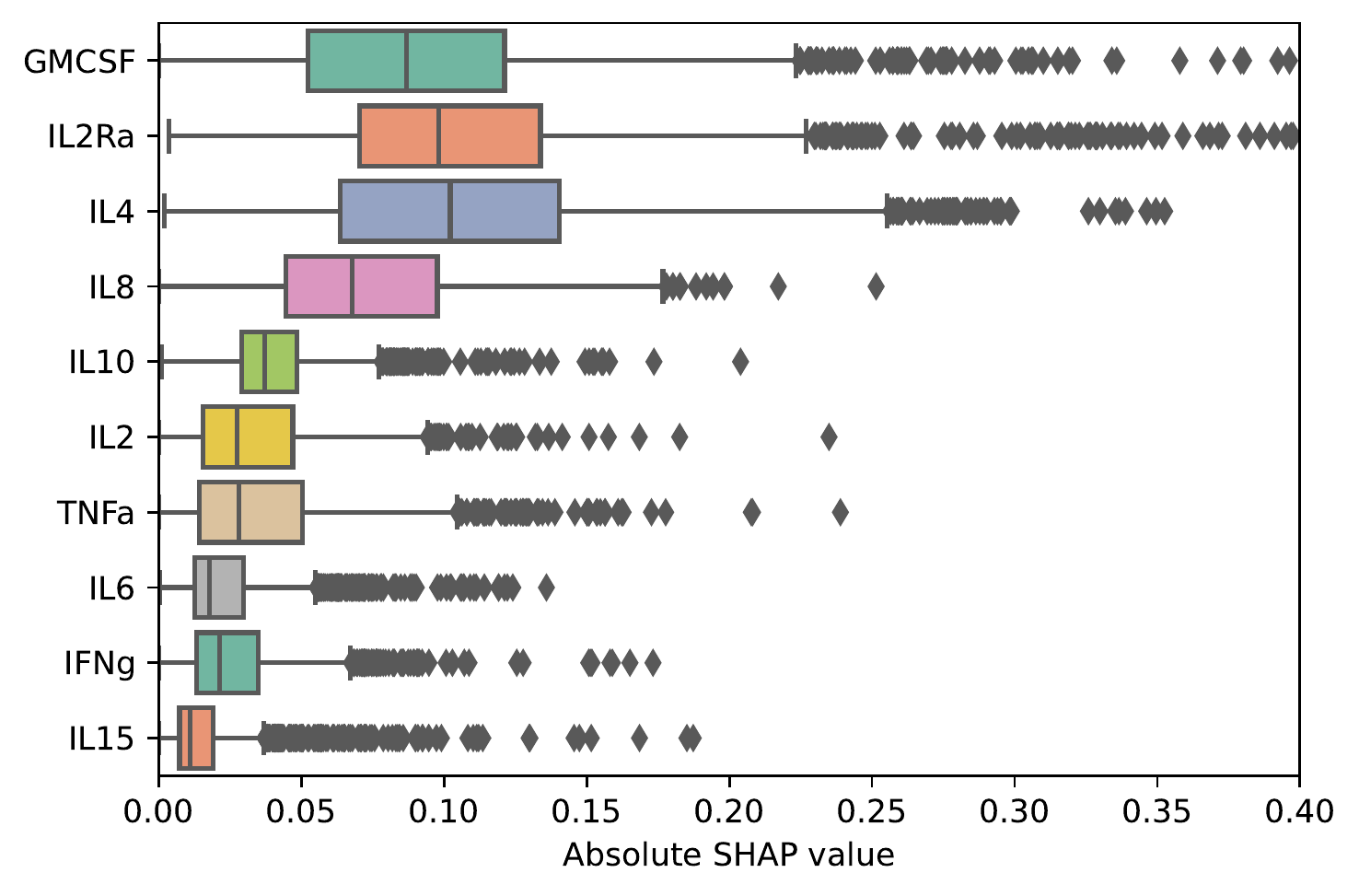}
\caption{SHAP explanation for RF model with binary scaling variant of KB integration}
\label{fig:boxplots_shap_binary}
\end{figure}

%%%%%%%%%%%%%%%%%%%%%%% SHAP dependency plots

\begin{figure}
\centering
\includegraphics[width= .8\textwidth]{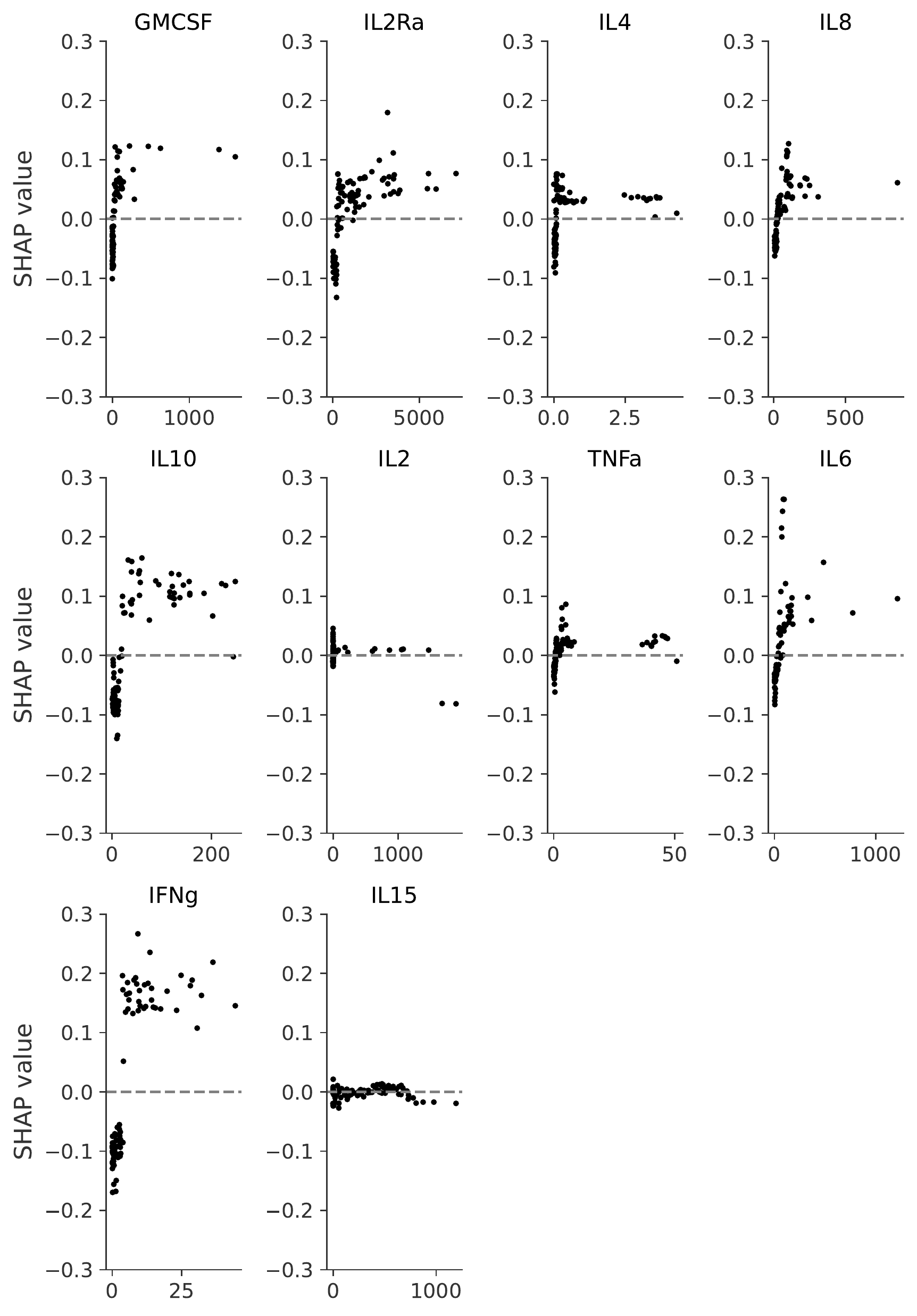}
\caption{SHAP explanation dependency plot for probabilistic scaling}
\label{fig:boxplots_shap_prob}
\end{figure}

\begin{figure}
\centering
\includegraphics[width= .8\textwidth]{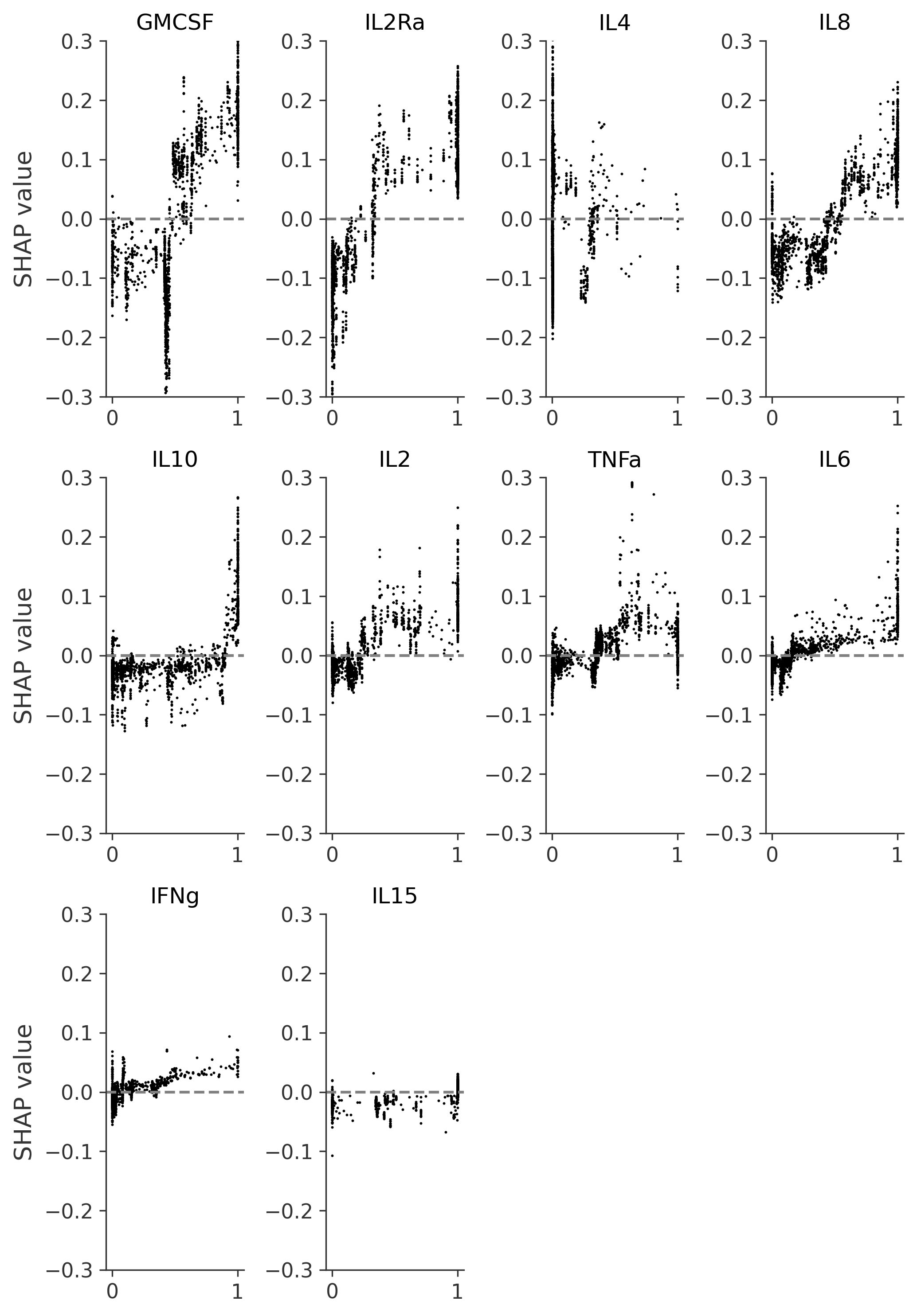}
\caption{SHAP explanation dependency plot  for probabilistic scaling}
\label{fig:boxplots_shap_prob}
\end{figure}

\begin{figure}
\centering
\includegraphics[width= .8\textwidth]{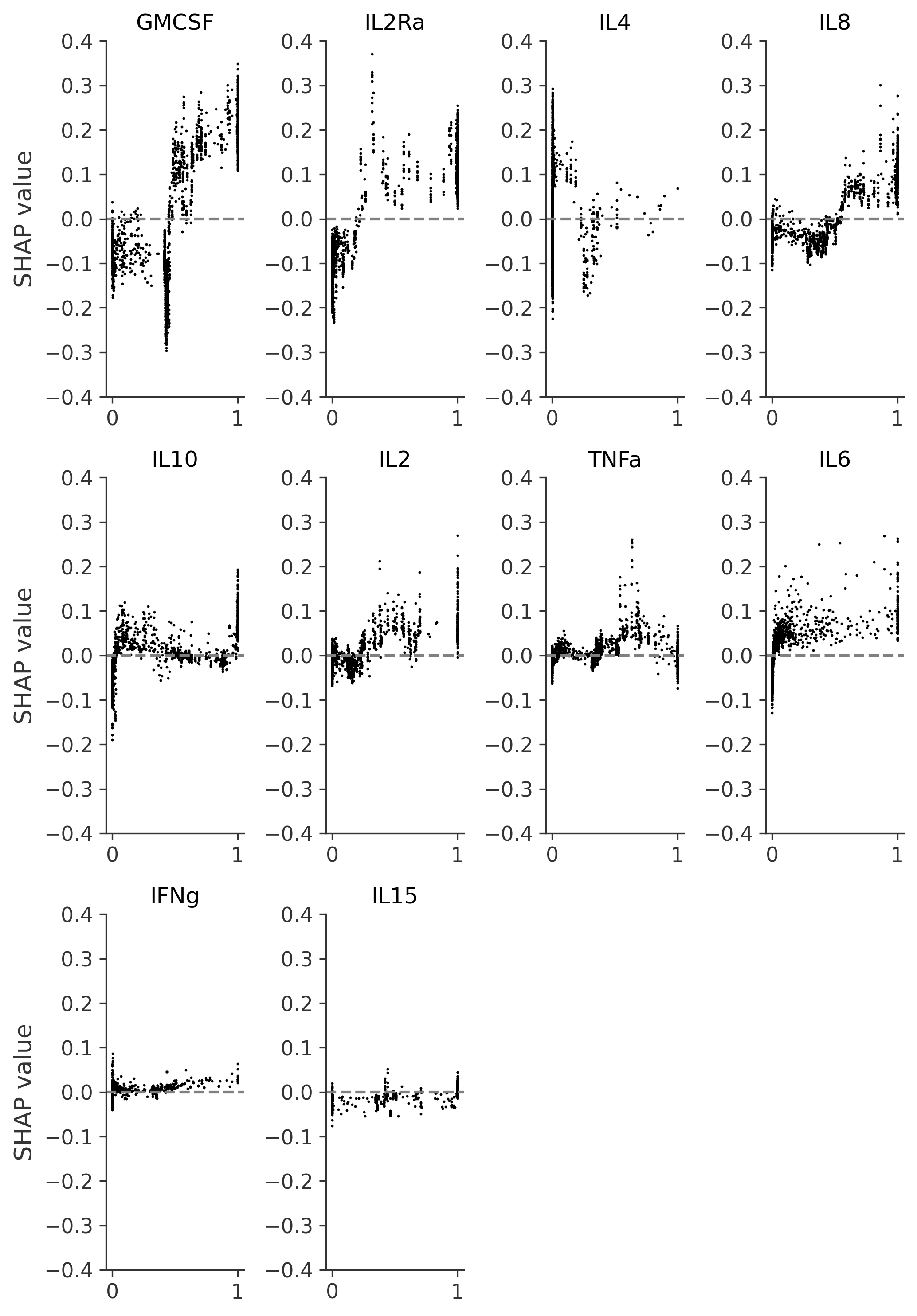}
\caption{SHAP explanation dependency plot  for continuous scaling}
\label{fig:boxplots_shap_prob}
\end{figure}

\begin{figure}
\centering
\includegraphics[width= .8\textwidth]{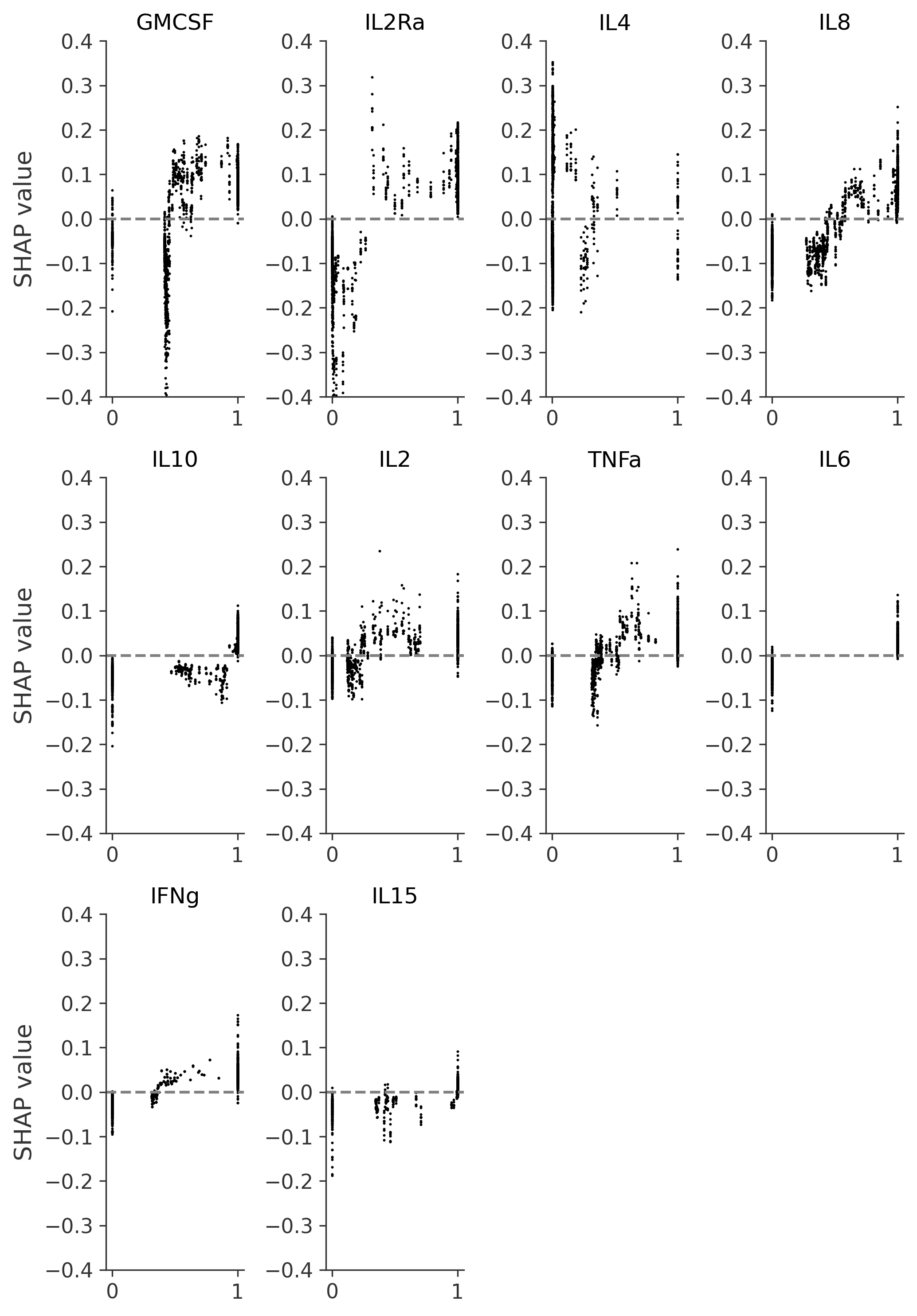}
\caption{SHAP explanation dependency plot for binary scaling}
\label{fig:boxplots_shap_prob}
\end{figure}

\end{document}